# Implication of modelling choices on connectivity estimation: A comparative analysis

Marie Soret[a,b], Maxime Lenormand[a]*, Sylvain Moulherat[b], Sandra Luque[a]

[a] *TETIS, Univ Montpellier, AgroParisTech, CIRAD, CNRS, INRAE, Montpellier, France*

[b]*OïkoLab TerrOïko, 2 place Dom Devic, Sorèze, 81540, Occitanie, France*

* *Corresponding author: maxime.lenormand@inrae.fr*

# Abstract

Landscape connectivity is an important field with important conservation implications. Connectivity modelling is a useful tool to inform and guide landscape planning. However, it involves assumptions and methodological decisions which ultimately impact connectivity outcomes. In order to understand the implications of modelling choices on final connectivity estimations, we compare two landscape characterisation approaches – expert knowledge and species distribution models – and three movements models – least-cost paths, circuit theory and an individual-based movement simulator. The implementation of the models and the construction of the analyses scope highlighted conceptual and methodological differences that made the comparison difficult. Landscape characterisation appears as the principal factor determining connectivity outcomes. Therefore, the confrontation between expert knowledge and species distribution models is critical to leverage points of convergence and complementarity between these two approaches. Conceptual differences between movement models are reported on connectivity map and habitat patch contribution estimations. In the want of data and protocol design specifically to validate connectivity models, approaches that integrate stochastic and behavioural processes, bring a more realistic perspective to connectivity estimation.

# 1. Introduction

Changes in land use significantly contribute to global biodiversity erosion (IPBES 2019). Anthropic land planning such as urbanisation, infrastructure development, and agriculture intensification induce a progressive artificialisation of natural and semi-natural landscapes (Gibbs et al. 2010; Hooke and Martín-Duque 2012; Cox and Gaston 2023). Resultant landscape fragmentation jeopardises the survival of wild species (Foley et al. 2005; Haddad et al. 2015) in three ways: loss of habitat areas (Hanski 2011), habitat quality degradation (European Environment Agency 2020; Yohannes et al. 2021), and barriers to dispersal which reduce gene flow between populations (Caplat et al. 2016; Berger-Tal and Saltz 2019). In this context, suitable habitats tend to be reduced to small patches, spread within an unfavourable matrix (Haddad et al. 2017). To ensure species long-term survival, the landscape needs to provide shelter, food, and breeding resources across a network of interconnected habitats (Gyllenberg and Hanski 1992; Taylor et al. 1993; Fahrig 2001). Planning measures aiming to avoid and mitigate land development, or to restore and manage natural lands and ecological corridors, can support landscape and species conservation (Bennett 2003; Keeley et al. 2019; Hilty et al. 2020).

Connectivity studies provide a framework to explore the spatial interaction of individuals and genes in fragmented landscapes and inform landscape planning and conservation action (Hanski 1999; Cushman et al. 2013; Drake et al. 2022). Modelling approaches provide insights into potential connectivity under restrictive assumptions (Kool et al. 2013; Hunter-Ayad et al. 2020). It involves methodological decisions that either reflect documented ecological processes or a lack of knowledge or resources such as data, methodological framework, tools, computational power or memory (Calabrese and Fagan 2004; Grimm et al. 2006; Bodin and Norberg 2007; Urban et al. 2022). These choices can introduce ecological assumptions or rely on partial representations of species' requirements and the local environment (Urban et al. 2016). In both cases, the interpretative scope of the outcomes such as movement estimation and connectivity indices is restricted to the model purview (Hortal et al. 2015). However, these results play a crucial role in landscape planning; they guide the orientation and implementation of development projects, restoration initiatives and conservation actions (Sahraoui et al. 2021; Moulherat et al. 2023). For efficient and transparent landscape planning, in-depth research is needed to better understand the implications of modelling choices on connectivity assessments.

To model landscape connectivity, the landscape must first be described from the species' ecological perspective (Pearson 2002). *Landscape characterisation* aims to identify landscape elements and characterise their contribution to species' local subsistence (Pearson et al. 1996). Areas purported to meet the habitat requirements are designated *potential suitable habitats* (Wiens 1976). S*uitable habitats* comprise a combination of biotic and abiotic factors which determine the presence of the studied species (Chase and Leibold 2003; Soberón and Nakamura 2009). The remaining landscape represents *the matrix*, which can be composed of a single homogeneous land type or of a patchwork of diverse land types. Its composition influences the ability of individuals to move between suitable habitats (Forman and Godron 1981; 1986). The elements of the matrix are characterised by resistance costs which represent the willingness, the physiological cost and the reduction in survival for an individual to cross the landscape unit (Sawyer, Epps, and Brashares 2011; Zeller et al. 2012). The definition and delimitation of the suitable habitats and matrix varies depending on the species and the spatiotemporal scale of the study (Turner, et al. 1989; Wiens 1989; Pearson et al. 1996). *Landscape characterisation* methods can be classified as either *knowledge-* or *data-driven* approaches (Kumar et al. 2022). We introduce a synthesis of definitions below to help frame our study:

*Knowledge-driven approaches* rely on the expertise of taxonomic experts, gathered from published literature or specifically for the study following expert opinion protocols (Perera et al. 2012; McBride and Burgman 2012). These approaches are advantageous as they use and consolidate existing information (Pearman-Gillman et al. 2020). They are also an efficient way to make up for a lack of data (Martin et al. 2005) and are easy to implement using spreadsheets and basic GIS processing (Reza et al. 2013). Their downsides include subjectivity, and equivocality, and may lack relevance to the reality of the field (Drescher et al. 2013). Their usefulness may also be limited when subject-specific expertise is sparse or unavailable (Johnson and Gillingham 2004).

*Data-driven approaches* exploit the relationship between species and environmental data acquired in the studied area (Guisan and Zimmermann 2000; Foltête et al. 2012; Préau et al. 2022). These approaches provide evidence-based insights into landscape characterisation, which is argued to enhance the efficiency and effectiveness of conservation actions. However, the quality of results lies in the quality and quantity of species and environmental data. Incomplete, biassed, or unrepresentative data can lead to inaccurate predictions (Guillera-Arroita 2017; Botella et al. 2020). Additionally, data-driven approaches may lack explanatory mechanisms for underlying ecological processes and instead focus on statistical correlations that do not necessarily represent the species' ecology (Pearson and Dawson 2003; Lee-Yaw et al. 2022).

Comparative analyses suggest that knowledge- and data-driven approaches yield different outcomes in landscape characterisation (Aizpurua et al. 2015; Di Febbraro et al. 2018; Godet and Clauzel 2021). However, these studies often rely on non-equivalent environmental data, complicating their ability to draw conclusions about the methods independently of the data sources. Despite this, the studies noted some convergence, indicating a relative compatibility between the two approaches. The lack of validation data and standardised protocols limits the accuracy assessment and comparison of both approaches (Holthausen et al. 1994; Godet and Clauzel 2021). Authors generally favoured data-driven methods when possible, but recommended using knowledge-driven approaches when empirical data is scarce or unavailable (Godet and Clauzel 2021). Knowledge-driven outcomes depend on the expert involved (Doswald, Zimmermann, and Breitenmoser 2007; Hurtado et al. 2023), whilst data-driven depend on the models used (Araújo and New 2007). In both cases, the use of a combination of several experts or models is recommended to ensure a more robust characterisation, reducing the impact of variability from any single expert or model.

Once landscape characterisation is completed, *connectivity estimation* predicts spatial interactions between suitable habitats (Rayfield et al. 2011). This relates to the movement ecology of animals (Baguette and Van Dyck 2007; Holyoak et al. 2008; Nathan et al. 2008), including daily movements associated with resource foraging, seasonal migrations and other significant dispersal events upon which the model assumptions are contingent (Morales et al. 2004; Bélisle 2005). Dispersal events emerge as important drivers of individual and gene flow, establishing a connection between landscape connectivity, metapopulation dynamics, and landscape genetics (Travis and Dytham 1999; Bowler and Benton 2005; Storfer et al. 2007; Clobert et al. 2012). The processes underlying connectivity hinge upon a combination of individuals' movement capacities, landscape structural elements (Kool et al. 2013), and external mortality risks such as predation and vehicular collisions (Medrano-Vizcaíno et al. 2023; Su and Yang 2023). Connectivity estimation can be approached through either *ecological isolation* (Moilanen and Nieminen 2002) or *individual-based modelling* (Grimm and Railsback 2005).

*Isolation-based* estimations suppose that the probability of connectivity between suitable habitats is related to the inverse of the ecological distance between them (Moilanen and Nieminen 2002). An ecological distance is an index of the habitat's remoteness relevant to the distance experienced by the studied species (Cushman et al. 2006). A straightforward measure of ecological distance is the *Euclidean distance* (Urban and Keitt 2001). It is used in fundamental theories, prerequisites of landscape ecology, such as *Isolation by distance* (Wright 1943; Slatkin 1993) and the insular biogeography model (MacArthur and Wilson 1963; Wilson and Willis 1975). However, the uniform landscape paradigm behind these theories is challenged by landscape ecologists (Taylor et al. 2006). The *minimal cumulative resistance*, also known as *least-cost path* (LCP), introduced by Knaapen et al. (1992) is an alternative measure which accounts for landscape variability (Adriaensen et al. 2003). While improving the Euclidean approach (Milanesi et al. 2017), this popular approach is limited by the restrictiveness of its ecological assumptions (Knaapen et al. 1992) and sensitivity to relative cost-surface values (Rayfield et al. 2010). Drawing from electrical circuit theory, McRae (2006) introduced *Isolation by resistance;* a resistance-based measure of isolation encompassing all possible paths. Resistance distance significantly outperformed Euclidean and LCP as predictor of genetics differentiation (McRae and Beier 2007). Additionally, current theory produces movement pattern convergent with aggregated bias random walkers (Doyle and Snell 1984), a more relevant movement assumption in connectivity applications (Nathan et al. 2008; Codling, Plank, and Benhamou 2008); and presents a low sensitivity to relative cost-surface values (Bowman et al. 2020).

*Individual-based models* (IBM) simulate animal movements between suitable habitats according to a set of rules. For instance, Palmer et al. (2011) presents the *stochastic movement simulator* (SMS), a biassed random walk involving additional parameters such as memory, perceptual range and movement auto-correlation. Other IBM explicitly incorporates other parameters such as mortality, departure, and settlement (Fletcher et al. 2023). IBM improves connectivity estimation (Coulon et al. 2015) and can be coupled with metapopulation modelling to perform viable population analysis (Moulherat 2014; Bocedi et al. 2014; Drake et al. 2022). The downsides of such models are their complexity given their increased demand for data and expert availability (Calabrese and Fagan 2004; Hunter-Ayad et al. 2020).

Kumar et al. (2022) used IBM as reference movement data to compare connectivity maps computed from LCP and circuit theory. They found out that circuit theory outperformed LCP but emphasised the limitation of these models regarding their simplistic movement assumptions. Despite their differences, LCP and circuit theory produce similar outcomes in conservation such as habitat prioritisation (Avon and Bergès 2016) which raises the question of complexity trade-offs regarding studies objective. Diniz et al. (2020) advocated for the use of more realistic movement models such as IBM whenever possible.

In this paper, we aim to compare the influence of landscape characterisation and connectivity estimation approaches on connectivity indices used in landscape planning. To our knowledge, we are the first study to compare the implication of these modelling choices consecutively in an integrative quantitative comparative scope. We used spatially explicit graphs to represent the spatial interactions between the reproductive habitats of the midwife toad across a heterogeneous agricultural landscape in rural western France. The midwife toad correspond to genus (*Alytes*) in the family *Alytidae* (formerly *Discoglossidae*), found in most of continental Europe and Northwestern Africa. We implemented two landscape characterisation approaches and four connectivity estimations models to address the various implications of these modelling choices on connectivity maps and indices. We anticipate that our results will offer insights into expected connectivity patterns and support the rationale for developing a connectivity modeling framework.

# 2. Methods

## 2.1 Study area

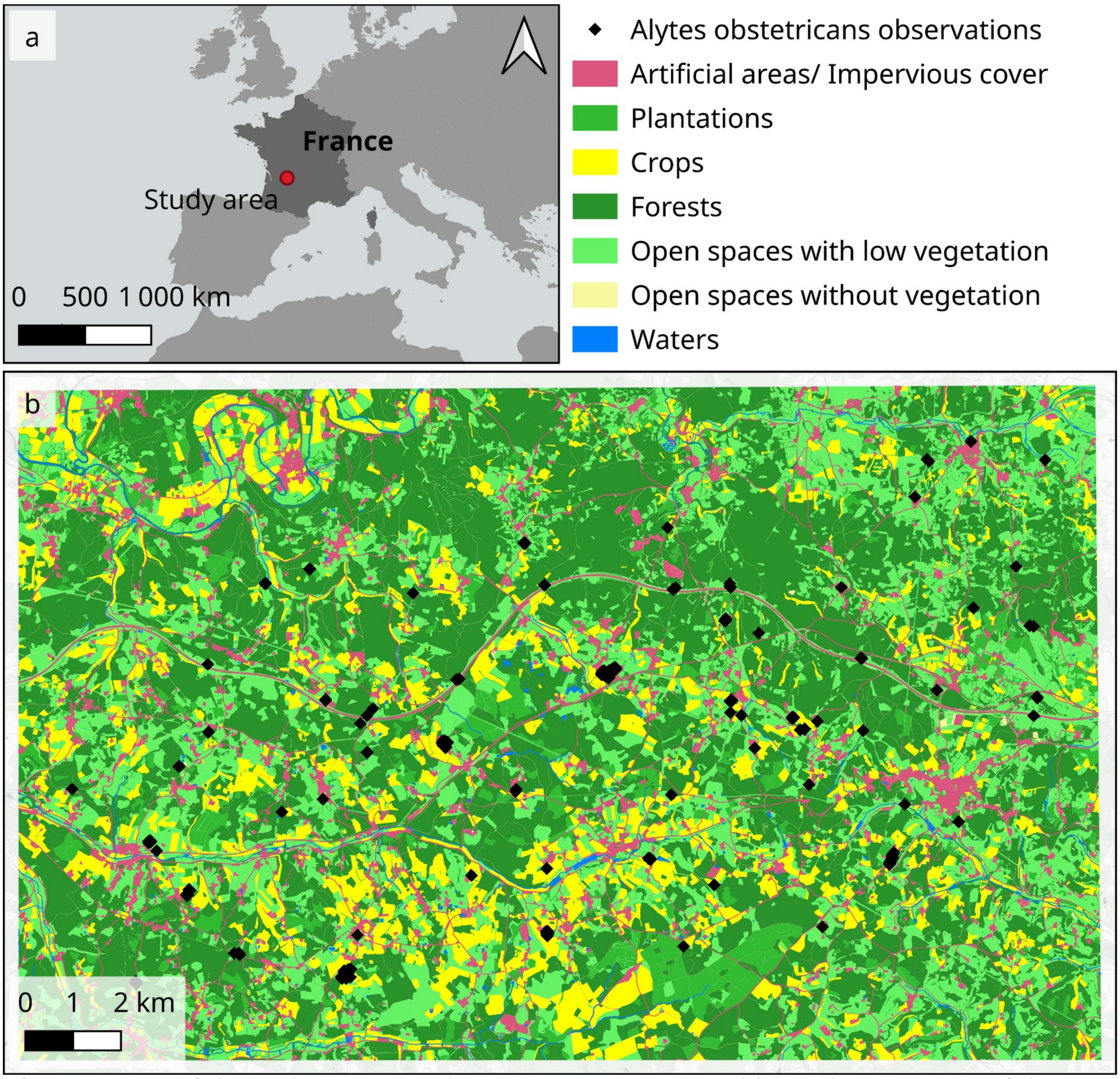

***Fig. 1*** *Location of the study area in South-western France; Périgord (a), and presentation of the main land-cover classes of the study area (b), with Alytes obstetricans observation points.*

To test the outcomes associated with different combinations of landscape characterisation and connectivity estimation, we selected a study site in rural western France (Dordogne) covering an area of 300km$^2$ on a limestone plateau. The landscape is composed of a patchwork of grasslands, cereal crops, walnut and chestnut orchards, deciduous forests, small streams, and a dense pond network, as well as a few urban areas and several linear transportation infrastructures.

Remon et al. (2022) produced a typological land-cover map using the 2012 European Nature Information System (EUNIS), a classification system introduced to harmonise habitat classification at the European scale (Davies and Moss 1999; Davies et al. 2004). The resultant EUNIS habitat map was reclassified into seven principal classes and 23 subclasses (Fig. 1). For modelling purposes, pre-processed vector data was rasterised at 10 metres resolution.

## 2.2 Focal species and observation data

The study was conducted on the Common Midwife Toad (*Alytes obstetricans*) which inhabits northern of Portugal and Spain, France, southern Belgium, southeastern Netherlands, Luxembourg, western Germany, and northern and western Switzerland (Frost 2024; AmphibiaWeb 2024). The midwife toad is listed in The IUCN Red List of Threatened species as Least Concern (IUCN SSC 2022). The IUCN SSC (2022) assessment reported a decrease in its population and a continuing decline in the area, extent, and quality of its habitat. Its habitats are principally threatened by the development of housing and urban areas, as well as agricultural and livestock farming practices (Luedtke et al. 2023). The survival of the midwife toad is also threatened by invasive and viral diseases and pollution from agricultural and forestry effluents (IUCN SSC 2022).

The midwife toad is a generalist species that can settle in and move through a variety of natural and semi-natural habitats, including inland wetlands, forests, shrublands, and artificial areas near freshwater (MNHN and OFB [Ed] 2024). Unlike more specialist species, its habitat niche does not fit a particular land-cover class (Siffert et al. 2022); instead, it results from more subtle land-cover combinations.

Adult midwife toads are reportedly not very mobile with a home range and maximal migration distance of only a few hundred metres (Frei 1990; Jehle and Sinsch 2007; Tobler et al. 2013). However, there is evidence to suggest that its maximal dispersal distance can be up to a few kilometres in exceptional dispersal events (Frei 1990; Legros et al. 2015). The analysis of genetic distances in the studied population revealed a significant isolation by distance of 5700 m, as reported by Remon et al. (2022).

Our observation dataset comprises 712 occurrence points acquired by Remon et al. (2022) (Fig. 1). Midwife toads were initially sighted at night and then captured for genetic surveying between 2015 and 2017 (Remon et al. 2022). The resultant presence points were rasterised into a binary 10 m resolution map for species distribution modelling, resulting in 292 presence pixels.

## 2.3. Methodological framework

Our methodological framework (Fig. 2) consists of three steps: (a) landscape characterisation, (b) connectivity estimation and (c) patch contribution to overall connectivity. Landscape characterisation involves taking our understanding of the landscape from the perspective of the midwife toad and translating it into maps of potentially suitable habitat patches (hereafter patches) and costs of resistance to movement (resistance map; Hanski 1998). Connectivity estimation consists of computing the potential paths between patches (either one-dimensional single-line paths or two-dimensional surface-wide corridors) and evaluating the intensity of potential interactions (Kool et al. 2013). Potential paths are spatially represented in movement maps and the intensity of potential interactions is denoted by the estimated probability of connectivity. Patch contribution involves estimating the importance of each patch in the network by computing their impact on global connectivity measures (Saura and Pascual-Hortal 2007).

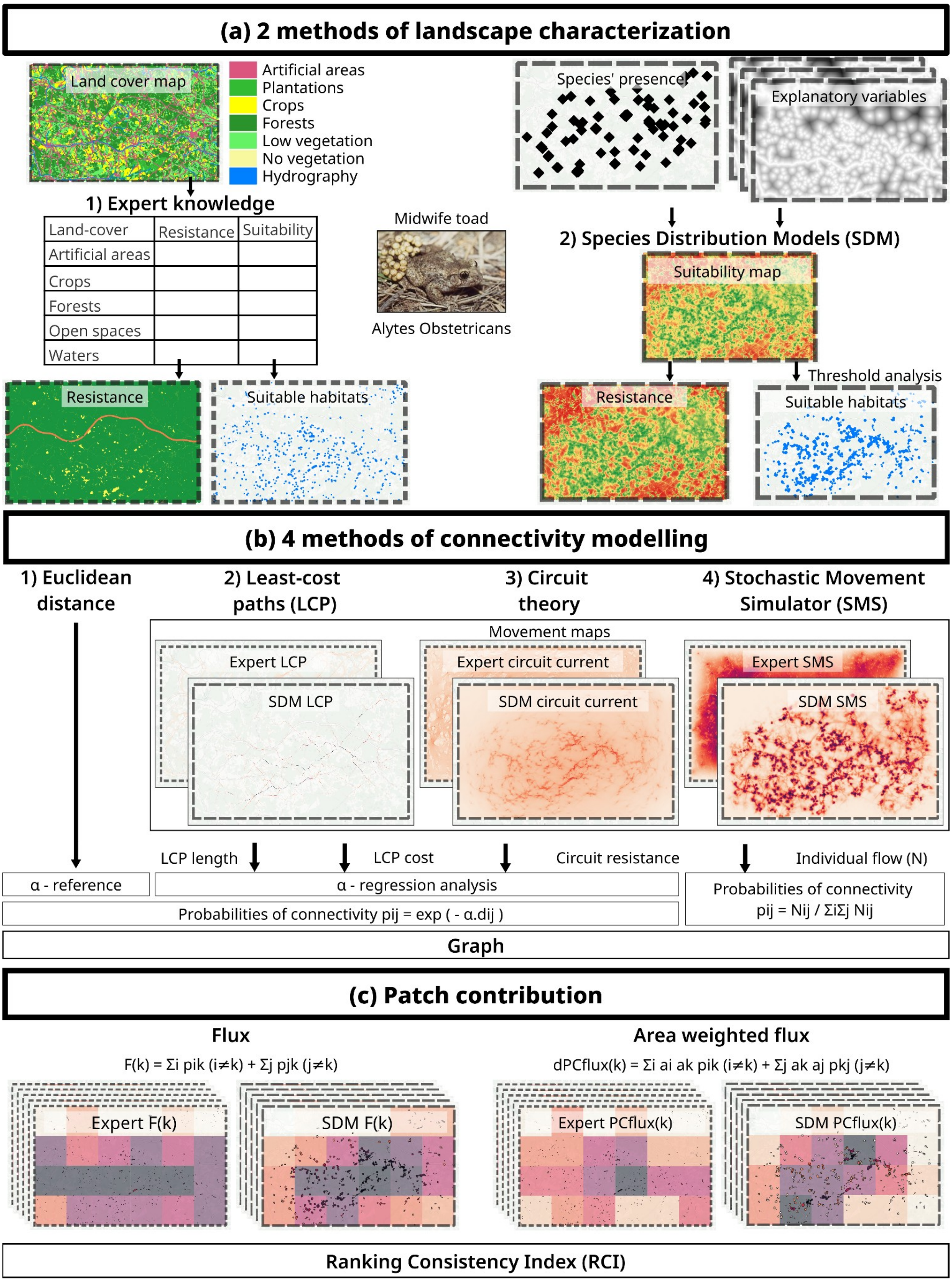

***Fig. 2*** *Methodological framework. landscape characterisation approaches: expert opinion based on land cover typologies and SDM based on explanatory variables and occurrence points (a); connectivity estimation: connectivity probabilities and movement map according to Euclidean distance, LCP, circuit theory and SMS (b); patch contribution: spatial and ranking consistency according to F and dPCflux (c).*

### 2.3.1. Landscape characterisation

We implemented and compared two landscape characterisation approaches: expert opinion on habitat requirements and movement resistance based on landscape typology, and estimation of habitat suitability through SDM based on landscape variables and observation points.

#### Expert opinion

Initial estimates of suitable habitats and resistance coefficients were extracted from the French National Natural Heritage Inventory portal (MNHN and OFB [Ed] 2024) and amphibian data from (Trochet et al. 2014). The estimates were sent to a group of 6 herpetologists for refinement. The final decision was made by unanimous agreement of the expert group. The experts selected the midwife toad reproductive habitat, standing surface water, as patches. They assigned resistances ranging from 1 to 100 to each land typology according to juvenile toads' movement abilities (Koen et al. 2012). Forests and shrublands, which are terrestrial habitats of the midwife toad, were assigned a low resistance. Croplands, open areas without vegetation, and low-density areas were allocated a moderate coefficient. A medium value was assigned to bodies of water as the midwife toad prefers terrestrial displacements despite its ability to swim. Finally, the highway was associated with a high resistance value as experts predicted it would act as a barrier to movement. See supplementary information S1 for detailed expert estimates.

#### Species distribution models

We computed an ensemble of SDM to estimate habitat suitability across the study area (Farashi and Alizadeh-Noughani 2023). We used distance to land cover class as predictors, and a binary presence raster from occurrence points as the response variable. The spatial resolution of rasters was 10 metres by 10 metres. Based on the species' ecology, we expected the following covariables to have a positive influence on the presence of *Alytes obstetricans*: distance to grassland, distance to forest, distance to standing or running water, and distance to low-density artificial area. We also expected agricultural land to incur a negative effect on midwife toad presence due to human disturbance and pollution.

We parameterised 100 random points of pseudo-absence data and set a 50% presence/absence prevalence (Barbet-Massin et al. 2012). We predicted suitability using an ensemble of 100 models to take advantage of inter-model agreement (Grenouillet et al. 2011). Our ensemble consisted of 10 models of 10 algorithms available in the R package biomod2 (Thuiller et al. 2024). We applied a selection threshold (ROC>0.8) and computed the weighted sum of their probabilities according to their evaluation score (Jiménez-Valverde 2012). Zurell et al. (2020) ODMAP protocol is available in supplementary information S2.

We scaled the resulting suitability map from 0 to 100 to fit the same range as the expert resistance and applied the transformation $y = 100 - x$ to invert the direction of variation while preserving the derivative coefficient. We extracted the binary map of patches by applying a suitability threshold which was selected following the calculation of the number of patches, the total patch area, and two Jaccard indices regarding the area and number of intersections (see supplementary information S4 for details on Jaccard indices computation).

### 2.3.2. Connectivity estimation

For each landscape approach, we prepared four connectivity estimation models, comprising three isolation-based: Euclidean distance, LCP, and circuit theory; and one individual-based movement: SMS. We then constructed corresponding complete graphs of connectivity probabilities. The connectivity probabilities refer to the likelihood of a connection between pairs of patches across the landscape. For isolation-based approaches, these probabilities are calculated based on an inverse exponential function of the ecological distance. For SMS, probabilities are calculated based on the fraction of migrants between two patches among the total number of migrants. We also computed movement maps resulting from LCP, circuit resistance, and SMS. The movement maps represent the predicted spatial distribution of movement across the landscape. We did not compute it for Euclidean distance because this approach does not account for heterogeneity and movement abilities. We used Spearman correlations to compare the results (Pearson and Kendall correlations are available in supplementary information S5).

## Isolation-based models

We computed four estimations of ecological distance $d$ used in connectivity literature as isolation proxies: $d_{euclid}$ the Euclidean distance in metres, $d_{lcp.length}$ the LCP length in pixels, $d_{lcp.cost}$ the LCP cost in cost units, and $d_{circuit}$ the circuit resistance from circuit theory in ohms. We used Python geopandas tools (Jordahl et al. 2020) to compute the Euclidean distance. We used Python rasterio V1.3.4 (Mapbox 2018), pandas V1.1.5 (pandas' contributors 2024) and skimage V0.19.3 (Van Der Walt et al. 2014) to compute LCPs. The resistance distance was computed using Circuitscape in Julia (Hall et al. 2021).

We converted the ecological distances to connectivity probabilities using an inverse exponential function of the distance (Bunn et al. 2000):

$$p(d) = e^{-ad} \qquad \text{(Eq. 1)}$$

With $p$ the connectivity probability and $d$ their ecological distance. $a$ is a parameter calibrated from a known probability:

$$a = - \log p_{ref} \ / \ d_{ref} \qquad \text{(Eq. 2)}$$

With $p_{ref} = p(d_{ref})$ the reference probability at $d_{ref}$. We used the maximal Euclidean dispersal distance estimated by Remon et al. (2022) $d_{euclid.ref} = 5700\ m$ and $p_{euclid.ref} = 0.01$.

We identify equivalents $\alpha$ for LCP and circuit theory by fitting regression models to their relationship with Euclidean distance.

## SMS

SMS (Palmer et al. 2011) can be used to model juvenile dispersal phases (Pittman and Semlitsch 2013). We simulated the dispersal of midwife toads using SimOïko; an agent-based, spatially explicit population dynamics simulator based on MetaConnect (Moulherat 2014). SimOïko features an embedded SMS algorithm which stops on three conditions: the simulated individual has either reached a suitable destination, went outside the limit of the study area, or consumed its displacement capacity. Step cost is proportional to the pixel size and resistance.

We used the following parameters to fit the movement ecology of the midwife toad: the perceptual range was set to the minimum possible value of 1 pixel (10 m) to account for the small perceptual range of juvenile amphibians (Semlitsch 2008). Using a biassed random walk as a reference (Pittman and Semlitsch 2013) we set the path memory size to 1 and the directional persistence to 1. Assuming $d_{euclid.ref}$ in an environment suitable for midwife toad movement, we computed the maximal displacement capacity $d_{sms.ref}$ as Eq. 3:

$$d_{sms.ref} = \left( \frac{d_{euclid.ref}}{pixel\ size} \right)^2 \times R_{mean} \qquad \text{(Eq. 3)}$$

With *pixel size* the spatial resolution and $R_{mean}$ the landscape mean resistance.

We ran the algorithm for 10,000 dispersal individuals per patch and constructed corresponding Origin-Destination matrices. We converted individual flows in dispersal probabilities as the fraction of individuals reaching $j$ from $i$ among the total number of successful dispersers:

$$p_{ij} = \frac{N_{ij}}{\sum_{i=1}^{n} \sum_{j=1}^{n} N_{ij}} \qquad \text{(Eq. 4)}$$

See supplementary information S8 for more insights on SMS.

### Movement maps

We constructed the movement maps for LCP, circuit current and SMS as the sum of all movement across the landscape. We compared the resulting movement maps with Spearman, Person and Kendall correlation, both with and without blurring kernels, to account for small spatial variability, as the kernel helps to reduce noise and smooth out minor spatial fluctuations (Wang *et al.* 2020). Supplementary information S5 presents more correlation insights.

## 2.3.3. Evaluation of patch importance

### Connectivity indices

We estimated individual patch's contribution to overall landscape connectivity regarding two connectivity indices; flux $F$ (Bunn et al. 2000), and area-weighted dispersal flux *dPCflux* (Saura and Rubio 2010):

$$F_k = \sum_{i=1,\ i\neq k}^{n} p_{ik} + \sum_{j=1,\ j\neq k}^{n} p_{kj} \quad \text{(Eq. 5)}$$

$$dPCflux_k = \sum_{i=1,\ i\neq k}^{n} a_i a_k p_{ik} + \sum_{j=1,\ j\neq k}^{n} a_k a_j p_{kj} \quad \text{(Eq. 6)}$$

With $p$ the pairwise probability of direct connectivity (see supplementary information S3 for details on *dPCflux* calculation) and $a$ the patch area. These indices were selected for their board usage among connectivity study and their straightforward set of parameters.

Since expert and SDM patches differ, we extracted a 4x6 grid from the study area and computed the average connectivity indices within each grid cell to perform inter-landscape characterisation comparison. We used Spearman, Pearson, and Kendall correlation to compare patch-based and zonal indices (see supplementary information S7). Additionally, we calculated the correlation with the patch area to evaluate its influence on *dPCflux*.

### Ranking Consistency Index

To compare the approaches regarding the patch hierarchy, the patches were ranked according to their importance in the landscape. As a result, 20 rankings were obtained and pooled according to the landscape approach and connectivity index. The pairwise connectivity model rankings were compared two-by-two by plotting the fraction of common patches in the top fraction of the rankings. The Ranking Consistency Index (RCI) was used to measure consistency between rankings, corresponding to the area under the curve defined above. The index ranges from RCI = 0.31 (no common patches in the top half of the ranking) to RCI = 1 (perfect accordance) (see supplementary information S9 for examples).

# 3. Results

## 3.1. Landscape characterisation

## 3.1.1. Potential resistance to movement

The experts considered the generalist aspect of midwife toad ecology. They attributed most terrestrial natural classes and sparsely artificialised areas with a low resistance to movement. The experts associated intensive cropland with an intermediate resistance due to probable pollution and unfavourable agricultural practices. They also identified the transportation network as a potential barrier with an increased probability of collision. Following the same logic, they predicted a high resistance value for the highway, considering the high risk of collision and the probable presence of physical barriers. They estimated low overall resistance across the studied landscape as it is predominantly rural (Fig. 3a). The average resistance value across the landscape was 6.91 and the median value was 1.00. Their predictions attribute 76.99% of the landscape with no resistance (1) and 22.73% with intermediate resistance (between 10 and 50). The remaining landscape, corresponding to the highway, presents the only potential barrier with a high resistance (80).

The resistance estimations derived from SDM differ greatly from the expert predictions (Fig. 2b). The SDMs are based on the hypothesis that the distance to different land cover classes explains the distribution of midwife toad observations. The most important factor affecting the suitability estimation was the distance to standing water, with an importance score ranging from 0.25 to 0.30. The second most significant variable was the distance to sparsely artificial areas (0.09), followed by the road and the rail network (0.03 and 0.02 respectively). All other variables scored lower than 0.01 and need not be considered. The importance scores attributed to ponds and sparsely artificial areas are consistent with expert insights. The relative importance attributed to the transportation network may elucidate a sampling bias based on the accessibility of observation points (Stolar and Nielsen 2015; Inman et al. 2021) — but more likely indicate the suitability of road embankments and excavations for midwife toad burrows (Lange et al. 2020). Contrary to expert predictions, the highway may act as a low-resistance longitudinal corridor rather than a barrier. The resulting map shows a high resistance overall, with an average of 75.75, a median of 82, and only 11.6% of the landscape presenting a resistance below 50.

### 3.1.2. Suitable patches

Experts focused on the reproductive habitat of the midwife toad. They defined ponds, where males lay their eggs, as patches (Fig. 3a). The experts identified 645 patches with a total area of 1.4 km². Meanwhile, the SDM approach utilised the occurrence data and corresponding environmental conditions as criteria for suitable patches (Fig. 3b). A suitability threshold of 751 was selected following the threshold analysis (Fig. 4). Consequently, 639 patches were identified, covering a total area of 8.6 km². See supplementary information S4 for further comparison of suitable patches.

The threshold analysis shows that the structure of SDM patches evolves as a function of the suitability threshold (Fig. 4). The number of patches obtained with SDM follows a bell curve, as seen in Fig. 4a. When the threshold is equal to 1, the landscape forms a single, large patch. As the threshold is increased, the mono-patch breaks down and the number of patches increases. The evolution of the patch number peaks at 450 and then decreases as fewer suitability values meet the threshold. The curve intersects the expert number of patches twice; at 165 and 751. The total patch area (Fig. 4b) decreases from the size of the landscape to zero, intersecting the expert total patch area at 858. After an initial drop, the Jaccard of the number of intersections (Fig. 4c) increases slowly, peaking at 740. The Jaccard of the area of intersection (Fig. 4d) peaks at 754. A suitability threshold of 751 was chosen as it most closely corresponds to the number of patches identified through expert opinion. The Jaccard indices are close to their respective peaks. The total patch area is roughly 10 times the experts' patch area estimation.

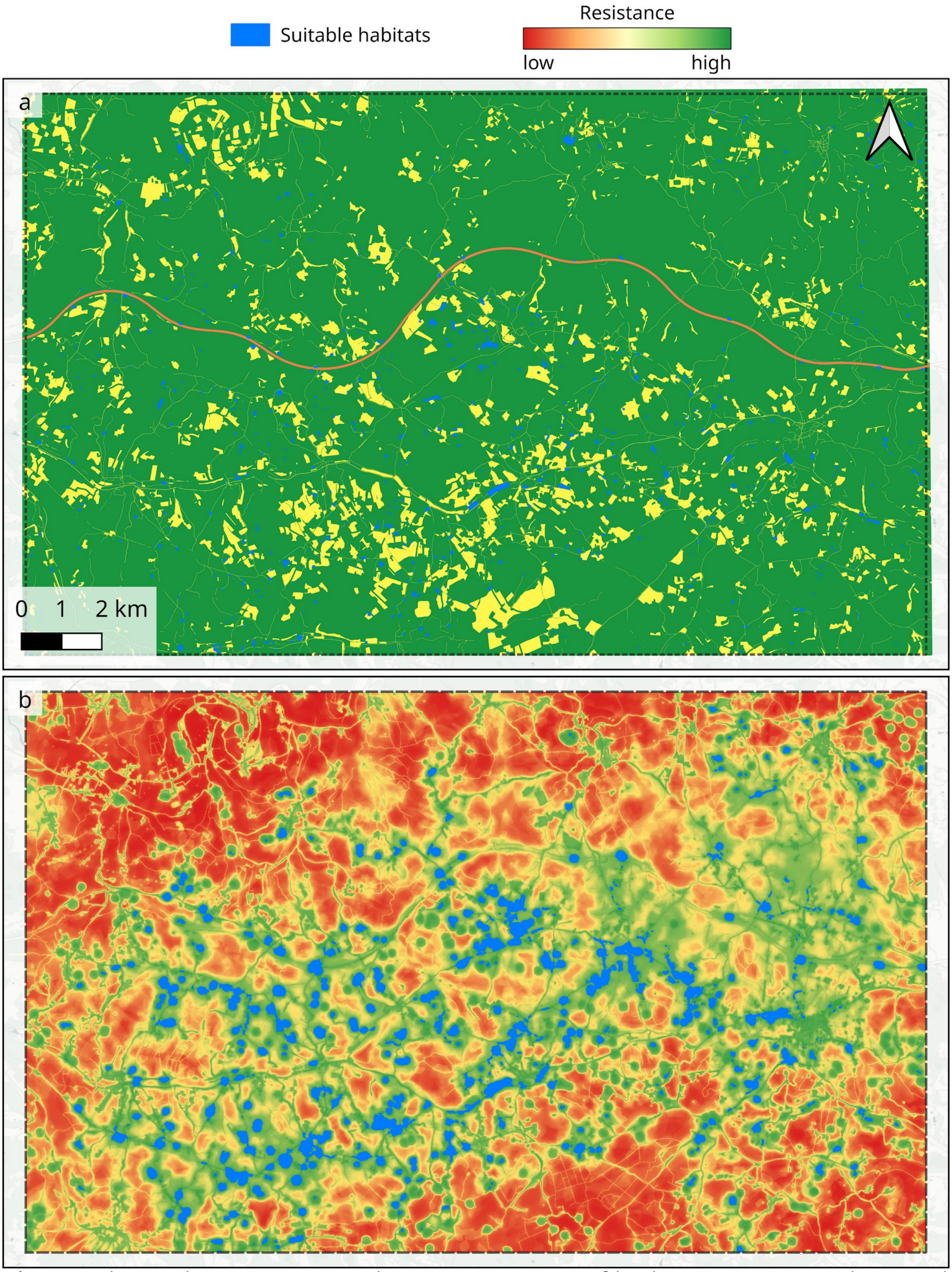

***Fig. 3*** *Landscape characterisation results. Expert estimation of landscape resistance and potential suitable habitats (a), see supplementary information S1 for the expert opinion table. SDM estimation of landscape resistance and potential suitable patches (b).*

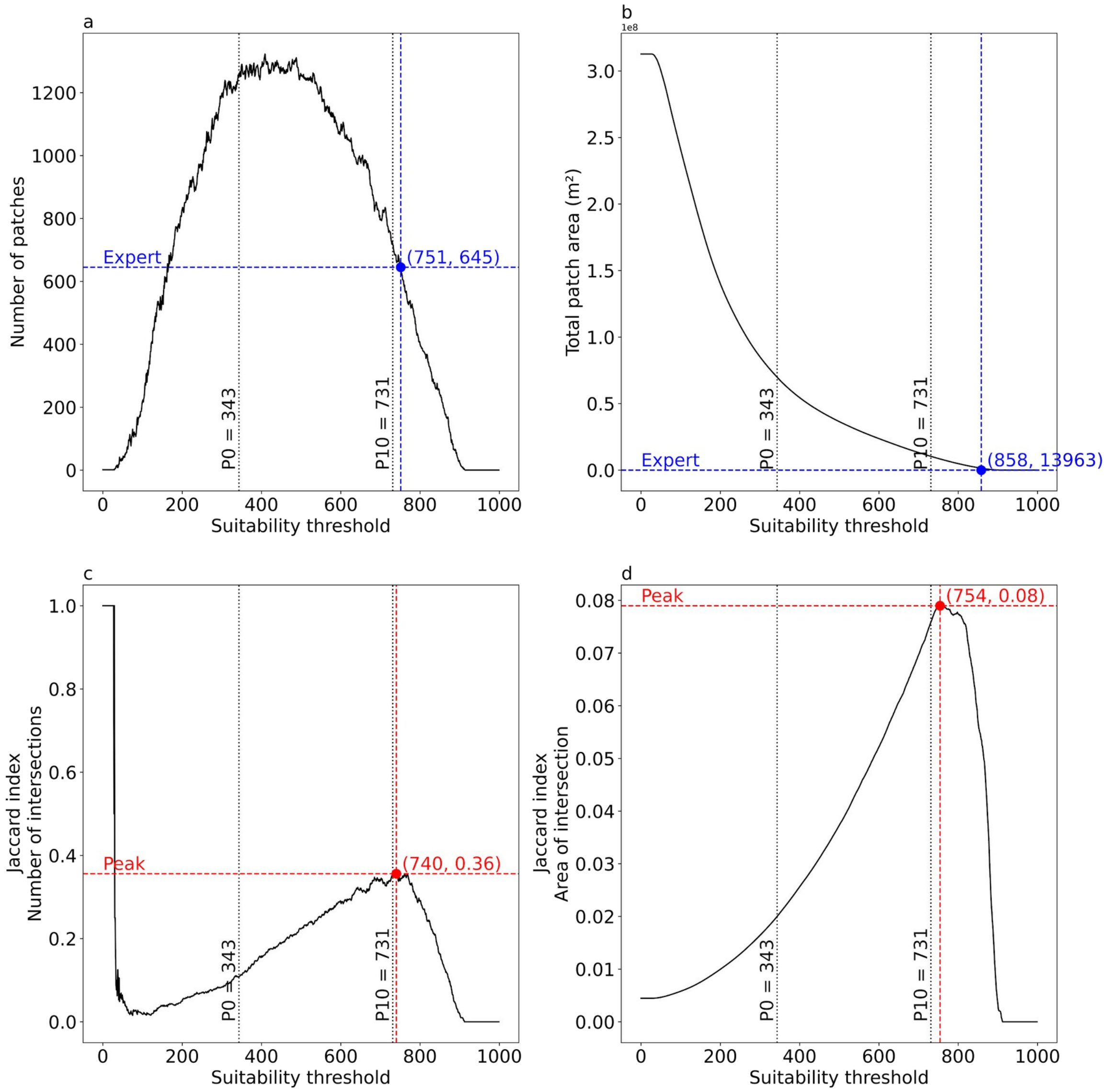

***Fig. 4*** *Suitability threshold analysis: evolution of the number of patches (a), patch area (b), Jaccard of the number of patch intersections (c), and Jaccard of the area of intersection (d).*

## 3.2. Connectivity modelling

### 3.2.1. Movement estimation

The patterns derived from movement maps differed between landscape and connectivity approaches (Fig. 5). The movement patterns derived from the expert approach appear more diffuse than those derived from SDM. The expert resistance map is homogeneous; it indicates a low overall resistance to movement and the probable existence of equivalent alternative paths. The SDM is characterised by a high resistance and a restricted range of continuous optimal paths that are constrained to narrower beams. This observation rings true across all connectivity approaches, especially in the case of LCP, whereby all paths of the SDM overlap to form a single, pixel-wide path. Concerning the same landscape approach, it was observed that areas of high connectivity coincide. This is particularly pronounced in the SDM given its high resistance contrast, but is also noticeable in the expert approach. The correlations quantitatively highlight the convergence of the movement (Table 1).

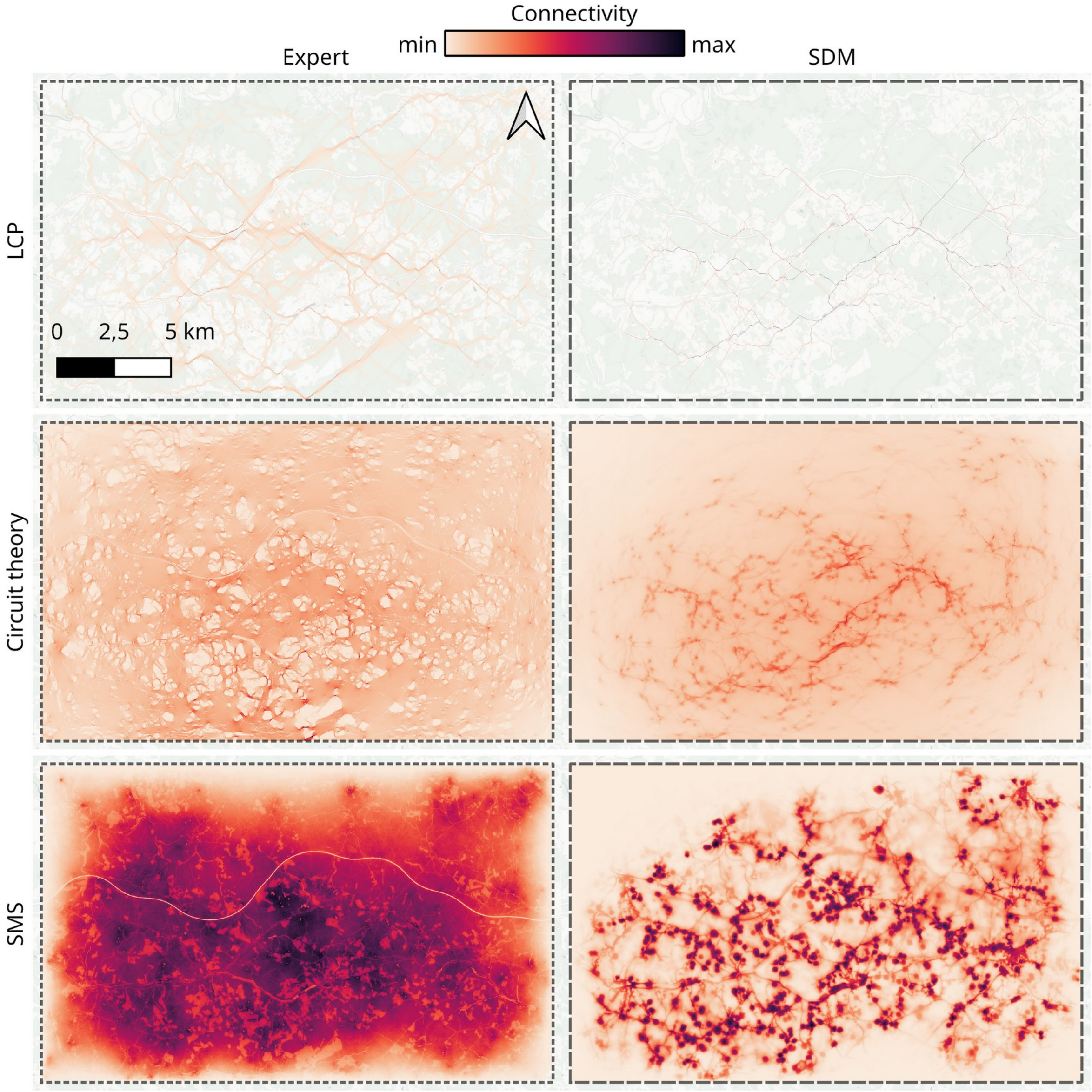

***Fig. 5*** *Movement maps. The first column corresponds to the expert approach and the second to SDM. The first row represents the cumulative LCP, the second circuit current distribution, and the third cumulative SMS.*

The expert correlations show a relatively high coefficient between LCP and circuit (0.63). The other expert coefficients indicate lower correlations (0.25 for SMS *vs.* LCP and 0.33 for SMS *vs.* circuit). The SDM correlations are overall lower except for circuit *vs.* SMS (0.37). SDM Circuit *vs.* LCP coefficient indicate a moderate correlation (0.39). SDM LCP and SMS coefficient supports a low correlation (0.11).

The inter-landscape characterisation's correlations present contrasting values. SDM LCP presents no correlation with expert LCP (0.06) or circuit (0.09); and a low correlation with expert SMS (0.18). Expert LCP coefficients are a little higher with SDM SMS (0.15) and circuit (0.27). The coefficient between SDM SMS and expert circuit is also relatively low (0.19). The three remaining coefficients stand out as remarkably high: SDM circuit *vs.* expert circuit (0.47), SDM circuit *vs.* expert SMS (0.53), and SDM SMS *vs.* expert SMS (0.65). The narrow scope of least-cost path analysis restrains its correlation with broader spatial movement approaches.

***Table 1*** *Spearman correlation coefficients of movement maps blurred with a 5-by-5 kernel. See supplementary information S5 for corresponding plots, different kernels, and Pearson and Kendall correlations.*

| | | Expert | | | SDM | | |
|---|---|---|---|---|---|---|---|
| | | LCP | Circuit | SMS | LCP | Circuit | SMS |
| Expert | LCP | - | | | | | |
| | Circuit | 0.63 | - | | | | |
| | SMS | 0.25 | 0.33 | - | | | |
| SDM | LCP | 0.06 | 0.09 | 0.18 | - | | |
| | Circuit | 0.27 | 0.47 | 0.53 | 0.39 | - | |
| | SMS | 0.15 | 0.19 | 0.65 | 0.11 | 0.37 | - |

## 3.2.2. Probabilities of connectivity

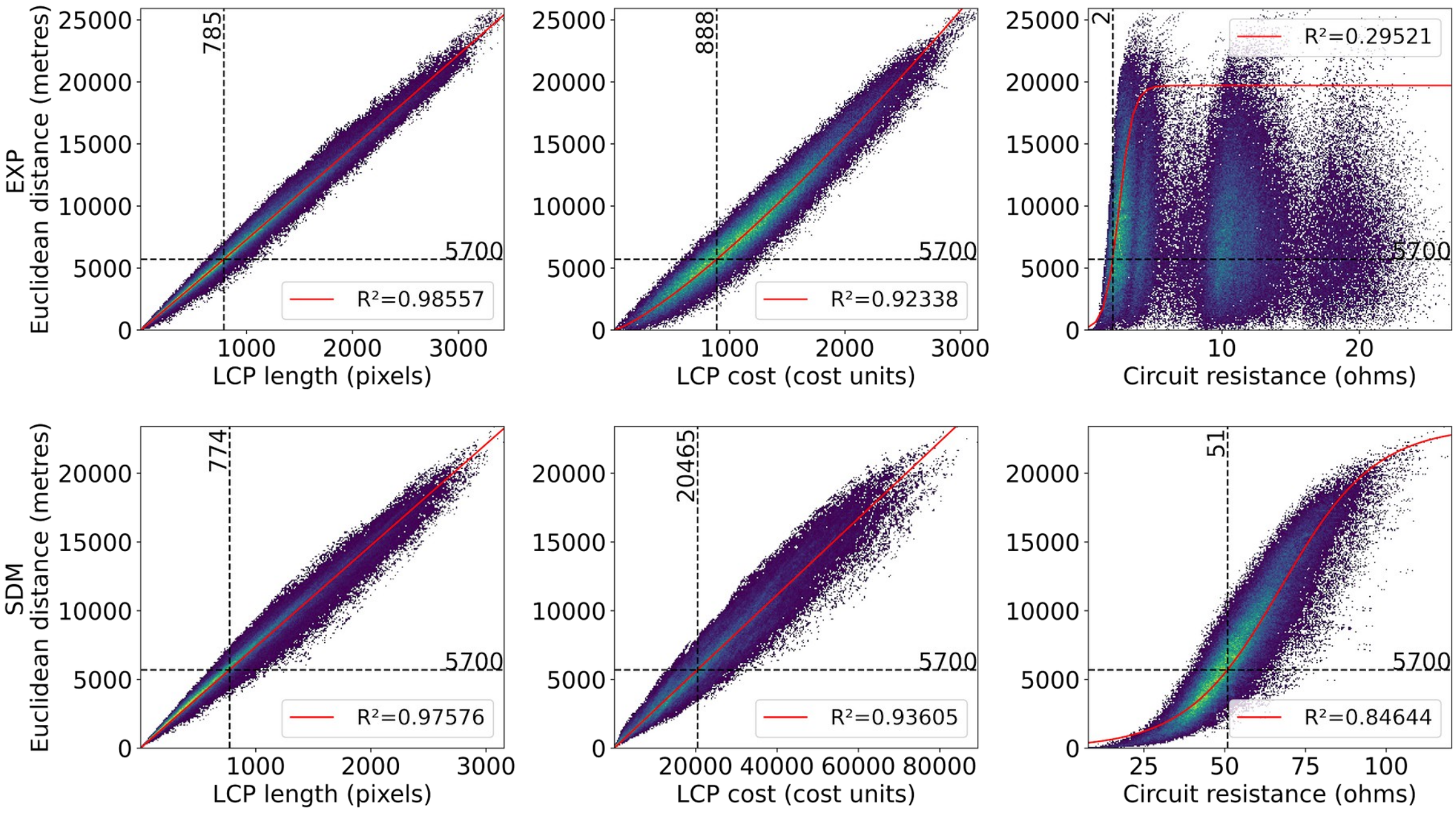

***Fig. 6*** *Ecological distances relationship to Euclidean distance. Each point represents a pair of patches. The projection of the maximal dispersal distance of 5,700 metres along a fitted regression line gives its conversion to the corresponding ecological distance unit.*

The regression analysis reveals how different relationships emerge depending on the landscape and distance approaches (Fig. 6); the regression curve between the LCPs and Euclidean distances varies depending on the landscape approach used. In the expert approach, the points form a slightly curved function that is characteristic of a power relationship. In the SDM approach, the points follow a straight line, characteristic of a linear relationship. The $R^2$ values close to 1 indicate narrow point beams around the regression curve. In both expert and SDM approaches, a logistic relationship emerges with circuit resistance. The SDM approach yields a single point cloud, which allows the corresponding distance to be explicitly identified ($R^2$ = 0.85). In the expert approach, the logistic pattern seems to repeat itself through several point clouds, which makes it difficult to fit a regression. We chose to arbitrarily limit the fit to the first, denser cloud ($R^2$=0.30).

The correlation of pairwise connectivity probabilities elucidates similarities between connectivity approaches (Table 2). In the expert approach, all correlations support a high degree of similarity — except circuit resistance, whereby the coefficients indicate a total absence of correlation. As previously mentioned, it was difficult to fit a regression model (Fig. 6). Euclid, LCP length, and LCP cost form a cluster of highly correlated values (>0.98). Probabilities derived from the SMS approach give slightly lower coefficients (between 0.64 and 0.66). In the SDM approach, we may supplement the high correlation between Euclid, LCP length, and LCP cost (>0.96) with circuit-based probabilities (between 0.92 and 0.95). The coefficient values derived from SMS correlations are slightly lower than the expert approach (between 0.42 and 0.45).

***Table 2*** *Spearman correlation coefficients between the estimation of connectivity probabilities. ns indicates the statistical non-significance of the coefficient (p-value > 0.01). See supplementary information S6 for Pearson and Kendall correlations and corresponding scatter plots.*

| | Expert | | | | | SDM | | | | |
|---|---|---|---|---|---|---|---|---|---|---|
| | Euclid | LCP length | LCP cost | Circuit | SMS | Euclid | LCP length | LCP cost | Circuit | SMS |
| Euclid | - | | | | | - | | | | |
| LCP length | 0.99 | - | | | | 0.99 | - | | | |
| LCP cost | 0.98 | 0.98 | - | | | 0.97 | 0.96 | - | | |
| Circuit | -0.00 ns | 0.01 | 0.10 | - | | 0.93 | 0.92 | 0.95 | - | |
| SMS | 0.66 | 0.66 | 0.64 | 0.06 | - | 0.45 | 0.45 | 0.43 | 0.42 | - |

## 3.3. Connectivity indices variation and evaluation of patches importance

### 3.3.1. Patch scale

dPCflux exhibits high correlation coefficients except for expert circuit which consistently presents lower correlation scores (between 0.13 and 0.29). Euclid and LCPs dPCflux present very high correlations in both expert (between 0.97 and 1.00) and SDM (between 0.99 and 1.00) approaches. SDM circuit is also highly correlated with Euclid and LCPs between (0.99 and 1.00). SMS presents slightly lower coefficients (between 0.90 and 0.93 for expert and between 0.80 and 0.82 for SDM). Except for expert circuit, all dPCflux estimations present a high correlation with the area.

F correlations show greater contrast. In the expert approach, F Euclid *vs.* LCP length presents the higher coefficient (0.97), followed by LCP cost *vs.* LCP length (0.71), SMS (0.70) and Euclid (0.66). SMS coefficients indicate a moderate correlation with LCP length (0.60), Euclid (0.58) and circuit (0.47). F circuit and LCP cost presents a correlation of 0.40. Finally, F expert circuit correlation with Euclid and LCP length present negative values, consolidating the anomaly of expert circuit. In the SDM approach, F coefficients show very high correlations for all combinations of Euclid, LCP length, and LCP cost (>0.95). F SDM circuit shows a high correlation with Euclid and LCP (between 0.84 and 0.86). F SDM SMS coefficients indicate a low correlation with isolation-based models (between 0.16 and 0.27) the higher score occurring with circuit.

Cross-index correlations are significantly lower (expert <0.43 and SDM <0.55). In the expert approach, circuit correlations are negative or null except between F circuit and dPCflux circuit (0.91) and F SMS and dPCflux circuit (0.43). The other coefficients present an intermediate correlation between F Euclid and LCPs with dPCflux Euclid and LCPs (between 0.23 and 0.39) and a lower correlation for SMS F and dPCflux (between 0.14 and 0.27). In the SDM approach, the higher correlations occur between F circuit and all dPCflux estimations (between 0.44 and 0.55), followed by F LCP cost correlation (between 0.15 and 0.29). F Euclid and LCP length are approximately in the same range (between 0.09 and 0.24). F SMS coefficients indicate no correlation with dPCflux estimations.

For both dPCflux and F, SDM coefficients are more exaggerated than the expert's (higher and lower, lower, lower).

***Table 3*** *Spearman correlation coefficients between connectivity indices. Bottom-left corner presents correlations related to the expert approach. Top-right corner those related to SDM. ns — 'non-significant' — indicates a p-value greater than the significance threshold, 0.01.*

| | area | *dPCflux Euclid* | *dPCflux lcp length* | *dPCflux lcp cost* | *dPCflux circuit* | *dPCflux SMS* | $F_{euclid}$ | $F_{lcp.len}$ | $F_{lcp.cos}$ | $F_{circuit}$ | $F_{SMS}$ |
|---|---|---|---|---|---|---|---|---|---|---|---|
| area | - | *SDM 0.96* | *SDM 0.96* | *SDM 0.93* | *SDM 0.97* | *SDM 0.93* | *SDM -0.05 ns* | *SDM -0.04 ns* | *SDM -0.01 ns* | *SDM 0.29* | *SDM -0.10 ns* |
| *dPCflux Euclid* | *expert 0.91* | - | *SDM 1.00* | *SDM 0.99* | *SDM 1.00* | *SDM 0.96* | *SDM 0.17* | *SDM 0.17* | *SDM 0.21* | *SDM 0.49* | *SDM -0.06 ns* |
| *dPCflux lcp length* | *expert 0.91* | *expert 1.00* | - | *SDM 0.99* | *SDM 1.00* | *SDM 0.96* | *SDM 0.17* | *SDM 0.17* | *SDM 0.21* | *SDM 0.49* | *SDM 0.085 ns* |
| *dPCflux lcp cost* | *expert 0.91* | *expert 0.97* | *expert 0.97* | - | *SDM 0.99* | *SDM 0.95* | *SDM 0.24* | *SDM 0.24* | *SDM 0.29* | *SDM 0.55* | *SDM -0.04 ns* |
| *dPCflux circuit* | *expert 0.20* | *expert 0.13* | *expert 0.14* | *expert 0.29* | - | *SDM 0.96* | *SDM 0.15* | *SDM 0.15* | *SDM 0.19* | *SDM 0.49* | *SDM -0.04 ns* |
| *dPCflux SMS* | *expert 0.86* | *expert 0.93* | *expert 0.93* | *expert 0.90* | *expert 0.16* | - | *SDM 0.09* | *SDM 0.09* | *SDM 0.15* | *SDM 0.44* | *SDM -0.14 ns* |
| $F_{euclid}$ | *expert 0.07 ns* | *expert 0.39* | *expert 0.37* | *expert 0.29* | *expert -0.15 ns* | *expert 0.25* | - | *SDM 0.99* | *SDM 0.96* | *SDM 0.84* | *SDM 0.16* |
| $F_{lcp.len}$ | *expert 0.07 ns* | *expert 0.39* | *expert 0.38* | *expert 0.32* | *expert -0.10 ns* | *expert 0.27* | *expert 0.97* | - | *SDM 0.95* | *SDM 0.84* | *SDM 0.16* |
| $F_{lcp.cos}$ | *expert -0.03 ns* | *expert 0.23* | *expert 0.23* | *expert 0.32* | *expert 0.39* | *expert 0.14* | *expert 0.66* | *expert 0.71* | - | *SDM 0.86* | *SDM 0.18* |
| $F_{circuit}$ | *expert -0.12* | *expert -0.17* | *expert -0.15 ns* | *expert 0.00 ns* | *expert 0.91* | *expert -0.12* | *expert -0.18 ns* | *expert -0.12 ns* | *expert 0.40* | - | *SDM 0.27* |
| $F_{SMS}$ | *expert 0.00 ns* | *expert 0.21* | *expert 0.20* | *expert 0.23* | *expert 0.43* | *expert 0.21* | *expert 0.58* | *expert 0.60* | *expert 0.70* | *expert 0.47 ns* | - |

### 3.3.2. Grid-cell scale

At the grid-cell scale (Table 4), the inter-landscape coefficients indicate good correlations between Euclid and LCPs for both dPCflux (between 0.64 and 0.72) and F (between 0.63 and 0.73). SDM circuit coefficients with expert Euclid and LCPs are close to this range (between 0.61 and 0.66 for dPCflux and between 0.58 and 0.61 for F). SDM SMS and expert circuit coefficients are non-significants for all correlations. Expert SMS p-values are close to the significance threshold and indicate moderate correlations with SDM Euclid, LCPs and circuit (between 0.52 and 0.58 for dPCflux and between 0.49 and 0.52 for F).

***Table 4*** *Spearman correlation coefficients between mean dPCflux and F per cell grid. ns — 'non-significant' — indicates a p-value greater than the significance threshold, 0.01.*

| | | Expert | | | | | | | | | |
|---|---|---|---|---|---|---|---|---|---|---|---|
| | | dPCflux | | | | | F | | | | |
| | | Euclid | LCP length | LCP cost | Circuit | SMS | Euclid | LCP length | LCP cost | Circuit | SMS |
| SDM | Euclid | 0.69 | 0.67 | 0.64 | 0.33 ns | 0.55 ns | 0.69 | 0.71 | 0.72 | -0.18 ns | 0.52 ns |
| | LCP length | 0.72 | 0.69 | 0.66 | 0.29 ns | 0.58 | 0.71 | 0.72 | 0.73 | -0.18 ns | 0.52 ns |
| | LCP cost | 0.69 | 0.67 | 0.64 | 0.27 ns | 0.57 | 0.63 | 0.63 | 0.65 | -0.19 ns | 0.46 ns |
| | Circuit | 0.66 | 0.63 | 0.61 | 0.27 ns | 0.52 ns | 0.58 | 0.60 | 0.61 | -0.15 ns | 0.49 ns |
| | SMS | 0.49 ns | 0.47 ns | 0.41 ns | 0.27 ns | 0.38 ns | 0.23 ns | 0.25 ns | 0.14 ns | 0.05 ns | 0.35 ns |

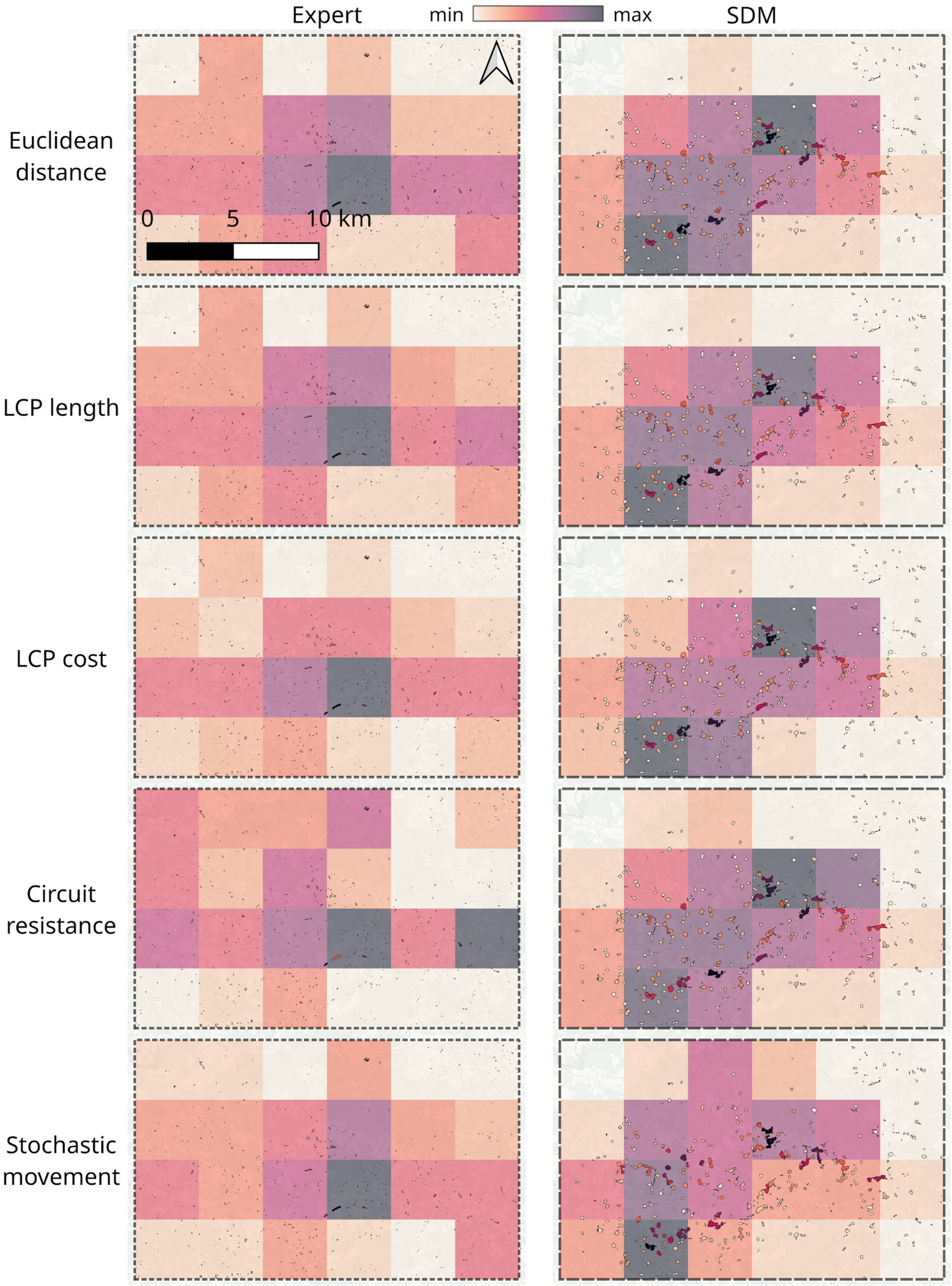

***Fig. 7*** *Suitable habitats according to dPCflux. The background colour represents the dPCflux average per grid cell.*

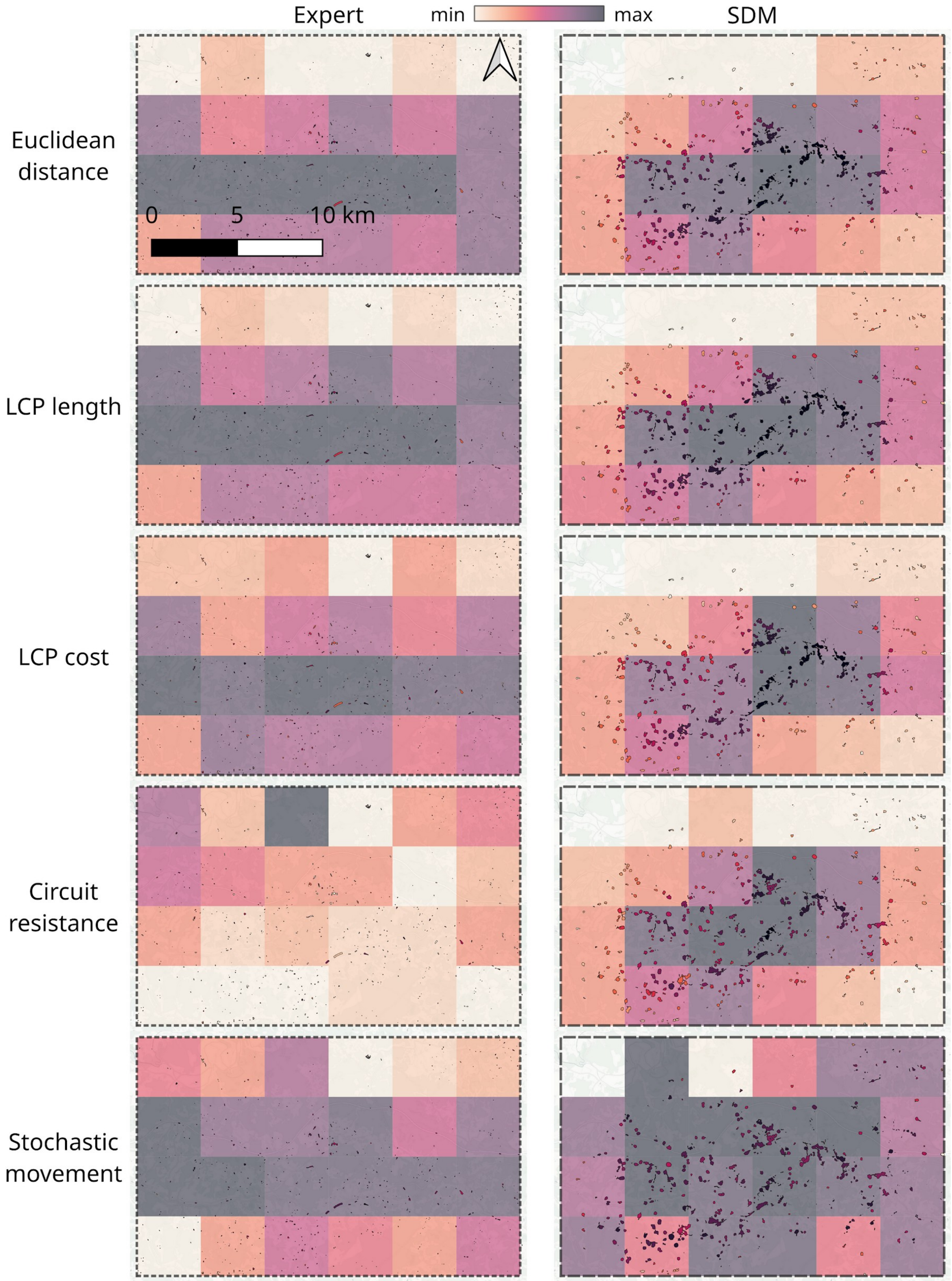

***Fig. 8*** *Suitable habitats according to F. The background colour represents the F average per grid cell.*

## 3.4. Patch prioritisation

| RCI | | *dPCflux* | | | | | *F* | | | | |
|---|---|---|---|---|---|---|---|---|---|---|---|
| | | Euclid | LCP length | LCP cost | Circuit | SMS | Euclid | LCP length | LCP cost | Circuit | SMS |
| Expert | Euclid | - | | | | | - | | | | |
| | LCP length | *0.98* | - | | | | *0.86* | - | | | |
| | LCP cost | *0.91* | *0.92* | - | | | *0.68* | *0.71* | - | | |
| | Circuit | *0.59* | *0.60* | *0.64* | - | | *0.46* | *0.49* | *0.64* | - | |
| | SMS | *0.88* | *0.89* | *0.57* | *0.59* | - | *0.64* | *0.65* | *0.74* | *0.68* | - |
| SDM | Euclid | - | | | | | - | | | | |
| | LCP length | *0.99* | - | | | | *0.91* | - | | | |
| | LCP cost | *0.97* | *0.96* | - | | | *0.86* | *0.87* | - | | |
| | Circuit | *0.97* | *0.97* | *0.95* | - | | *0.76* | *0.76* | *0.77* | - | |
| | SMS | *0.88* | *0.88* | *0.88* | *0.88* | - | *0.53* | *0.53* | *0.54* | *0.57* | - |

***Table 5*** *Ranking Consistency Index.*

The rankings derived from dPCflux are in greater agreement than those derived from flux (dPCflux RCI > 0.57, while F RCI > 0.46). This is especially true of the SDM approach (dPCflux SDM RCI > 0.88). The RCIs representing the comparison of Euclidean distance and LCP length ranking are consistently the highest scores. Expert F circuit presents consistently low RCIs, with anomalies observed since the calibration of α. SDM F SMS results are close to the random RCI reference of 0.50.

# 4. Discussion

## 4.1. Landscape characterisation

Since the definition of suitable habitats and the estimation of the matrix resistance determines the modelling outcomes, landscape characterisation appears to be the main driver of the differences reported in our case study. Despite their different assumptions, both landscape characterisation approaches present converging outcomes, revealed by the suitable habitat overlap and the correlation observed between circuit theory and SMS movement maps. This suggests that the expert opinion and SDM are both compatible and complementary.

In a recent study, Hurtado et al. (2023) found that expert habitat suitability assessment can be improved by integrating feedback and estimate revision into the protocol. They also reported that a diverse expert group of up to five persons results in a more reliable aggregate prediction. In their study, they found that results contrasted significantly between the study of specialist and generalist species; camera trap observations of specialist species correlated with expert predictions to a greater extent. This is in keeping with the findings of Keeley et al. (2016). According to Broekman et al. (2022), in most cases, IUCN expert habitat suitability information corresponds with suitability models based on GPS-tracking data. They suggest that, whilst IUCN habitat assessment may prove sufficient, the database may be improved by integrating tracking data. Cushman and Lewis (2010) note that GPS tracking data can also be used to calibrate the resistance to movement coefficient. Unlike Di Febbraro et al. (2018), who advocate an expert-based approach only when SDM data is unavailable, we propose that expert opinion and species data be integrated into landscape characterisation protocols (Arfan et al. 2018). Current applications of expert opinion and SDM are adversely affected by bias. They also overlook functional aspects of the focal species, such as the availability of food resources, mating, shelter, predation pressures, ecological competition, and human disturbances. Given that the approaches complement one another, an integrated approach may improve our understanding of species ecology through enhanced connectivity assessment reliability and the advancement of associated conservation measures.

Matrix resistance characterisation methodologies are similarly limited; they rarely consider fundamental movement drivers, such as spatiotemporal variability, intra and interspecific interaction, human influence and individual internal states (Kumar et al. 2022). Promising approaches include the differentiation of movement drivers into different layers (Fletcher et al. 2019; Kumar et al. 2022) and the increasing integration of telemetry and genetics into characterisation methods. The accessibility of input data must however be improved before they become widespread practices.

The reliability of environmental input data is of critical importance irrespective of the landscape characterisation approach used. The land cover classification is often unsuitable to match species' habitat requirements. Recently, vegetation indices from remote sensing have been promoted as reliable and objective spatiotemporal alternatives. Their use in species distribution models has been proven to perform at least as well as variables derived from land cover classification (Suiza-Oliveira *et al.*, in prep). The increasing temporal and spatial resolution of their acquisition promotes the development of processing routines for their integration into ecological studies (Morin et al. 2022). Given the criticality of the landscape characterisation for the reliability of the results, we emphasised the importance of intermediary field validation of the landscape characterisation outcomes before the movement modelling step.

## 4.2. Connectivity estimation

The choice of modelling approach has the most impact on the spatial distribution of movement. The outcomes from the least-cost path approaches are spatially restricted, limiting their relevance to a small fraction of the study area. Some important connectivity areas, such as bottleneck and connectivity "holes'', are not interpretable from this movement analysis which may prove problematic to some conservation applications. Regarding the other connectivity outcomes, the correlation analysis reveals significant similarities between LCP and Euclidean distance. The assumptions behind the least-cost path algorithm refer to a specific movement ecology. More stochastic approaches have proven to better explain genetic distances (McRae and Beier 2007; Coulon et al. 2015). Despite its limitations for movement ecology applications (Moilanen and Hanski 2001; Taylor et al. 2006) and the existence of suitable alternatives, LCP analysis remains prevalent in research and operational applications given its accessibility.

Explicit movement models such as SMS are conceptually very different from isolation-based models (Kumar and Cushman 2022). Movement predictions in isolation-based models are almost exclusively determined by the resistance surface of the landscape. Whilst explicit movement models take landscape characteristics into account, they also consider a set of predefined processes pertaining to the simulated individual. Their respective definitions of connectivity probability reflect this distinction. In the case of isolation-based models, pairwise probability is determined by pair remoteness, whereas explicit movement models define pairwise probability as the portion of all successfully dispersed individuals that move from one specified patch to another. Moreover, in the latter case, the number of successful dispersals is limited by the number of initiated dispersals which is fixed by the simulation parameters. No equivalent limitation or normalisation is incurred in isolation-based approaches. This affects the comparability of the two approaches with connectivity indices such as the d*PCflux* and *F*, and correlations.

Connectivity modelling approaches are based on somewhat ill-defined characterisation steps, such as the choice of the suitability threshold used to determine potentially suitable habitats. While the subject has been discussed in the relevant literature (Liu et al. 2015), it ultimately remains an arbitrary choice which risks error. Another potential issue lies in the calibration of alpha from a *known probability of connectivity*. Used as a reference for calibration, the maximal or mean distance is often deduced from literature, or from controlled experiments that hardly represent the reality of the process. Once the reference distance has been decided and the probability of connectivity calculated, the conversion of the Euclidean distance into another unit of distance presents yet more room for error. Although this step is often overlooked, the example of expert landscape characterisation combined with circuit theory demonstrates the consequences of poor alpha calibration on connectivity probabilities and subsequent indices. Systematic analyses of the regression plot may improve calibration, a step which is often absent from connectivity studies, regardless of the connectivity approach.

## 4.3. Connectivity assessment

The impact of the modelling approach on the final connectivity estimation depends on the connectivity index. All measures of dPCflux present high correlation scores, particularly in the SDM, whereby the estimation derived from the Euclidean distance is highly correlated with LCP and circuit current estimations. In this instance, a direct measurement of the inter-patch Euclidean distance would provide the same outcomes as a more complex and time-consuming modelling framework. Patch area variation generally appears to be a strong driver of dPCflux variation. The flux, which presents more contrasting results, focuses on the variability of the potential overall flux of individuals. This provides more insight into inter-patch connectivity but neglects dispersal potential and intra-patch connectivity. The choice of connectivity index should reflect the conservation objective of the study. At an earlier stage in the study, an additional step is needed to characterise the relevant aspects of connectivity to be targeted. In some cases, the choice of relevant connectivity metrics might be oriented toward a set of indices instead of a single one; a strategy that may better capture the various functions of landscape elements.

In the absence of field validation protocols, studies employing comparative analysis to research the connectivity of modelling frameworks are at an impasse (McClure, Hansen, and Inman 2016; Makwana et al. 2023). Other comparison studies, such as Godet and Clauzel (2021) and Avon and Bergès (2016), similarly struggled to attest the reliability of connectivity outcomes.

# 5. Conclusion

To conclude, we demonstrated through a real case study how modeling choices can influence connectivity estimation. Connectivity maps and patch importance indices are usully used in landscape planning to support the location of contruction projects or conservation actions. These results are sensible to landscape characterisation hypothesis and movement estimation models. We emphasise that, when choosing a methodology, it is important to take into account the ecological assumptions and limitations of the modelling approach (economical, temporal, data, computer, etc.) at each stage of the study. *Landscape characterisation* determines the modelling output irrespective of pairwise connectivity. We, therefore, advocate landscape characterisation methods that conciliate species data and expert knowledge, and that the potential suitable areas be subject to field validation where applicable. Further research efforts are needed to develop modelling and field protocols in this regard. *Pairwise connectivity modelling* relates to animal movement capability. Distance-based approaches, such as Euclidean distance and least-cost path, are constrained by their ecological assumptions. We argue that their use should be limited to specific case studies where such simplifications can be justified. Stochastic approaches such as circuit theory and SMS present more holistic alternatives - but calibrating corresponding connectivity probabilities can prove difficult, as experienced in the expert approach. Individual-based models predict movement patterns more flexibly and intuitively and can integrate specific movement drivers. Whilst they present a promising alternative to connectivity modelling, their implementation can prove challenging. Connectivity assessment should reflect the complexity and multiplicity of interactions between the focal species and their landscape. A set of indices that capture the diverse functionality of the landscape appears more fit for purpose than a single connectivity index. We emphasise the need for field validation and protocol within the modelling framework, particularly for the systematic validation of potentially suitable habitat patches.

# Statements & Declarations

## Funding

Marie Soret and Sylvain Moulherat have received research support from TerrOïko, EDF, and the French National Association for Research and Development (ANRT) (CIFRE 2019/1738).

## Competing Interests

Marie Soret is employed by TerrOïko. Sylvain Moulherat is General Director and CTO of TerrOïko. Maxime Lenormand and Sandra Luque have no relevant financial or non-financial interests to disclose.

## Author Contributions

All authors contributed to the study's conception and design. S.M. provided the data. M.S. prepared the data, implemented the models, performed the analysis, and prepared the figures. M.S. wrote the first draft of the manuscript. All authors commented on previous versions and approved the final manuscript.

## Data and Code Availability

The datasets analysed during the current study are available at: https://cloud.terroiko.fr/index.php/s/GRaj DormmpRsrqS. The Python notebooks used to compute analysis are available at https://oikolab.terroiko.fr: 10001/publications/implication-of-modelling-choices-on-connectivity-estimation/-/tree/main.

# Supplementary Information

## S1. Expert opinion

***Table S1*** *expert opinion on suitable habitat and resistance cost for the* Alytes obstetricans *based on land cover classification used to produce the expert opinion landscape characterisation (Fig. 2a)*

| Habitat typology | Reproductive habitat | Resistance |
|---|---:|---:|
| **Forest** | | |
| Broad-leaved forest | no | 1 |
| Needle-leaved forest | no | 1 |
| Mixed forest | no | 1 |
| Riparian forest | no | 1 |
| **Plantation** | | |
| Orchard | no | 1 |
| Recently felled area | no | 1 |
| Hedgerow | no | 1 |
| Forestry plantation | no | 1 |
| **Open space with low vegetation** | | |
| Grassland | no | 1 |
| Shrub and herbaceous flooded | no | 1 |
| **Open space without vegetation** | | |
| Sandy Shore | no | 1 |
| Rock | no | 20 |
| Bare soil | no | 20 |
| Pathway | no | 20 |
| **Crop** | | |
| Vineyard | no | 1 |
| Mixed crop of market garden | no | 20 |
| Intensive unmixed crop | no | 30 |
| **Artificial area** | | |
| Sparsely artificialized | no | 10 |
| Rail network | no | 20 |
| Road network | no | 30 |
| Highway | no | 80 |
| **Waters** | | |
| Running waters | no | 50 |
| Standing waters | **yes** | 50 |

# S2. ODMAP protocol for species distribution models

The ODMAP protocol presented by Zurell et al. (2020) encompasses the general specifications and technical details involved in the species distribution models (SDM). The first part of the protocol covers the overview/conceptualisation of the modelling framework, including the model objective, the eco-spatio-temporal scope, the hypothesis and assumptions, the software used, and the availability of code and data. The second part develops the data aspect; their source and eventual pre-processing. The third part establishes the technical specifications of model fitting, including variable selection, model characterisation, model estimates, model section, and threshold selection. The "Assessment" discusses the performance and the plausibility of model outcomes. The final part concerns model predictions and uncertainty quantification. The protocol corresponds to the models used to produce the SDM landscape characterisation (Main Figure 2B).

## S2.1. Overview

### 1.1. Authorship

#### 1.1.1. Study title

Implication of modelling choices on connectivity estimation: a comparative analysis

#### 1.1.2. Authors' names

Marie Soret
Sylvain Moulherat
Maxime Lenormand
Sandra Luque

#### 1.1.3. Contact

marie.soret@terroiko.fr

#### 1.1.4. Study link

https://oikolab.terroiko.fr:10001/publications/implication-of-modelling-choices-on-connectivity-estimation

### 1.2. Model objective

#### 1.2.1. Model objective

Mapping and interpolation

#### 1.2.2. Target output

Binary map of potential suitable habitat patches
Resistance map for movement modelling

### 1.3. Focal taxon

Midwife toad (*Alytes obstetricans*)

### 1.4. Location

Périgord, Dordogne, France
See Figure 1 from the main document for the presentation of the study area and its location in France.

### 1.5. Scale of the analysis

#### 1.5.1. spatial extent

| | |
|---|---|
| Upper Left Corner | (0d49’21.06”E, 45d12’35.62”N) |
| Lower Left Corner | (0d49’38.66”E, 45d 5’ 3.61”N) |
| Upper Right Corner | (1d 6’28.89”E, 45d12’54.34”N) |
| Lower Right Corner | (1d 6’44.18”E, 45d 5’22.29”N) |

#### 1.5.2. Type of extent boundary
Rectangle

#### 1.5.3. Spatial resolution
10m

#### 1.5.4. Temporal extent
2015-2016

#### 1.5.5. Temporal resolution
One shot

### 1.6. Biodiversity data

#### 1.6.1. Observation type
Field Survey

#### 1.6.2. Response data type
presence-only

### 1.7. Predictors

#### 1.7.1. Predictor types
23 distance to land-cover classes
NB: Topographic and climatic variables are assumed constant in the range of the study area.
NB: To maintain the same level of landscape input, we used the same typology level as the one given to the experts.

### 1.8. Hypothesis
Ponds (reproductive habitat) have a positive influence on habitat suitability. Potential adult habitat classes such as forest, grassland, and shrubland, as well as sparsely artificial areas, have a positive influence on habitat suitability. Agricultural areas have a negative impact on habitat suitability due to agricultural practices.

### 1.9. Model assumptions
Ecological equilibrium between species and their environment;
No sampling bias;
Independence of species observation;
Availability of all important predictors;
Error-free predictors.

### 1.10. Algorithms

#### 1.10.1 Modeling techniques
Generalized linear model (GLM),
Generalized Boosting Regression (GBM),
Generalized Additive Model (GAM),
Classification tree (CTA),
Artificial Neural Network (ANN),

Surface Range Envelop (SRE),
Flexible Discriminant Analysis (FDA),
Multiple Adaptive Regression Splines (MARS),
Breiman and Cutler random forests for classification and regression (RF),
Maximum entropy modelling (MAXNET)

**1.10.2. Model Complexity**
We used 10 algorithms to account for model uncertainty. We used 5 sets of random pseudo-absences to account for the incomplete observation dataset. We repeated all training processes two times to account for the stochasticity at the initialisation. Therefore, each model was trained 10 times, resulting in 100 models.

**1.10.3 Model averaging**
We selected the models for the ensemble model based on a ROC threshold of 0,8. We combined the selected models with a weighted average rule based on the models' ROC scores.

## 1.11. Workflow

Step 1: Data formatting
Step 2: Pseudo-absence selection
Step 3: Data split in train-set and test-set
Step 4: Model training
Step 5: Model evaluation
Step 6: Model selection
Step 7: Ensemble model weight
Step 8: Ensemble model evaluation
Step 9: Individual model projection
Step 10: Ensemble model projection
Step 11: Binary threshold selection
Step 12: Conservation from suitability to resistance

## 1.12. Software

**1.12.1. Software**
R package biomod2

**1.12.2. Code availability**
https://oikolab.terroiko.fr:10001/publications/implication-of-modelling-choices-on-connectivity-estimation/-/tree/main/A_Landscape_characterisation?ref_type=heads

**1.12.3. Data Availability**
https://cloud.terroiko.fr/index.php/apps/files/files/998132?dir=/implication_of_modelling_choices_data

# S2.2. DATA

## 2.1. Biodiversity data

**2.1.1. Taxon names**
Midwife toad (*Alytes obstetricans)*

**2.1.2. Taxonomic reference system**
Laurenti, 1768

**2.1.3. Ecological level**
Individual

#### 2.1.4. Data sources
ITTECOP 2014 project : CIRFE - Cumul d'infrastructures linéaires de transports terrestres et relations écologiques fonctionnelles

#### 2.1.5. Data sources DOI
10.13140/RG.2.2.26710.60482

#### 2.1.6. Sampling design
Systematic night prospection of pre-identified potential suitable areas, opportunistic observations, MRR survey

#### 2.1.7. Sample size
712 vector points, 288 pixels

#### 2.1.8. Scaling
Observation points rasterized at 10m

#### 2.1.9. Absence data
No real absence data, pseudo-absence generated with biomod2

### 2.2. Data partitioning

#### 2.2.1. Training data
Depends on the model

#### 2.2.2. Validation data
Depends on the model

#### 2.2.3. Test data
No test data.
NB: option not available with biomod2 when using pseudo-absence

### 2.3. Predictors data

#### 2.3.1. Predictor variables
Distance to the rail network,
Distance to the sparsely artificialized areas,
Distance to the mixed crops of market garden,
Distance to bare soil,
Distance to hedgerow,
Distance to highway,
Distance to mixed forest,
Distance to needle-leaved forest,
Distance to surface standing waters,
Distance to surface running waters,
Distance to shrubland and herbaceous flooded,
Distance to rock,
Distance to the riparian forests,
Distance to forestry plantations,
Distance to the recently felled areas,
Distance to sandy shores,
Distance to pathway,
Distance to orchards
Distance to broad-leaved deciduous forest,
Distance to intensive unmixed crop,

Distance to vineyards,
Distance to the road network,
Distance to grassland.

**2.3.2. Data sources**
BD TOPO 2016 compilation enriched with field survey and photo interpretation in 2016

**2.3.3. spatial extent**

| | |
|---|---|
| Upper Left Corner | (0d49'21.06"E, 45d12'35.62"N) |
| Lower Left Corner | (0d49'38.66"E, 45d 5' 3.61"N) |
| Upper Right Corner | (1d 6'28.89"E, 45d12'54.34"N) |
| Lower Right Corner | (1d 6'44.18"E, 45d 5'22.29"N) |

**2.3.4. Spatial resolution**
Original data 1/2000 rasterize at 10 metres

**2.3.5. Coordinate reference system**
EPSG:2154 - RGF93 / Lambert-93

**2.3.6. Temporal extent**
2015-2016

**2.3.7. Temporal resolution**
One shot

**2.3.8. Data processing**
We used gdal proximity at 10m resolution on vector layers of the land-use classes

**2.3.9. Errors and biases**
No correction of error or bias

## S2.3. Model

### 3.1. Variable selection

**3.1.1. Variable pre-selection**
We selected the landscape classes at the same typologic resolution as the expert grid to ensure the same level of precision

**3.1.2. Collinearity**
No co-linearity

**3.2. Model setting**
Settings obtain with biomod2 command get_option(myBiomodModelOut)

**3.2.1 GLM**
type = quadratic
interaction level = 0
myFormula = NULL
test = AIC
family = binomial(link=lmustart = 0.5
control epsilon = 1e-08
control maxit = 50
control trace = FALSE

**3.2.2. GBM 2.1.8**

distribution = bernoulli
n trees = 2500
interaction depth = 7
n minobsinnode = 5
shrinkage = 0.001
bag fraction = 0.5
train fraction = 1
cv folds = 3
keep data = FALSE
verbose = FALSE
perf method = cv
n cores = 1

**3.2.3. GAM**
algo = GAM_mgcv
type = s_smoother
k = -1
family = binomial(link = logit)
method = GCV.Cp
optimizer = c('outer', 'newton')
select = FALSE
knots = NULL
paraPen = NULL
control nthreads = 1
control irls reg = 0
control epsilon = 1e-07
control maxit = 200
control trace = FALSE
control mgcv tol = 1e-07
control mgcv half = 15
control rank tol = 1.49011611938477e-08
control nlm ndigit = 7
control nlm gradtol = 1e-06
control nlm stepmax = 2
control nlm steptol = 1e-04
control nlm iterlim = 200
control nlm check analytical = 0
control optim factor = 1e+07
control newton conv tol = 1e-06
control newton maxNstep = 5
control newton maxSstep = 2
control newton maxHalf = 30
control newton use svd = 0
control outerPIstep = 0
control idLinksBases = TRUE
control scalePenalty = TRUE
control efs.lspmax = 15
control efs tol = 0.1
control keepData = FALSE
control scale est = fletcher
control edge correct = FALSE

#### 3.2.4. CTA
method = class
parms = default
cost = NULL
control xval = 5
control minbucket = 5
control minsplit = 5
control cp = 0.001
control maxdepth = 25

#### 3.2.5. ANN
3 repetitions
NbCV = 5
size = NULL
decay = NULL
rang = 0.1
maxit = 200

#### 3.2.6. SRE
quant = 0.025

#### 3.2.7. FDA
method = mars
add args = NULL

#### 3.2.8. MARS
type = simple
interaction level = 0
myFormula = NULL
nk = NULL
penalty = 2
thresh = 0.001
nprune = NULL
pmethod = backward

#### 3.2.9. RF
do classif = TRUE
ntree = 500
mtry = default
sampsize = NULL
nodesize = 5
maxnodes = NULL

#### 3.2.10. MAXNET
myFormula = NULL
regmult = 1
regfun = <function>

### 3.5. Threshold selection
We performed an iterative threshold analysis and compared the resulting suitable habitat areas with the expert prediction based on structural indices. Untimely we selected the threshold for which the number of suitable patches was the closest to the number in the expert's landscape characterisation. See Figure 3 in the section Materials and Methods of the main document.

# S2.4. Assessment

## 4.2. Plausibility check

### 4.2.1. Expert judgement
Comparison with expert opinion equivalent results.

# S2.5. Prediction

## 5.1. Prediction outputs

### 5.1.1. Prediction unit
suitability ranging from 1 to 1000

### 5.1.2. Post-processing
Binary threshold for potential habitat patches

Transformation from suitability to resistance to movement

# S3. dPCflux

The dPCflux component of PC was described by Saura and Rubio (2010):

$$d\,PCflux_k = \sum_{i=1}^{n} \sum_{j=1,\ j\neq i}^{n} a_i \ a_j \ p_{ij}^{*} \qquad \text{(Eq.S1)}$$

With $n$ the number of patches, $a$ the patch area, and $p_{ij}^{*}$ the maximal product probability of all possible paths (direct and stepping stones).

Since we kept the complete graphs, without any distance threshold or pruning, all patches are linked with a direct connection. In this case, we demonstrate, that for any ecological distance respecting (Eq.S2), the maximal product probability $p_{ij}^{*}$ is always equal to the direct probability of connectivity $p_{ij}$ (Eq.S3) :

$\forall \ i,j,k \in [1,n]^3$

$$d_{ij} \leq d_{ik} + d_{kj} \qquad \text{(Eq.S2)}$$

$$\Leftrightarrow$$

$$p_{ij}^{*} = \ p_{ij} \qquad \text{(Eq.S3)}$$

Demonstration (Eq.S2) $\Leftrightarrow$ (Eq.S3):

$\forall \ i,j,k \in [1,n]^3$

$$d_{ij} \quad \leq \quad d_{ik} + d_{kj} \qquad \text{(Eq.S2)}$$

$$\Leftrightarrow$$

$$-\alpha\, d_{ij} \quad \geq \quad -\alpha\left(d_{ik} + d_{kj}\right)$$

$$\Leftrightarrow$$

$$e^{-\alpha\, d_{ij}} \quad \geq \quad e^{-\alpha\, d_{ik} - \alpha\, d_{kj}}$$

$$\Leftrightarrow$$

$$e^{-\alpha\, d_{ij}} \quad \geq \quad e^{-\alpha\, d_{ik}} \times e^{-\alpha\, d_{kj}}$$

We recall $p(d) = e^{-ad}$, therefor:

$$p\left(d_{ij}\right) \ \geq \quad p\left(d_{ik}\right) \times p\left(d_{kj}\right)$$

The direct probability of probability is always superior or equal to any product probabilities, and since $p_{ij}^{*}$ is the maximal product probability of all possible paths, we deduced $\forall \ i,j,k \in [1,n]^3$ :

$$\text{(Eq.S2)} \quad d_{ij} \leq d_{ik} + d_{kj} \ \Leftrightarrow p\left(d_{ij}\right) \ \geq \ p\left(d_{ik}\right) \times p\left(d_{kj}\right) \Leftrightarrow \ p_{ij}^{*} = \ p_{ij} \qquad \text{(Eq.S3)}$$

Since the shortest path between two points is a straight line, (Eq.S2) is true for Euclidean distance, LCP length, and LCP cost. Regarding the resistance distance, we know from McRae (2006) and McRae et al. (2008) that *"the resistance distance increases [...] with its log transformation in two-dimensional habitats"*. The logarithm is an increasing function on the domain of definition of the distance (the set of real positive non-zero numbers), thus, the log-transformation conserves the order in the inequations. (Eq.S2) is true for the resistance distance.

We can simplify the expression of *dPCflux* as :

$$d\,PCflux_k = \sum_{i=1}^{n} \sum_{j=1,\ j\neq i}^{n} a_i \ a_j \ p_{ij} \qquad \text{(Eq. 6)}$$

# S4. Suitable habitats comparison

**Jaccard index formulas between the sets of patches expert $P_{\exp}$ and SDM $P_{sdm}$**

Jaccard of the number of intersections:

$$P_{\exp} \cap P_{sdm} / P_{\exp} \cup P_{sdm}$$

Jaccard of the area of intersection:

$$area(P_{\exp} \cap P_{sdm}) / area(P_{\exp} \cup P_{sdm})$$

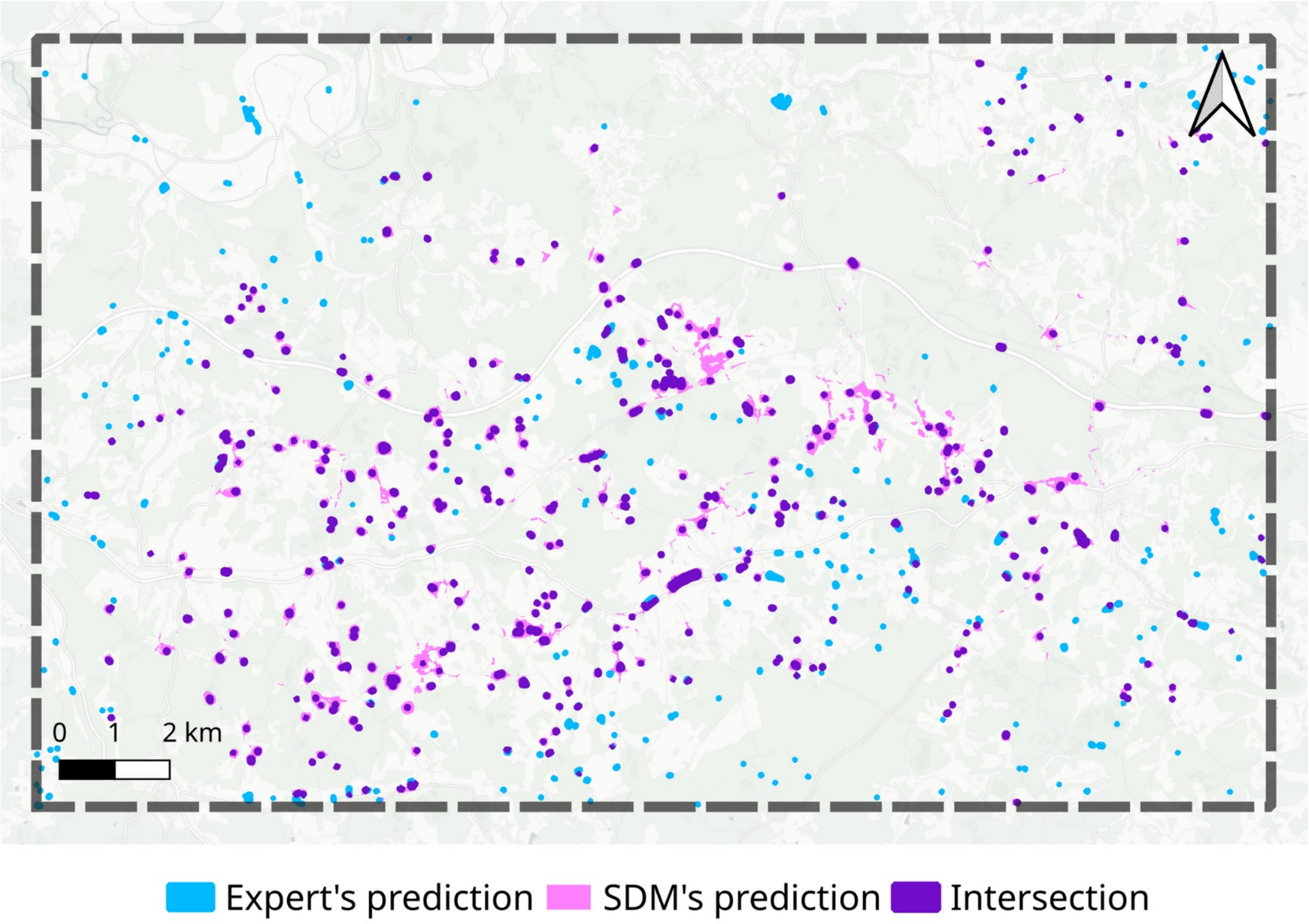

***Fig. S4*** *Comparison of suitable habitats prediction. Expert patches are represented in blue, SDM patches in pink, and their intersection is represented in purple.*

***Table S4*** *Comparison of patches structural indices.*

| Patch estimation | Expert | SDM | Intersection |
|---|---|---|---|
| Number of patches | 645 | 639 | 415 |
| Total area | 832 493 | 8 638 100 | 2 721 |
| Mean area | 1 289 | 13 518 | 1 041 |
| Min area | 17 | 100 | 1 |
| Area 1st quartile | 238 | 200 | 190 |
| Area 2nd quartile | 485 | 2 500 | 421 |
| Area 3rd quartile | 1 287 | 16 900 | 1 098 |
| Max area | 50 053 | 212 700 | 46 352 |

# S5. Movement maps correlations

## S5.a. No kernel

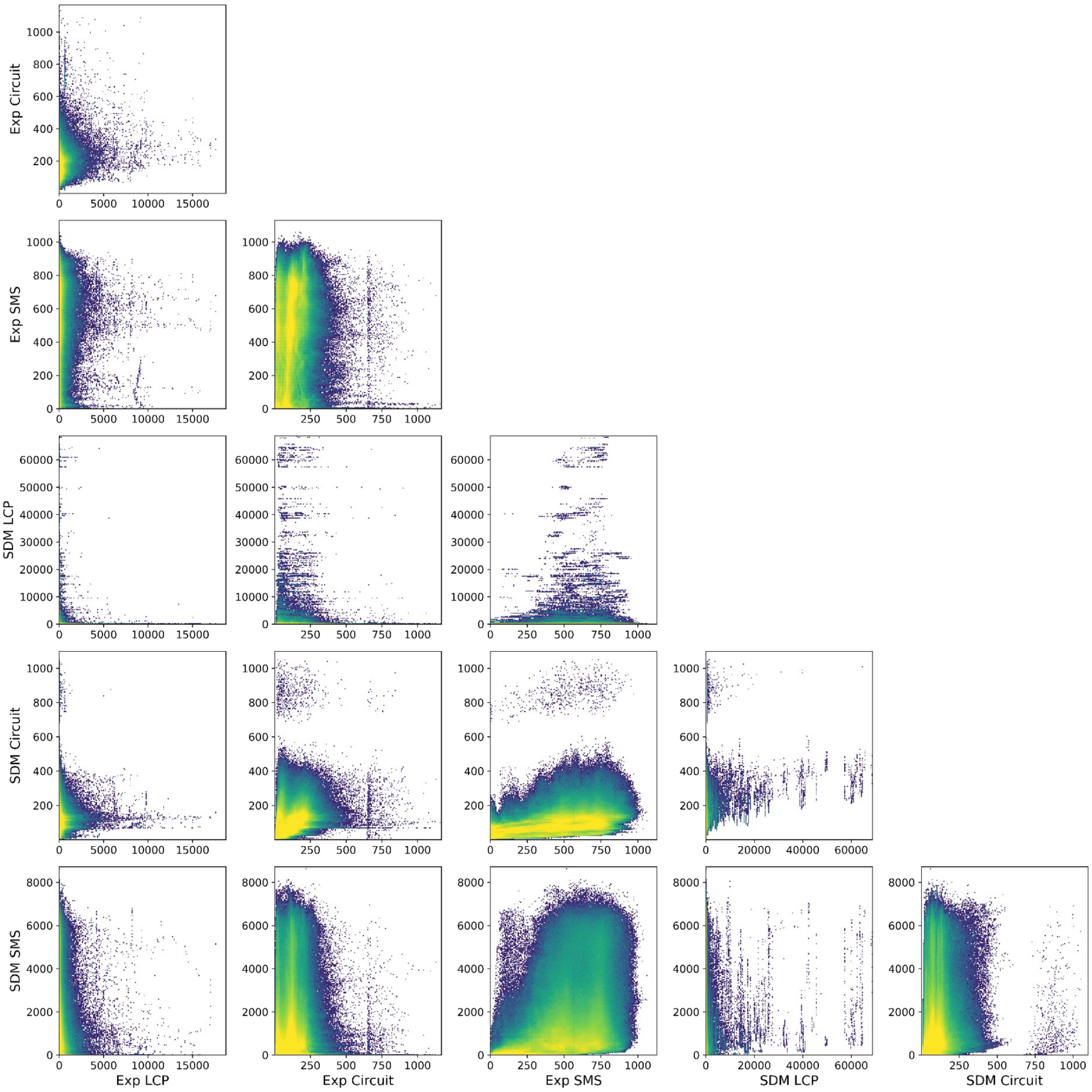

***Fig. S5a*** *Movement correlation plot. Each point represents a pixel of movement maps (main Fig. 4). See Table S5a below for corresponding correlation coefficients and p-values.*

***Table S5a*** *movement maps correlation coefficients and p-values.* $(\rho)$ *indicates Pearson correlation,* $(r)$ *Spearman and* $(\tau)$ *Kendall. Highlighted values correspond to p-values above the statistical significance threshold of 0.01. See Fig. S5a for corresponding scatterplots.*

| | | Expert | | | SDM | | |
|---|---|---|---|---|---|---|---|
| | | LCP | Circuit | SMS | LCP | Circuit | SMS |
| Expert | LCP | | *p-values*<br>$(\rho)$ 0.00<br>$(r)$ 0.00<br>$(\tau)$ 0.00 | *p-values*<br>$(\rho)$ 0.00<br>$(r)$ 0.00<br>$(\tau)$ 0.00 | *p-values*<br>$(\rho)$ 0.00<br>$(r)$ 0.00<br>$(\tau)$ 0.00 | *p-values*<br>$(\rho)$ 0.00<br>$(r)$ 0.00<br>$(\tau)$ 0.00 | *p-values*<br>$(\rho)$ 0.00<br>$(r)$ 0.00<br>$(\tau)$ 0.00 |
| | Circuit | *coefficients*<br>$(\rho)$ 0.28<br>$(r)$ 0.56<br>$(\tau)$ 0.42 | | *p-values*<br>$(\rho)$ 0.00<br>$(r)$ 0.00<br>$(\tau)$ 0.00 | *p-values*<br>$(\rho)$ 0.00<br>$(r)$ 0.00<br>$(\tau)$ 0.00 | *p-values*<br>$(\rho)$ 0.00<br>$(r)$ 0.00<br>$(\tau)$ 0.00 | *p-values*<br>$(\rho)$ 0.00<br>$(r)$ 0.00<br>$(\tau)$ 0.00 |
| | SMS | *coefficients*<br>$(\rho)$ 0.08<br>$(r)$ 0.23<br>$(\tau)$ 0.17 | *coefficients*<br>$(\rho)$ 0.25<br>$(r)$ 0.30<br>$(\tau)$ 0.21 | | *p-values*<br>$(\rho)$ 0.00<br>$(r)$ 0.00<br>$(\tau)$ 0.00 | *p-values*<br>$(\rho)$ 0.00<br>$(r)$ 0.00<br>$(\tau)$ 0.00 | *p-values*<br>$(\rho)$ 0.00<br>$(r)$ 0.00<br>$(\tau)$ 0.00 |
| SDM | LCP | *coefficients*<br>$(\rho)$ 0.00<br>$(r)$ -0.01<br>$(\tau)$ -0.01 | *coefficients*<br>$(\rho)$ 0.01<br>$(r)$ 0.01<br>$(\tau)$ 0.01 | *coefficients*<br>$(\rho)$ 0.03<br>$(r)$ 0.07<br>$(\tau)$ 0.06 | | *p-values*<br>$(\rho)$ 0.00<br>$(r)$ 0.00<br>$(\tau)$ 0.00 | *p-values*<br>$(\rho)$ 0.00<br>$(r)$ 0.00<br>$(\tau)$ 0.00 |
| | Circuit | *coefficients*<br>$(\rho)$ 0.10<br>$(r)$ 0.20<br>$(\tau)$ 0.15 | *coefficients*<br>$(\rho)$ 0.36<br>$(r)$ 0.42<br>$(\tau)$ 0.32 | *coefficients*<br>$(\rho)$ 0.47<br>$(r)$ 0.53<br>$(\tau)$ 0.37 | *coefficients*<br>$(\rho)$ 0.15<br>$(r)$ 0.18<br>$(\tau)$ 0.14 | | *p-values*<br>$(\rho)$ 0.00<br>$(r)$ 0.00<br>$(\tau)$ 0.00 |
| | SMS | *coefficients*<br>$(\rho)$ 0.02<br>$(r)$ 0.14<br>$(\tau)$ 0.11 | *coefficients*<br>$(\rho)$ 0.10<br>$(r)$ 0.18<br>$(\tau)$ 0.12 | *coefficients*<br>$(\rho)$ 0.47<br>$(r)$ 0.64<br>$(\tau)$ 0.46 | *coefficients*<br>$(\rho)$ 0.02<br>$(r)$ 0.04<br>$(\tau)$ 0.04 | *coefficients*<br>$(\rho)$ 0.22<br>$(r)$ 0.37<br>$(\tau)$ 0.25 | |

## S5.b. 5-by-5 blurring kernel

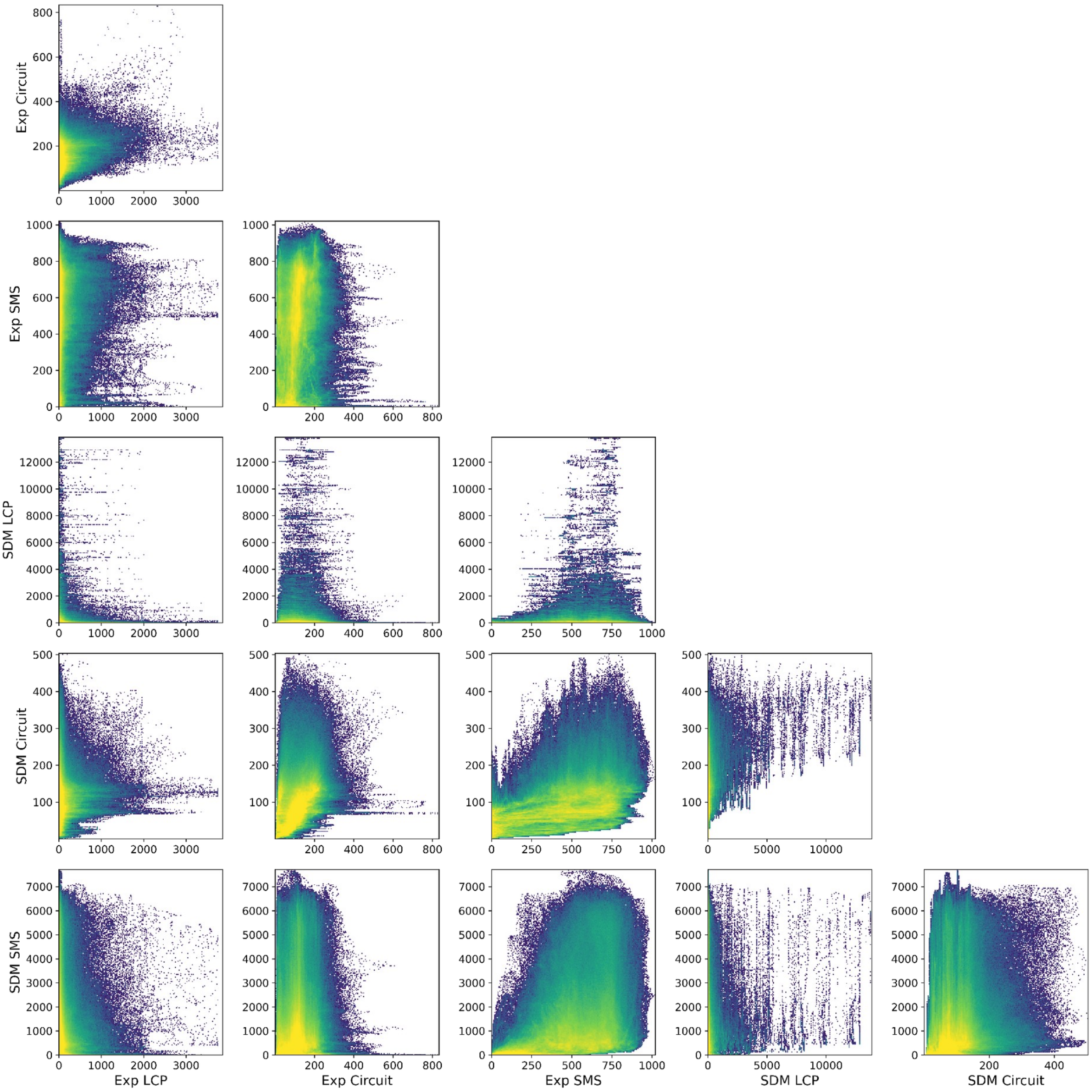

***Fig. S5b*** *Movement 5-by-5 blur correlation plot. Each point represents a pixel of movement maps. See Table S5b below for corresponding correlation coefficients and p-values.*

***Table S5b*** *movement maps 5-by-5 blur correlation coefficients and p-values.* $(\rho)$ *indicates Pearson correlation,* $(r)$ *Spearman and* $(\tau)$ *Kendall. Highlighted values correspond to correlation coefficients above 0.50. See Fig. S5b above for corresponding scatterplots.*

| | | Expert | | | SDM | | |
|---|---|---|---|---|---|---|---|
| | | LCP | Circuit | SMS | LCP | Circuit | SMS |
| Expert | LCP | | *p-values*<br>$(\rho)$ 0.00<br>$(r)$ 0.00<br>$(\tau)$ 0.00 | *p-values*<br>$(\rho)$ 0.00<br>$(r)$ 0.00<br>$(\tau)$ 0.00 | *p-values*<br>$(\rho)$ 0.00<br>$(r)$ 0.00<br>$(\tau)$ 0.00 | *p-values*<br>$(\rho)$ 0.00<br>$(r)$ 0.00<br>$(\tau)$ 0.00 | *p-values*<br>$(\rho)$ 0.00<br>$(r)$ 0.00<br>$(\tau)$ 0.00 |
| | Circuit | *coefficients*<br>$(\rho)$ 0.41<br>$(r)$ 0.63<br>$(\tau)$ 0.47 | | *p-values*<br>$(\rho)$ 0.00<br>$(r)$ 0.00<br>$(\tau)$ 0.00 | *p-values*<br>$(\rho)$ 0.00<br>$(r)$ 0.00<br>$(\tau)$ 0.00 | *p-values*<br>$(\rho)$ 0.00<br>$(r)$ 0.00<br>$(\tau)$ 0.00 | *p-values*<br>$(\rho)$ 0.00<br>$(r)$ 0.00<br>$(\tau)$ 0.00 |
| | SMS | *coefficients*<br>$(\rho)$ 0.13<br>$(r)$ 0.25<br>$(\tau)$ 0.17 | *coefficients*<br>$(\rho)$ 0.28<br>$(r)$ 0.33<br>$(\tau)$ 0.23 | | *p-values*<br>$(\rho)$ 0.00<br>$(r)$ 0.00<br>$(\tau)$ 0.00 | *p-values*<br>$(\rho)$ 0.00<br>$(r)$ 0.00<br>$(\tau)$ 0.00 | *p-values*<br>$(\rho)$ 0.00<br>$(r)$ 0.00<br>$(\tau)$ 0.00 |
| SDM | LCP | *coefficients*<br>$(\rho)$ 0.02<br>$(r)$ 0.06<br>$(\tau)$ 0.05 | *coefficients*<br>$(\rho)$ 0.05<br>$(r)$ 0.09<br>$(\tau)$ 0.07 | *coefficients*<br>$(\rho)$ 0.08<br>$(r)$ 0.18<br>$(\tau)$ 0.14 | | *p-values*<br>$(\rho)$ 0.00<br>$(r)$ 0.00<br>$(\tau)$ 0.00 | *p-values*<br>$(\rho)$ 0.00<br>$(r)$ 0.00<br>$(\tau)$ 0.00 |
| | Circuit | *coefficients*<br>$(\rho)$ 0.16<br>$(r)$ 0.27<br>$(\tau)$ 0.19 | *coefficients*<br>$(\rho)$ 0.41<br>$(r)$ 0.47<br>$(\tau)$ 0.35 | *coefficients*<br>$(\rho)$ 0.48<br>$(r)$ 0.53<br>$(\tau)$ 0.37 | *coefficients*<br>$(\rho)$ 0.32<br>$(r)$ 0.39<br>$(\tau)$ 0.31 | | *p-values*<br>$(\rho)$ 0.00<br>$(r)$ 0.00<br>$(\tau)$ 0.00 |
| | SMS | *coefficients*<br>$(\rho)$ 0.04<br>$(r)$ 0.15<br>$(\tau)$ 0.10 | *coefficients*<br>$(\rho)$ 0.11<br>$(r)$ 0.19<br>$(\tau)$ 0.13 | *coefficients*<br>$(\rho)$ 0.48<br>$(r)$ 0.64<br>$(\tau)$ 0.47 | *coefficients*<br>$(\rho)$ 0.04<br>$(r)$ 0.11<br>$(\tau)$ 0.09 | *coefficients*<br>$(\rho)$ 0.23<br>$(r)$ 0.37<br>$(\tau)$ 0.25 | |

## S5.c. 10-by-10 blurring kernel

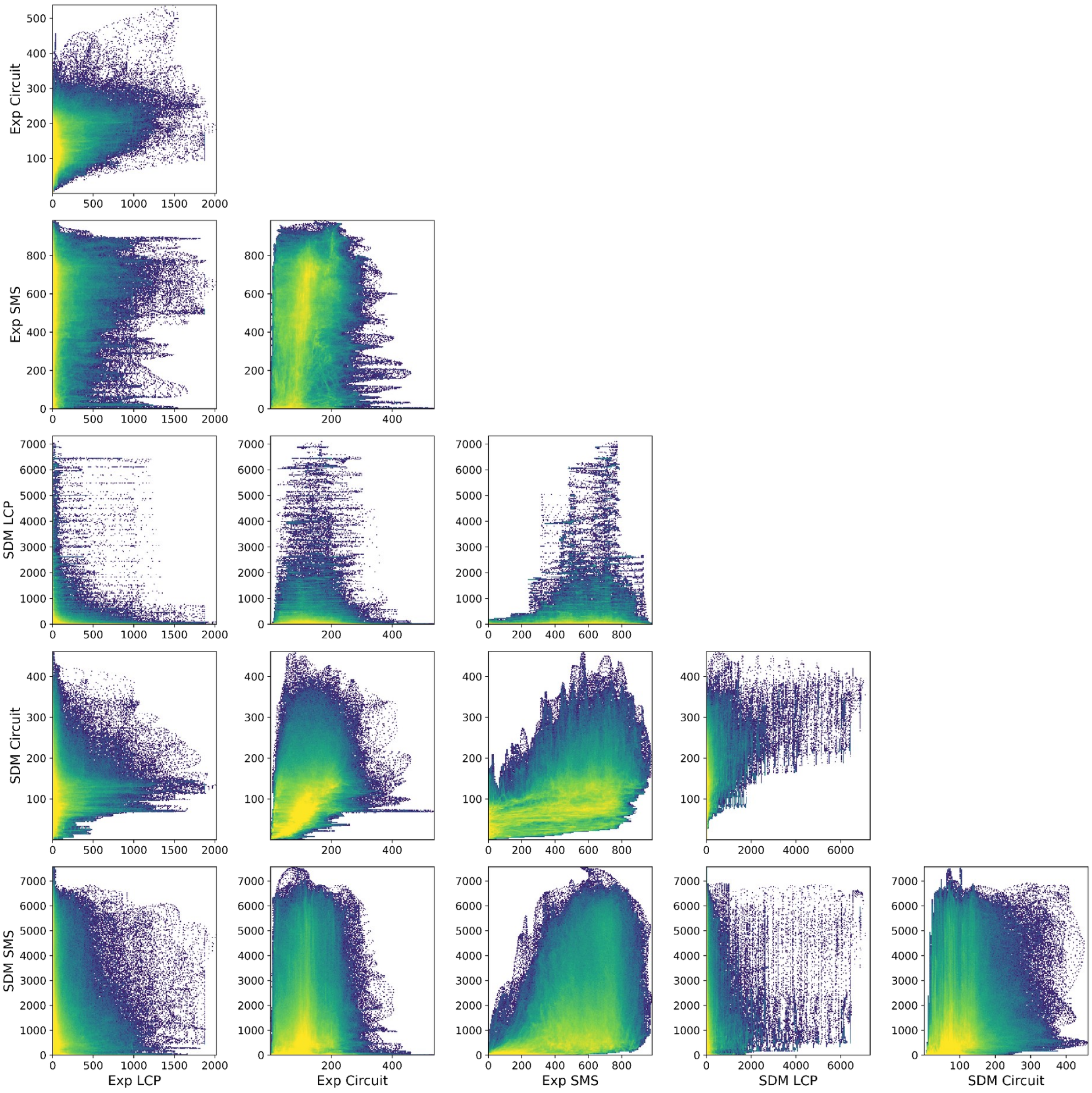

***Fig. S5c*** *Movement 10-by-10 blur correlation plot. Each point represents a pixel of movement maps. See Table S5c below for corresponding correlation coefficients and p-values.*

***Table S5c*** *movement maps 10-by-10 blur correlation coefficients and p-values.* ( $\rho$ ) *indicates Pearson correlation,* ( $r$ ) *Spearman and* ( $\tau$ ) *Kendall. Highlighted values correspond to correlation coefficients above 0.50. See Fig. S5c above for corresponding scatterplots.*

| | | Expert | | | SDM | | |
|---|---|---|---|---|---|---|---|
| | | LCP | Circuit | SMS | LCP | Circuit | SMS |
| Expert | LCP | | *p-values* ($\rho$) 0.00 ($r$) 0.00 ($\tau$) 0.00 | *p-values* ($\rho$) 0.00 ($r$) 0.00 ($\tau$) 0.00 | *p-values* ($\rho$) 0.00 ($r$) 0.00 ($\tau$) 0.00 | *p-values* ($\rho$) 0.00 ($r$) 0.00 ($\tau$) 0.00 | *p-values* ($\rho$) 0.00 ($r$) 0.00 ($\tau$) 0.00 |
| | Circuit | *coefficients* ($\rho$) 0.46 ($r$) 0.64 ($\tau$) 0.47 | | *p-values* ($\rho$) 0.00 ($r$) 0.00 ($\tau$) 0.00 | *p-values* ($\rho$) 0.00 ($r$) 0.00 ($\tau$) 0.00 | *p-values* ($\rho$) 0.00 ($r$) 0.00 ($\tau$) 0.00 | *p-values* ($\rho$) 0.00 ($r$) 0.00 ($\tau$) 0.00 |
| | SMS | *coefficients* ($\rho$) 0.15 ($r$) 0.27 ($\tau$) 0.19 | *coefficients* ($\rho$) 0.31 ($r$) 0.35 ($\tau$) 0.24 | | *p-values* ($\rho$) 0.00 ($r$) 0.00 ($\tau$) 0.00 | *p-values* ($\rho$) 0.00 ($r$) 0.00 ($\tau$) 0.00 | *p-values* ($\rho$) 0.00 ($r$) 0.00 ($\tau$) 0.00 |
| SDM | LCP | *coefficients* ($\rho$) 0.03 ($r$) 0.14 ($\tau$) 0.11 | *coefficients* ($\rho$) 0.08 ($r$) 0.17 ($\tau$) 0.12 | *coefficients* ($\rho$) 0.11 ($r$) 0.25 ($\tau$) 0.19 | | *p-values* ($\rho$) 0.00 ($r$) 0.00 ($\tau$) 0.00 | *p-values* ($\rho$) 0.00 ($r$) 0.00 ($\tau$) 0.00 |
| | Circuit | *coefficients* ($\rho$) 0.20 ($r$) 0.33 ($\tau$) 0.23 | *coefficients* ($\rho$) 0.45 ($r$) 0.51 ($\tau$) 0.38 | *coefficients* ($\rho$) 0.49 ($r$) 0.54 ($\tau$) 0.37 | *coefficients* ($\rho$) 0.40 ($r$) 0.51 ($\tau$) 0.39 | | *p-values* ($\rho$) 0.00 ($r$) 0.00 ($\tau$) 0.00 |
| | SMS | *coefficients* ($\rho$) 0.05 ($r$) 0.16 ($\tau$) 0.11 | *coefficients* ($\rho$) 0.12 ($r$) 0.20 ($\tau$) 0.14 | *coefficients* ($\rho$) 0.50 ($r$) 0.65 ($\tau$) 0.47 | *coefficients* ($\rho$) 0.06 ($r$) 0.16 ($\tau$) 0.12 | *coefficients* ($\rho$) 0.24 ($r$) 0.37 ($\tau$) 0.25 | |

## S5.d. 20-by-20 blurring kernel

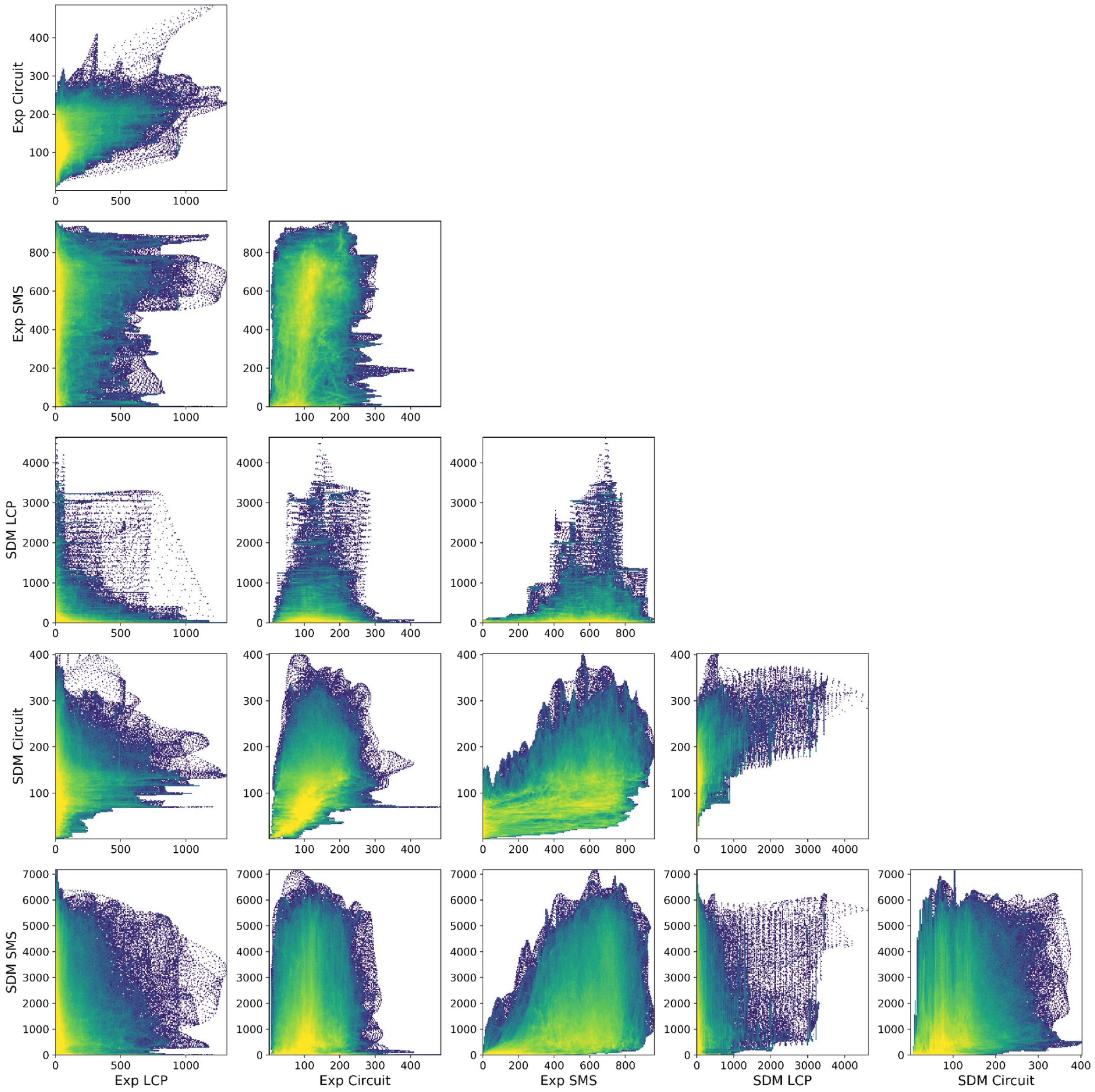

***Fig. S5d*** *Movement 20-by-20 blur correlation plot. Each point represents a pixel of movement maps. See Table S5d below for corresponding correlation coefficients and p-values.*

***Table S5d*** *movement maps 20-by-20 blur correlation coefficients and p-values. ( ρ ) indicates Pearson correlation, ( r ) Spearman and ( τ ) Kendall. Highlighted values correspond to correlation coefficients above 0.50. See Fig. S5d above for corresponding scatterplots.*

| | | Expert | | | SDM | | |
|---|---|---|---|---|---|---|---|
| | | LCP | Circuit | SMS | LCP | Circuit | SMS |
| Expert | LCP | | *p-values*<br>(ρ) 0.00<br>(r) 0.00<br>(τ) 0.00 | *p-values*<br>(ρ) 0.00<br>(r) 0.00<br>(τ) 0.00 | *p-values*<br>(ρ) 0.00<br>(r) 0.00<br>(τ) 0.00 | *p-values*<br>(ρ) 0.00<br>(r) 0.00<br>(τ) 0.00 | *p-values*<br>(ρ) 0.00<br>(r) 0.00<br>(τ) 0.00 |
| | Circuit | *coefficients*<br>(ρ) 0.53<br>(r) 0.66<br>(τ) 0.47 | | *p-values*<br>(ρ) 0.00<br>(r) 0.00<br>(τ) 0.00 | *p-values*<br>(ρ) 0.00<br>(r) 0.00<br>(τ) 0.00 | *p-values*<br>(ρ) 0.00<br>(r) 0.00<br>(τ) 0.00 | *p-values*<br>(ρ) 0.00<br>(r) 0.00<br>(τ) 0.00 |
| | SMS | *coefficients*<br>(ρ) 0.19<br>(r) 0.30<br>(τ) 0.21 | *coefficients*<br>(ρ) 0.34<br>(r) 0.33<br>(τ) 0.27 | | *p-values*<br>(ρ) 0.00<br>(r) 0.00<br>(τ) 0.00 | *p-values*<br>(ρ) 0.00<br>(r) 0.00<br>(τ) 0.00 | *p-values*<br>(ρ) 0.00<br>(r) 0.00<br>(τ) 0.00 |
| SDM | LCP | *coefficients*<br>(ρ) 0.06<br>(r) 0.25<br>(τ) 0.18 | *coefficients*<br>(ρ) 0.14<br>(r) 0.29<br>(τ) 0.21 | *coefficients*<br>(ρ) 0.16<br>(r) 0.34<br>(τ) 0.25 | | *p-values*<br>(ρ) 0.00<br>(r) 0.00<br>(τ) 0.00 | *p-values*<br>(ρ) 0.00<br>(r) 0.00<br>(τ) 0.00 |
| | Circuit | *coefficients*<br>(ρ) 0.26<br>(r) 0.40<br>(τ) 0.28 | *coefficients*<br>(ρ) 0.52<br>(r) 0.58<br>(τ) 0.42 | *coefficients*<br>(ρ) 0.51<br>(r) 0.54<br>(τ) 0.37 | *coefficients*<br>(ρ) 0.48<br>(r) 0.64<br>(τ) 0.49 | | *p-values*<br>(ρ) 0.00<br>(r) 0.00<br>(τ) 0.00 |
| | SMS | *coefficients*<br>(ρ) 0.06<br>(r) 0.19<br>(τ) 0.12 | *coefficients*<br>(ρ) 0.14<br>(r) 0.22<br>(τ) 0.15 | *coefficients*<br>(ρ) 0.53<br>(r) 0.67<br>(τ) 0.48 | *coefficients*<br>(ρ) 0.10<br>(r) 0.22<br>(τ) 0.16 | *coefficients*<br>(ρ) 0.26<br>(r) 0.37<br>(τ) 0.26 | |

# S6. Probabilities of connectivity correlations

## S6.a. Expert approach

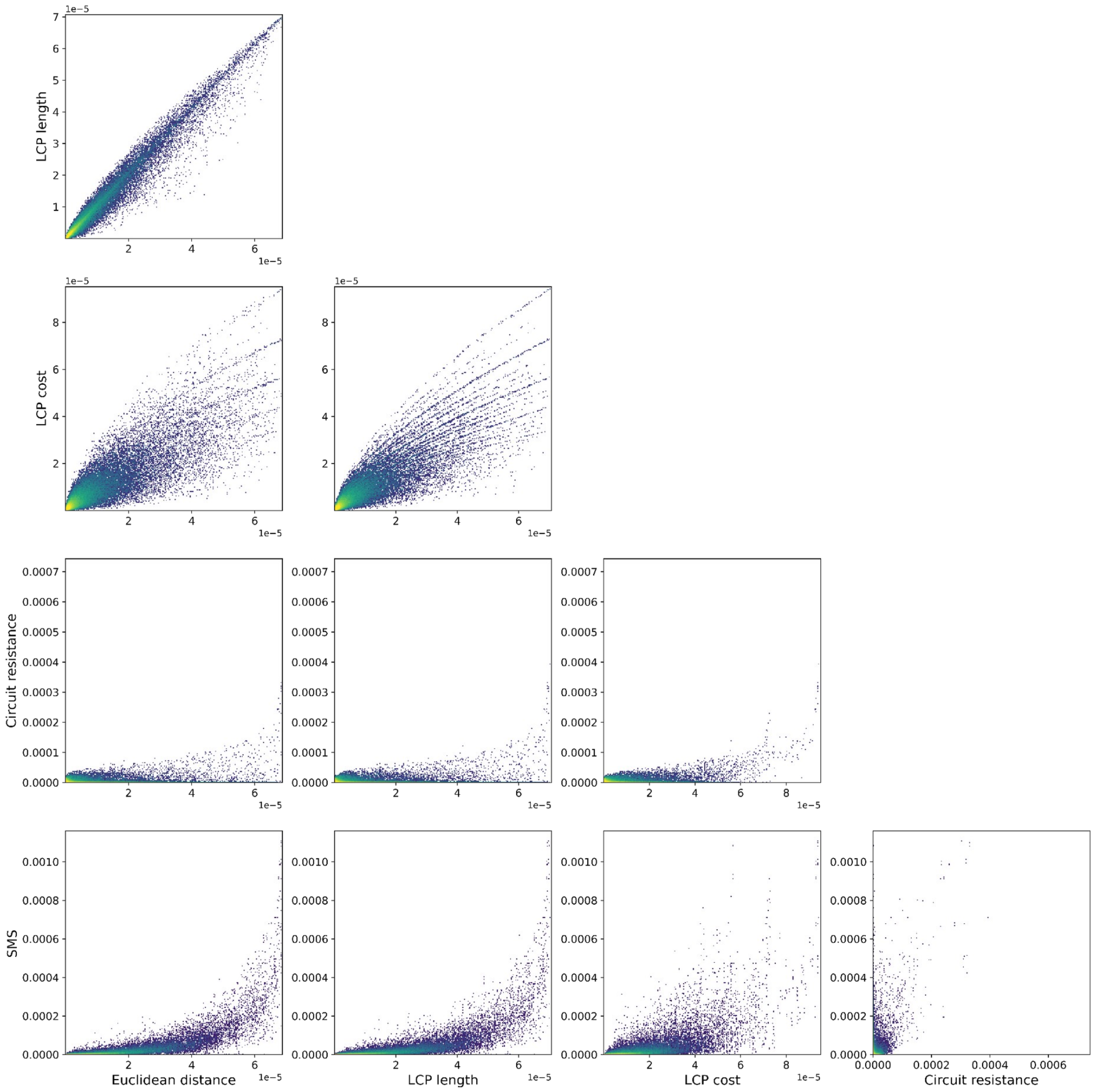

***Fig. S6a*** *Scatterplots of expert's probabilities of connectivity (Table 2) according to different connectivity estimations. Each point represents a pair of expert patches. See Table S6a for corresponding correlation parameters.*

***Table S6a*** *Expert connectivity probabilities correlation coefficients and p-values.* $(\rho)$ *indicates Pearson correlation,* $(r)$ *Spearman and* $(\tau)$ *Kendall. Highlighted values correspond to p-values above the statistical significance threshold of 0.01. See Figure S6a for corresponding scatterplots.*

| | Euclidean distance | LCP length | LCP cost | Circuit resistance | SMS flow |
|---|---|---|---|---|---|
| Euclidean distance | | *p-values*<br>$(\rho)$ 0.00<br>$(r)$ 0.00<br>$(\tau)$ 0.00 | *p-values*<br>$(\rho)$ 0.00<br>$(r)$ 0.00<br>$(\tau)$0.00 | *p-values*<br>$(\rho)$ 0.00<br>$(r)$ 0.19<br>$(\tau)$ 0.59 | *p-values*<br>$(\rho)$ 0.00<br>$(r)$ 0.00<br>$(\tau)$ 0.00 |
| LCP length | *coefficients*<br>$(\rho)$ 0.99<br>$(r)$ 0.99<br>$(\tau)$ 0.93 | | *p-values*<br>$(\rho)$ 0.00<br>$(r)$ 0.00<br>$(\tau)$ 0.00 | *p-values*<br>$(\rho)$ 0.00<br>$(r)$ 0.00<br>$(\tau)$ 0.00 | *p-values*<br>$(\rho)$ 0.00<br>$(r)$ 0.00<br>$(\tau)$ 0.00 |
| LCP cost | *coefficients*<br>$(\rho)$ 0.90<br>$(r)$ 0.98<br>$(\tau)$ 0.86 | *coefficients*<br>$(\rho)$ 0.92<br>$(r)$ 0.98<br>$(\tau)$ 0.86 | | *p-values*<br>$(\rho)$ 0.00<br>$(r)$ 0.00<br>$(\tau)$ 0.00 | *p-values*<br>$(\rho)$ 0.00<br>$(r)$ 0.00<br>$(\tau)$ 0.00 |
| Circuit resistance | *coefficients*<br>$(\rho)$ 0.41<br>$(r)$-0.00<br>$(\tau)$-0.00 | *coefficients*<br>$(\rho)$ 0.42<br>$(r)$ 0.01<br>$(\tau)$ 0.00 | *coefficients*<br>$(\rho)$ 0.58<br>$(r)$ 0.10<br>$(\tau)$ 0.07 | | *p-values*<br>$(\rho)$ 0.00<br>$(r)$ 0.00<br>$(\tau)$ 0.00 |
| SMS flow | *coefficients*<br>$(\rho)$ 0.64<br>$(r)$ 0.66<br>$(\tau)$ 0.54 | *coefficients*<br>$(\rho)$ 0.64<br>$(r)$ 0.66<br>$(\tau)$ 0.54 | *coefficients*<br>$(\rho)$ 0.56<br>$(r)$ 0.64<br>$(\tau)$ 0.53 | *coefficients*<br>$(\rho)$ 0.07<br>$(r)$ 0.06<br>$(\tau)$ 0.05 | |

## S6.b. SDM approach

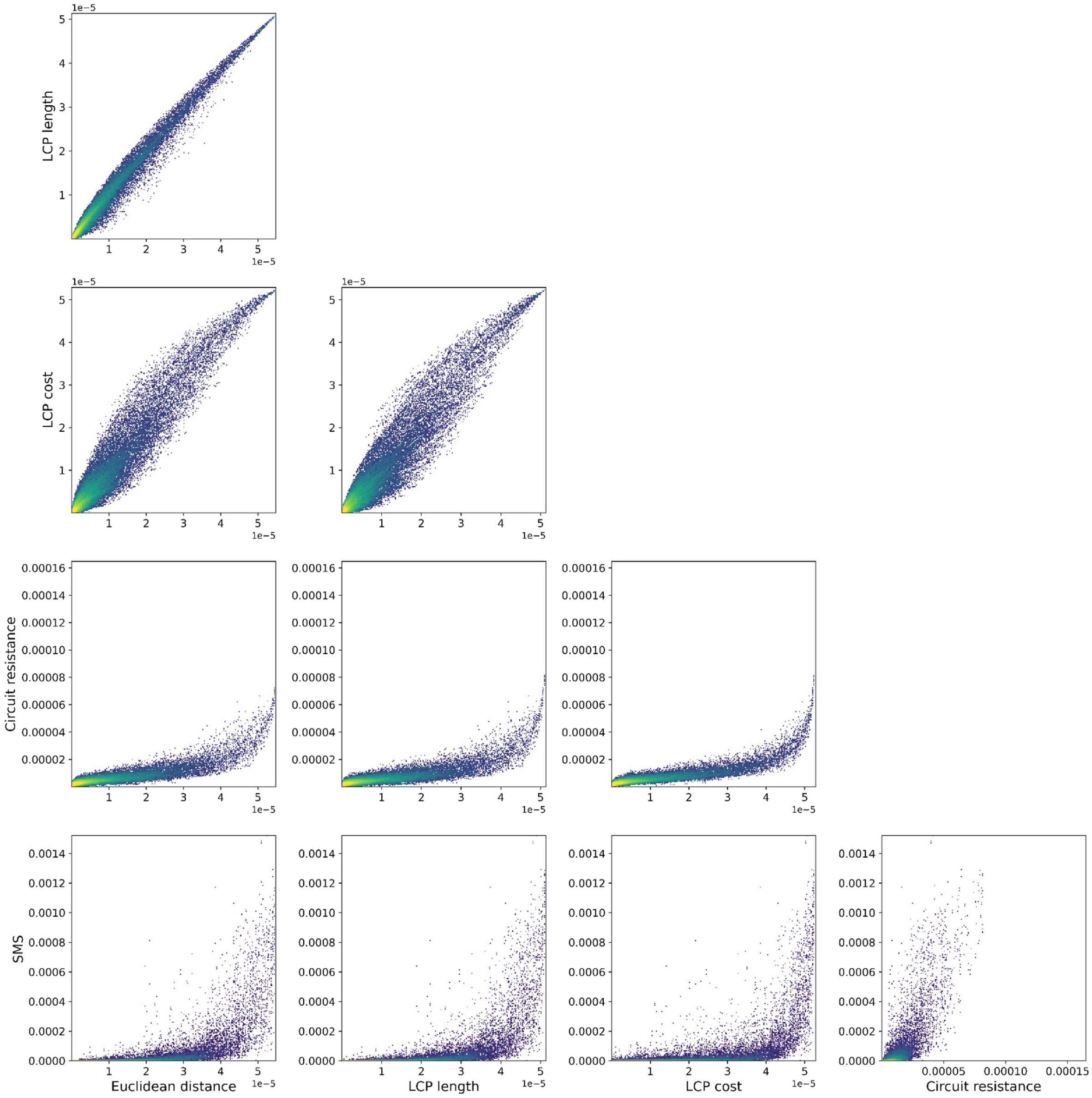

***Fig. S6b*** *Scatterplots of SDM connectivity probabilities (Table 2) according to different connectivity estimations. Each point represents a pair of SDM patches. See Table S6b for corresponding correlation parameters.*

***Table S6.B*** *SDM connectivity probabilities correlation coefficients and p-values.* $(\rho)$ *indicates Pearson correlation,* $(r)$ *Spearman and* $(\tau)$ *Kendall. See Figure S6.B for corresponding scatterplots.*

| | Euclidean distance | LCP length | LCP cost | Circuit resistance | SMS flow |
|---|---|---|---|---|---|
| Euclidean distance | | *p-values*<br>$(\rho)$ 0.00<br>$(r)$ 0.00<br>$(\tau)$ 0.00 | *p-values*<br>$(\rho)$ 0.00<br>$(r)$ 0.00<br>$(\tau)$ 0.00 | *p-values*<br>$(\rho)$ 0.00<br>$(r)$ 0.00<br>$(\tau)$ 0.00 | *p-values*<br>$(\rho)$ 0.00<br>$(r)$ 0.00<br>$(\tau)$ 0.00 |
| LCP length | *coefficients*<br>$(\rho)$ 0.99<br>$(r)$ 0.99<br>$(\tau)$ 0.92 | | *p-values*<br>$(\rho)$ 0.00<br>$(r)$ 0.00<br>$(\tau)$ 0.00 | *p-values*<br>$(\rho)$ 0.00<br>$(r)$ 0.00<br>$(\tau)$ 0.00 | *p-values*<br>$(\rho)$ 0.00<br>$(r)$ 0.00<br>$(\tau)$ 0.00 |
| LCP cost | *coefficients*<br>$(\rho)$ 0.97<br>$(r)$ 0.97<br>$(\tau)$ 0.85 | *coefficients*<br>$(\rho)$ 0.97<br>$(r)$ 0.96<br>$(\tau)$ 0.84 | | *p-values*<br>$(\rho)$ 0.00<br>$(r)$ 0.00<br>$(\tau)$ 0.00 | *p-values*<br>$(\rho)$ 0.00<br>$(r)$ 0.00<br>$(\tau)$ 0.00 |
| Circuit resistance | *coefficients*<br>$(\rho)$ 0.68<br>$(r)$ 0.93<br>$(\tau)$ 0.77 | *coefficients*<br>$(\rho)$ 0.67<br>$(r)$ 0.92<br>$(\tau)$ 0.76 | *coefficients*<br>$(\rho)$ 0.67<br>$(r)$ 0.95<br>$(\tau)$ 0.80 | | *p-values*<br>$(\rho)$ 0.00<br>$(r)$ 0.00<br>$(\tau)$ 0.00 |
| SMS flow | *coefficients*<br>$(\rho)$ 0.48<br>$(r)$ 0.45<br>$(\tau)$ 0.37 | *coefficients*<br>$(\rho)$ 0.48<br>$(r)$ 0.45<br>$(\tau)$ 0.37 | *coefficients*<br>$(\rho)$ 0.47<br>$(r)$ 0.43<br>$(\tau)$ 0.35 | *coefficients*<br>$(\rho)$ 0.31<br>$(r)$ 0.42<br>$(\tau)$ 0.34 | |

# S7. Connectivity indices correlations

## S7.a. Patch-based expert approach

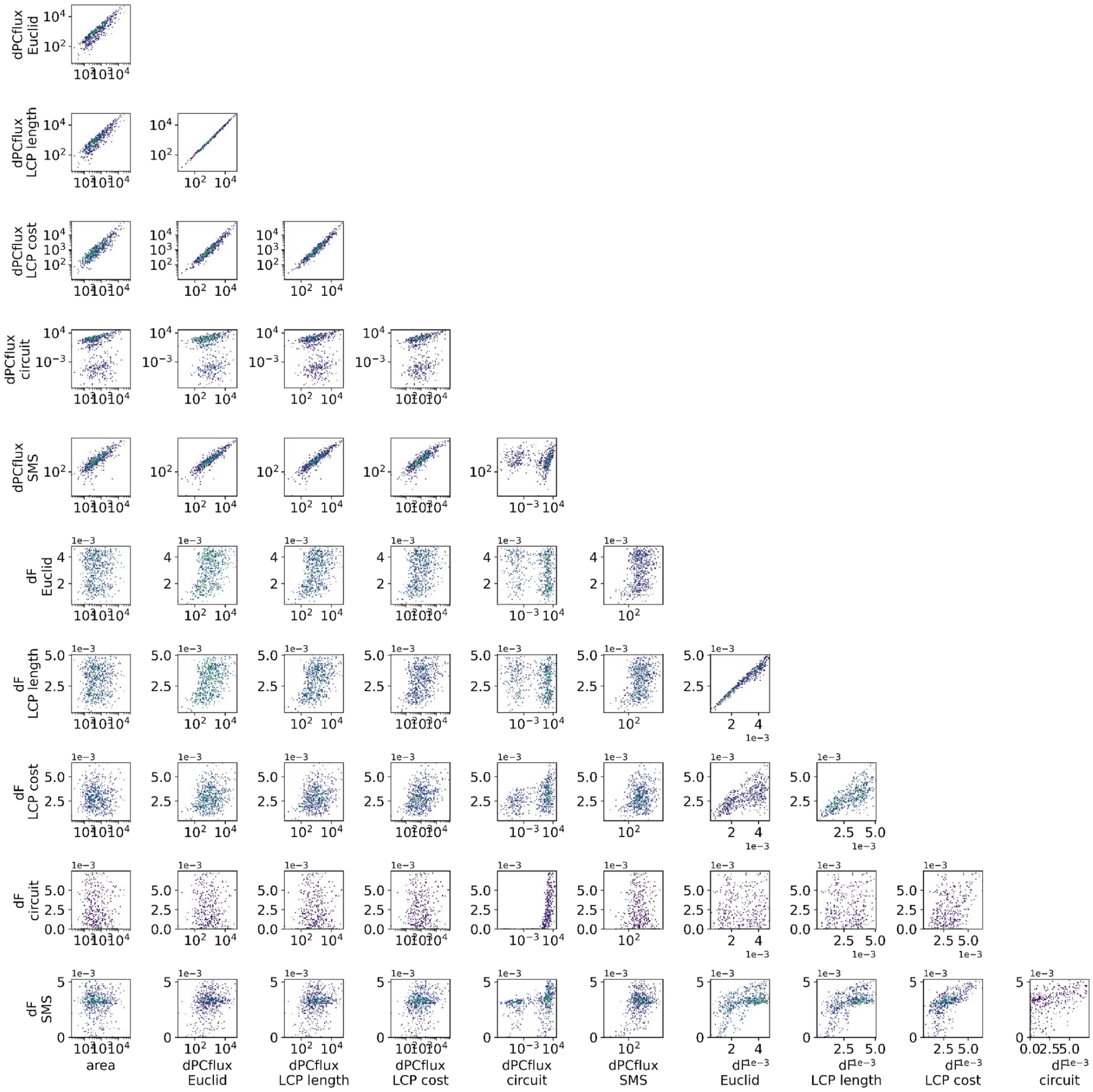

***Fig. S7a*** *expert area, dPCflux, and F scatterplots according to different connectivity estimations. Each point represents a pair of expert patches. Area and dPCflux are represented on a logarithm scale. See Table S7a for corresponding correlation parameters.*

***Table S7a*** *Expert area, dPCflux, and $F$ correlation coefficients and p-values. $(\rho)$ indicates Pearson correlation, $(r)$ Spearman, and $(\tau)$ Kendall. Highlighted values correspond to p-values above the statistical significance threshold of 0.01. See Figures S7a for corresponding scatterplots.*

| **Expert** | area | *dPCflux Euclid* | *dPCflux lcp length* | *dPCflux lcp cost* | *dPCflux circuit* | *dPCflux SMS* | $F_{euclid}$ | $F_{lcp.length}$ | $F_{lcp.cost}$ | $F_{circuit}$ | $F_{SMS}$ |
|---|---|---|---|---|---|---|---|---|---|---|---|
| area | | *p-values*<br>(ρ) 0.00<br>(r) 0.00<br>(τ) 0.00 | *p-values*<br>(ρ) 0.00<br>(r) 0.00<br>(τ) 0.00 | *p-values*<br>(ρ) 0.00<br>(r) 0.00<br>(τ) 0.00 | *p-values*<br>(ρ) 0.00<br>(r) 0.00<br>(τ) 0.00 | *p-values*<br>(ρ) 0.00<br>(r) 0.00<br>(τ) 0.00 | *p-values*<br>(ρ) 0.64<br>(r) 0.09<br>(τ) 0.09 | *p-values*<br>(ρ) 0.49<br>(r) 0.09<br>(τ) 0.09 | *p-values*<br>(ρ) 0.29<br>(r) 0.47<br>(τ) 0.45 | *p-values*<br>(ρ) 0.99<br>(r) 0.00<br>(τ) 0.00 | *p-values*<br>(ρ) 0.29<br>(r) 0.94<br>(τ) 0.96 |
| *dPCflux Euclid* | *coef*<br>(ρ) 0.89<br>(r) 0.91<br>(τ) 0.75 | | *p-values*<br>(ρ) 0.00<br>(r) 0.00<br>(τ) 0.00 | *p-values*<br>(ρ) 0.00<br>(r) 0.00<br>(τ) 0.00 | *p-values*<br>(ρ) 0.00<br>(r) 0.00<br>(τ) 0.00 | *p-values*<br>(ρ) 0.00<br>(r) 0.00<br>(τ) 0.00 | *p-values*<br>(ρ) 0.00<br>(r) 0.00<br>(τ) 0.00 | *p-values*<br>(ρ) 0.00<br>(r) 0.00<br>(τ) 0.00 | *p-values*<br>(ρ) 0.00<br>(r) 0.00<br>(τ) 0.00 | *p-values*<br>(ρ) 0.76<br>(r) 0.00<br>(τ) 0.00 | *p-values*<br>(ρ) 0.00<br>(r) 0.00<br>(τ) 0.00 |
| *dPCflux lcp length* | *coef*<br>(ρ) 0.89<br>(r) 0.91<br>(τ) 0.75 | *coef*<br>(ρ) 1.00<br>(r) 1.00<br>(τ) 0.97 | | *p-values*<br>(ρ) 0.00<br>(r) 0.00<br>(τ) 0.00 | *p-values*<br>(ρ) 0.00<br>(r) 0.00<br>(τ) 0.00 | *p-values*<br>(ρ) 0.00<br>(r) 0.00<br>(τ) 0.00 | *p-values*<br>(ρ) 0.00<br>(r) 0.00<br>(τ) 0.00 | *p-values*<br>(ρ) 0.00<br>(r) 0.00<br>(τ) 0.00 | *p-values*<br>(ρ) 0.00<br>(r) 0.00<br>(τ) 0.00 | *p-values*<br>(ρ) 0.99<br>(r) 0.00<br>(τ) 0.00 | *p-values*<br>(ρ) 0.00<br>(r) 0.00<br>(τ) 0.00 |
| *dPCflux lcp cost* | *coef*<br>(ρ) 0.90<br>(r) 0.91<br>(τ) 0.74 | *coef*<br>(ρ) 0.96<br>(r) 0.97<br>(τ) 0.84 | *coef*<br>(ρ) 0.97<br>(r) 0.97<br>(τ) 0.85 | | *p-values*<br>(ρ) 0.00<br>(r) 0.00<br>(τ) 0.00 | *p-values*<br>(ρ) 0.00<br>(r) 0.00<br>(τ) 0.00 | *p-values*<br>(ρ) 0.00<br>(r) 0.00<br>(τ) 0.00 | *p-values*<br>(ρ) 0.00<br>(r) 0000<br>(τ) 0.00 | *p-values*<br>(ρ) 0.00<br>(r) 0.00<br>(τ) 0.00 | *p-values*<br>(ρ) 0.06<br>(r) 0.94<br>(τ) 0.94 | *p-values*<br>(ρ) 0.00<br>(r) 0.00<br>(τ) 0.00 |
| *dPCflux circuit* | *coef*<br>(ρ) 0.63<br>(r) 0.20<br>(τ) 0.15 | *coef*<br>(ρ) 0.64<br>(r) 0.13<br>(τ) 0.09 | *coef*<br>(ρ) 0.65<br>(r) 0.14<br>(τ) 0.10 | *coef*<br>(ρ) 0.75<br>(r) 0.29<br>(τ) 0.21 | | *p-values*<br>(ρ) 0.00<br>(r) 0.00<br>(τ) 0.00 | *p-values*<br>(ρ) 0.86<br>(r) 0.00<br>(τ) 0.00 | *p-values*<br>(ρ) 0.61<br>(r) 0.01<br>(τ) 0.01 | *p-values*<br>(ρ) 0.00<br>(r) 0.00<br>(τ) 0.00 | *p-values*<br>(ρ) 0.00<br>(r) 0.00<br>(τ) 0.00 | *p-values*<br>(ρ) 0.00<br>(r) 0.00<br>(τ) 0.00 |
| *dPCflux SMS* | *coef*<br>(ρ) 0.88<br>(r) 0.86<br>(τ) 0.69 | *coef*<br>(ρ) 0.96<br>(r) 0.93<br>(τ) 0.77 | *coef*<br>(ρ) 0.96<br>(r) 0.93<br>(τ) 0.78 | *coef*<br>(ρ) 0.93<br>(r) 0.90<br>(τ) 0.74 | *coef*<br>(ρ) 0.59<br>(r) 0.16<br>(τ) 0.12 | | *p-values*<br>(ρ) 0.00<br>(r) 0.00<br>(τ) 0.00 | *p-values*<br>(ρ) 0.00<br>(r) 0.00<br>(τ) 0.00 | *p-values*<br>(ρ) 0.00<br>(r) 0.00<br>(τ) 0.00 | *p-values*<br>(ρ) 0.80<br>(r) 0.00<br>(τ) 0.00 | *p-values*<br>(ρ) 0.00<br>(r) 0.00<br>(τ) 0.00 |
| $F_{euclid}$ | *coef*<br>(ρ) 0.02<br>(r) 0.07<br>(τ) 0.04 | *coef*<br>(ρ) 0.19<br>(r) 0.39<br>(τ) 0.27 | *coef*<br>(ρ) 0.18<br>(r) 0.37<br>(τ) 0.25 | *coef*<br>(ρ) 0.13<br>(r) 0.29<br>(τ) 0.20 | *coef*<br>(ρ)-<br>0.01<br>(r)-0.15<br>(τ)-0.10 | *coef*<br>(ρ) 0.12<br>(r) 0.25<br>(τ) 0.17 | | *p-values*<br>(ρ) 0.00<br>(r) 0.00<br>(τ) 0.00 | *p-values*<br>(ρ) 0.00<br>(r) 0.00<br>(τ) 0.00 | *p-values*<br>(ρ) 0.00<br>(r) 0.00<br>(τ) 0.00 | *p-values*<br>(ρ) 0.00<br>(r) 0.00<br>(τ) 0.00 |
| $F_{lcp.length}$ | *coef*<br>(ρ) 0.03<br>(r) 0.07<br>(τ) 0.05 | *coef*<br>(ρ) 0.21<br>(r) 0.39<br>(τ) 0.27 | *coef*<br>(ρ) 0.20<br>(r) 0.38<br>(τ) 0.26 | *coef*<br>(ρ) 0.15<br>(r) 0.32<br>(τ) 0.22 | *coef*<br>(ρ) 0.02<br>(r)-0.10<br>(τ)-0.07 | *coef*<br>(ρ) 0.14<br>(r) 0.27<br>(τ) 0.18 | *coef*<br>(ρ) 0.98<br>(r) 0.97<br>(τ) 0.88 | | *p-values*<br>(ρ) 0.00<br>(r) 0.00<br>(τ) 0.00 | *p-values*<br>(ρ) 0.06<br>(r) 0.00<br>(τ) 0.00 | *p-values*<br>(ρ) 0.00<br>(r) 0.00<br>(τ) 0.00 |
| $F_{lcp.cost}$ | *coef*<br>(ρ) 0.04<br>(r)-0.03<br>(τ)-0.02 | *coef*<br>(ρ)0.18<br>(r) 0.23<br>(τ) 0.16 | *coef*<br>(ρ)0.19<br>(r) 0.23<br>(τ) 0.16 | *coef*<br>(ρ) 0.23<br>(r) 0.32<br>(τ) 0.22 | *coef*<br>(ρ) 0.21<br>(r) 0.39<br>(τ) 0.27 | *coef*<br>(ρ)-<br>0.13<br>(r) 0.14<br>(τ) 0.10 | *coef*<br>(ρ) 0.66<br>(r) 0.66<br>(τ) 0.48 | *coef*<br>(ρ) 0.70<br>(r) 0.71<br>(τ) 0.52 | | *p-values*<br>(ρ) 0.00<br>(r) 0.00<br>(τ) 0.00 | *p-values*<br>(ρ) 0.00<br>(r) 0.00<br>(τ) 0.00 |
| $F_{circuit}$ | *coef*<br>(ρ)-<br>0.00 | *coef*<br>(ρ)-0.01<br>(r)-0.17 | *coef*<br>(ρ)-<br>0.00 | *coef*<br>(ρ) 0.07<br>(r) 0.00 | *coef*<br>(ρ)<br>0.34 | *coef*<br>(ρ)-<br>0.01 | *coef*<br>(ρ)-<br>0.13 | *coef*<br>(ρ)-<br>0.07 | *coef*<br>(ρ) 0.42<br>(r) 0.40 | | *p-values*<br>(ρ) 0.00<br>(r) 0.00 |
| | $(r)$-0.12<br>$(\tau)$-0.08 | $(\tau)$-0.11 | $(r)$-0.15<br>$(\tau)$-0.10 | $(\tau)$ 0.00 | $(r)$ 0.91<br>$(\tau)$ 0.77 | $(r)$-0.12<br>$(\tau)$-0.08 | $(r)$-0.18<br>$(\tau)$-0.12 | $(r)$-0.12<br>$(\tau)$-0.08 | $(\tau)$ 0.29 | | $(\tau)$ 0.00 |
| $F_{SMS}$ | *coef*<br>$(\rho)$ 0.04<br>$(r)$-0.00<br>$(\tau)$-0.00 | *coef*<br>$(\rho)$ 0.14<br>$(r)$ 0.21<br>$(\tau)$ 0.14 | *coef*<br>$(\rho)$ 0.14<br>$(r)$ 0.20<br>$(\tau)$ 0.14 | *coef*<br>$(\rho)$ 0.15<br>$(r)$ 0.23<br>$(\tau)$ 0.16 | *coef*<br>$(\rho)$ 0.17<br>$(r)$ 0.43<br>$(\tau)$ 0.29 | *coef*<br>$(\rho)$ 0.15<br>$(r)$ 0.21<br>$(\tau)$ 0.15 | *coef*<br>$(\rho)$ 0.60<br>$(r)$ 0.58<br>$(\tau)$ 0.40 | *coef*<br>$(\rho)$ 0.62<br>$(r)$ 0.60<br>$(\tau)$ 0.42 | *coef*<br>$(\rho)$ 0.62<br>$(r)$ 0.70<br>$(\tau)$ 0.51 | *coef*<br>$(\rho)$ 0.41<br>$(r)$ 0.47<br>$(\tau)$ 0.33 | |

# S7.b. Patch-based SDM approach

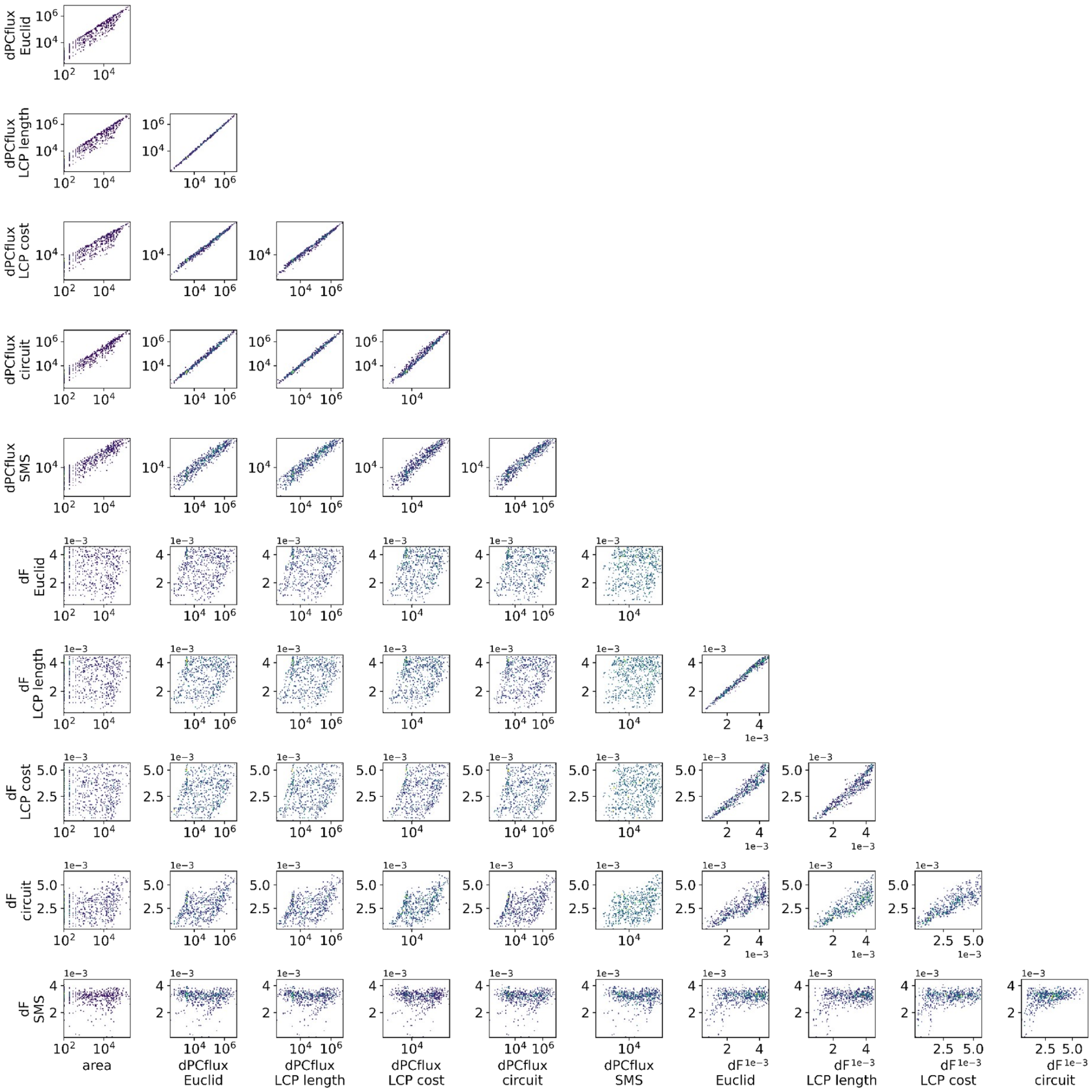

***Fig. S7b*** *SDM area, dPCflux, and F scatterplots according to different connectivity estimations. Each point represents a pair of SDM patches. Area and dPCflux are represented on a logarithm scale. See Table S7b for corresponding correlation parameters.*

***Table S7b*** *SDM area, $dPCflux_{\square}$, and $F_{\square}$ correlation coefficients and p-values. $(\rho)$ indicates Pearson correlation, $(r)$ Spearman, and $(\tau)$ Kendall. Highlighted values correspond to p-values above the statistical significance threshold of 0.01. See Figures S7b for corresponding scatterplots.*

| **SDM** | Area | *dPCflu x Euclid* | *dPCflu x lcp length* | *dPCflu x lcp cost* | *dPCflu x circuit* | *dPCflu x SMS* | $F_{euclid}$ | $F_{lcp.length}$ | $F_{lcp.cost}$ | $F_{circuit}$ | $F_{SMS}$ |
|---|---|---|---|---|---|---|---|---|---|---|---|
| Area | | *p-values*<br>$(\rho)$ 0.00<br>$(r)$ 0.00<br>$(\tau)$ 0.00 | *p-values*<br>$(\rho)$ 0.00<br>$(r)$ 0.00<br>$(\tau)$ 0.00 | *p-values*<br>$(\rho)$ 0.00<br>$(r)$ 0.00<br>$(\tau)$ 0.00 | *p-values*<br>$(\rho)$ 0.00<br>$(r)$ 0.00<br>$(\tau)$ 0.00 | *p-values*<br>$(\rho)$ 0.00<br>$(r)$ 0.00<br>$(\tau)$ 0.00 | *p-values*<br>$(\rho)$ 0.02<br>$(r)$ 0.22<br>$(\tau)$ 0.23 | *p-values*<br>$(\rho)$ 0.01<br>$(r)$ 0.28<br>$(\tau)$ 0.31 | *p-values*<br>$(\rho)$ 0.00<br>$(r)$ 0.75<br>$(\tau)$ 0.78 | *p-values*<br>$(\rho)$ 0.00<br>$(r)$ 0.00<br>$(\tau)$ 0.00 | *p-values*<br>$(\rho)$ 0.37<br>$(r)$ 0.01<br>$(\tau)$ 0.01 |
| *dPCflu x Euclid* | *coef*<br>$(\rho)$ 0.96<br>$(r)$ 0.96<br>$(\tau)$ 0.83 | | *p-values*<br>$(\rho)$ 0.00<br>$(r)$ 0.00<br>$(\tau)$ 0.00 | *p-values*<br>$(\rho)$ 0.00<br>$(r)$ 0.00<br>$(\tau)$ 0.00 | *p-values*<br>$(\rho)$ 0.00<br>$(r)$ 0.00<br>$(\tau)$ 0.00 | *p-values*<br>$(\rho)$ 0.00<br>$(r)$ 0.00<br>$(\tau)$ 0.00 | *p-values*<br>$(\rho)$ 0.00<br>$(r)$ 0.00<br>$(\tau)$ 0.00 | *p-values*<br>$(\rho)$ 0.00<br>$(r)$ 0.00<br>$(\tau)$ 0.00 | *p-values*<br>$(\rho)$ 0.00<br>$(r)$ 0.00<br>$(\tau)$ 0.00 | *p-values*<br>$(\rho)$ 0.00<br>$(r)$ 0.00<br>$(\tau)$ 0.00 | *p-values*<br>$(\rho)$ 0.03<br>$(r)$ 0.11<br>$(\tau)$ 0.10 |
| *dPCflu x lcp length* | *coef*<br>$(\rho)$ 0.96<br>$(r)$ 0.96<br>$(\tau)$ 0.84 | *coef*<br>$(\rho)$ 1.00<br>$(r)$ 1.00<br>$(\tau)$ 0.98 | | *p-values*<br>$(\rho)$ 0.00<br>$(r)$ 0.00<br>$(\tau)$ 0.00 | *p-values*<br>$(\rho)$ 0.00<br>$(r)$ 0.00<br>$(\tau)$ 0.00 | *p-values*<br>$(\rho)$ 0.00<br>$(r)$ 0.00<br>$(\tau)$ 0.00 | *p-values*<br>$(\rho)$ 0.00<br>$(r)$ 0.00<br>$(\tau)$ 0.00 | *p-values*<br>$(\rho)$ 0.00<br>$(r)$ 0.00<br>$(\tau)$ 0.00 | *p-values*<br>$(\rho)$ 0.00<br>$(r)$ 0.00<br>$(\tau)$ 0.00 | *p-values*<br>$(\rho)$ 0.00<br>$(r)$ 0.00<br>$(\tau)$ 0.00 | *p-values*<br>$(\rho)$ 0.04<br>$(r)$ 0.11<br>$(\tau)$ 0.09 |
| *dPCflu x lcp cost* | *coef*<br>$(\rho)$ 0.95<br>$(r)$ 0.93<br>$(\tau)$ 0.79 | *coef*<br>$(\rho)$ 0.99<br>$(r)$ 0.99<br>$(\tau)$ 0.94 | *coef*<br>$(\rho)$ 0.99<br>$(r)$ 0.99<br>$(\tau)$ 0.93 | | *p-values*<br>$(\rho)$ 0.00<br>$(r)$ 0.00<br>$(\tau)$ 0.00 | *p-values*<br>$(\rho)$ 0.00<br>$(r)$ 0.00<br>$(\tau)$ 0.00 | *p-values*<br>$(\rho)$ 0.00<br>$(r)$ 0.00<br>$(\tau)$ 0.00 | *p-values*<br>$(\rho)$ 0.00<br>$(r)$ 0.00<br>$(\tau)$ 0.00 | *p-values*<br>$(\rho)$ 0.00<br>$(r)$ 0.00<br>$(\tau)$ 0.00 | *p-values*<br>$(\rho)$ 0.00<br>$(r)$ 0.00<br>$(\tau)$ 0.00 | *p-values*<br>$(\rho)$ 0.02<br>$(r)$ 0.23<br>$(\tau)$ 0.23 |
| *dPCflu x circuit* | *coef*<br>$(\rho)$ 0.95<br>$(r)$ 0.97<br>$(\tau)$ 0.86 | *coef*<br>$(\rho)$ 0.98<br>$(r)$ 1.00<br>$(\tau)$ 0.95 | *coef*<br>$(\rho)$ 0.97<br>$(r)$ 1.00<br>$(\tau)$ 0.95 | *coef*<br>$(\rho)$ 0.97<br>$(r)$ 0.99<br>$(\tau)$ 0.91 | | *p-values*<br>$(\rho)$ 0.00<br>$(r)$ 0.00<br>$(\tau)$ 0.00 | *p-values*<br>$(\rho)$ 0.00<br>$(r)$ 0.00<br>$(\tau)$ 0.00 | *p-values*<br>$(\rho)$ 0.00<br>$(r)$ 0.00<br>$(\tau)$ 0.00 | *p-values*<br>$(\rho)$ 0.00<br>$(r)$ 0.00<br>$(\tau)$ 0.00 | *p-values*<br>$(\rho)$ 0.00<br>$(r)$ 0.00<br>$(\tau)$ 0.00 | *p-values*<br>$(\rho)$ 0.01<br>$(r)$ 0.28<br>$(\tau)$ 0.25 |
| *dPCflu x SMS* | *coef*<br>$(\rho)$ 0.82<br>$(r)$ 0.93<br>$(\tau)$ 0.77 | *coef*<br>$(\rho)$ 0.85<br>$(r)$ 0.96<br>$(\tau)$ 0.82 | *coef*<br>$(\rho)$ 0.83<br>$(r)$ 0.96<br>$(\tau)$ 0.82 | *coef*<br>$(\rho)$ 0.82<br>$(r)$ 0.95<br>$(\tau)$ 0.80 | *coef*<br>$(\rho)$ 0.83<br>$(r)$ 0.96<br>$(\tau)$ 0.82 | | *p-values*<br>$(\rho)$ 0.00<br>$(r)$ 0.00<br>$(\tau)$ 0.02 | *p-values*<br>$(\rho)$ 0.00<br>$(r)$ 0.00<br>$(\tau)$ 0.02 | *p-values*<br>$(\rho)$ 0.00<br>$(r)$ 0.00<br>$(\tau)$ 0.00 | *p-values*<br>$(\rho)$ 0.00<br>$(r)$ 0.00<br>$(\tau)$ 0.00 | *p-values*<br>$(\rho)$ 0.38<br>$(r)$ 0.00<br>$(\tau)$ 0.00 |
| $F_{euclid}$ | *coef*<br>$(\rho)$ 0.10<br>$(r)$ -0.05<br>$(\tau)$ -0.03 | *coef*<br>$(\rho)$ 0.21<br>$(r)$ 0.17<br>$(\tau)$ 0.12 | *coef*<br>$(\rho)$ 0.22<br>$(r)$ 0.17<br>$(\tau)$ 0.11 | *coef*<br>$(\rho)$ 0.22<br>$(r)$ 0.24<br>$(\tau)$ 0.17 | *coef*<br>$(\rho)$ 0.20<br>$(r)$ 0.15<br>$(\tau)$ 0.10 | *coef*<br>$(\rho)$ 0.13<br>$(r)$ 0.09<br>$(\tau)$ 0.06 | | *p-values*<br>$(\rho)$ 0.00<br>$(r)$ 0.00<br>$(\tau)$ 0.00 | *p-values*<br>$(\rho)$ 0.00<br>$(r)$ 0.00<br>$(\tau)$ 0.00 | *p-values*<br>$(\rho)$ 0.00<br>$(r)$ 0.00<br>$(\tau)$ 0.00 | *p-values*<br>$(\rho)$ 0.00<br>$(r)$ 0.00<br>$(\tau)$ 0.00 |
| $F_{lcp.length}$ | *coef*<br>$(\rho)$ 0.10<br>$(r)$ -0.04<br>$(\tau)$ -0.03 | *coef*<br>$(\rho)$ 0.21<br>$(r)$ 0.17<br>$(\tau)$ 0.12 | *coef*<br>$(\rho)$ 0.22<br>$(r)$ 0.17<br>$(\tau)$ 0.12 | *coef*<br>$(\rho)$ 0.22<br>$(r)$ 0.24<br>$(\tau)$ 0.17 | *coef*<br>$(\rho)$ 0.20<br>$(r)$ 0.15<br>$(\tau)$ 0.11 | *coef*<br>$(\rho)$ 0.13<br>$(r)$ 0.09<br>$(\tau)$ 0.06 | *coef*<br>$(\rho)$ 0.99<br>$(r)$ 0.99<br>$(\tau)$ 0.91 | | *p-values*<br>$(\rho)$ 0.00<br>$(r)$ 0.00<br>$(\tau)$ 0.00 | *p-values*<br>$(\rho)$ 0.00<br>$(r)$ 0.00<br>$(\tau)$ 0.00 | *p-values*<br>$(\rho)$ 0.00<br>$(r)$ 0.00<br>$(\tau)$ 0000 |
| $F_{lcp.cost}$ | *coef*<br>$(\rho)$ 0.14<br>$(r)$ -0.01<br>$(\tau)$ - | *coef*<br>$(\rho)$ 0.25<br>$(r)$ 0.21<br>$(\tau)$ 0.15 | *coef*<br>$(\rho)$ 0.27<br>$(r)$ 0.21<br>$(\tau)$ 0.14 | *coef*<br>$(\rho)$ 0.27<br>$(r)$ 0.29<br>$(\tau)$ 0.20 | *coef*<br>$(\rho)$ 0.25<br>$(r)$ 0.19<br>$(\tau)$ 0.13 | *coef*<br>$(\rho)$ 0.17<br>$(r)$ 0.15<br>$(\tau)$ 0.10 | *coef*<br>$(\rho)$ 0.96<br>$(r)$ 0.96<br>$(\tau)$ 0.84 | *coef*<br>$(\rho)$ 0.95<br>$(r)$ 0.95<br>$(\tau)$ 0.81 | | *p-values*<br>$(\rho)$ 0.00<br>$(r)$ 0.00<br>$(\tau)$ 0.00 | *p-values*<br>$(\rho)$ 0.00<br>$(r)$ 0.00<br>$(\tau)$ 0.00 |
| | 0.01 | | | | | | | | | | |
| $F_{circuit}$ | *coef*<br>($\rho$) 0.39<br>($r$) 0.29<br>($\tau$) 0.20 | *coef*<br>($\rho$) 0.48<br>($r$) 0.49<br>($\tau$) 0.35 | *coef*<br>($\rho$) 0.47<br>($r$) 0.49<br>($\tau$) 0.35 | *coef*<br>($\rho$) 0.47<br>($r$) 0.55<br>($\tau$) 0.39 | *coef*<br>($\rho$) 0.50<br>($r$) 0.49<br>($\tau$) 0.35 | *coef*<br>($\rho$) 0.44<br>($r$) 0.44<br>($\tau$) 0.30 | *coef*<br>($\rho$) 0.84<br>($r$) 0.84<br>($\tau$) 0.65 | *coef*<br>($\rho$) 0.83<br>($r$) 0.84<br>($\tau$) 0.64 | *coef*<br>($\rho$) 0.85<br>($r$) 0.86<br>($\tau$) 0.67 | | *p-values*<br>($\rho$) 0.00<br>($r$) 0.00<br>($\tau$) 0.00 |
| $F_{SMS}$ | *coef*<br>($\rho$) 0.04<br>($r$)-<br>0.10<br>($\tau$)-<br>0.07 | *coef*<br>($\rho$) 0.08<br>($r$)-<br>0.06<br>($\tau$)-<br>0.04 | *coef*<br>($\rho$) 0.08<br>($r$)-<br>0.06<br>($\tau$)-<br>0.04 | *coef*<br>($\rho$) 0.09<br>($r$)-<br>0.04<br>($\tau$)-<br>0.03 | *coef*<br>($\rho$) 0.10<br>($r$)-<br>0.04<br>($\tau$)-<br>0.03 | *coef*<br>($\rho$) 0.03<br>($r$)-<br>0.14<br>($\tau$)-<br>0.10 | *coef*<br>($\rho$) 0.33<br>($r$) 0.16<br>($\tau$) 0.11 | *coef*<br>($\rho$) 0.34<br>($r$) 0.16<br>($\tau$) 0.11 | *coef*<br>($\rho$) 0.32<br>($r$) 0.18<br>($\tau$) 0.15 | *coef*<br>($\rho$) 0.40<br>($r$) 0.27<br>($\tau$) 0.19 | |

# S7.c. dPCflux zonal average

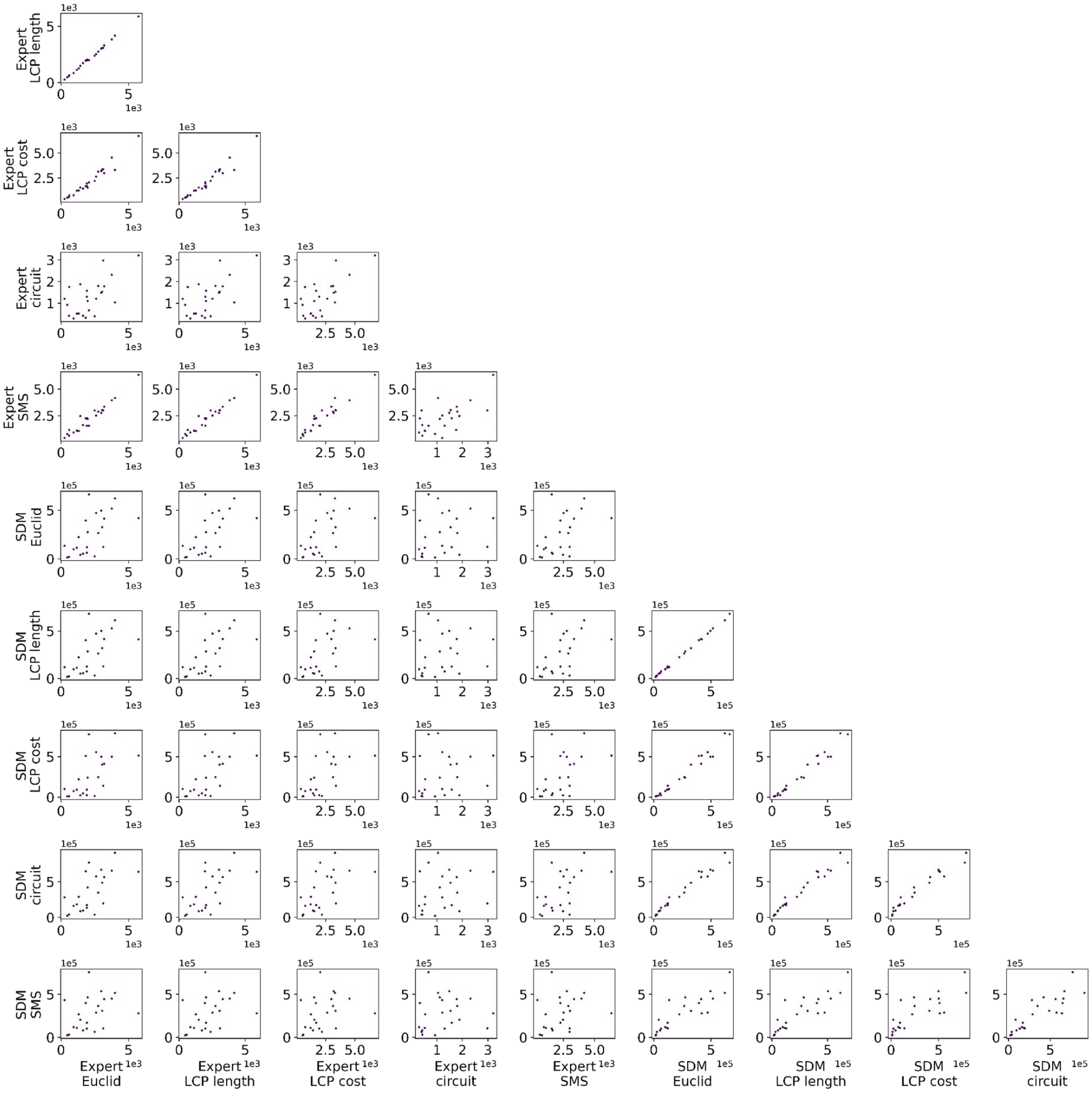

***Fig. S7c*** *Grid-cell average dPCflux correlations. Each point represents a grid cell.*

***Table S7c*** *Grid-average dPCflux correlation coefficients and p-values.* $(\rho)$ *indicates Pearson correlation,* $(r)$ *Spearman, and* $(\tau)$ *Kendall. Highlighted values correspond to p-values above the statistical significance threshold of 0.01 See Supplementary Information Fig. S7c for corresponding scatterplots. An extract of the Spearman correlation coefficients of this table is presented in Table 4 of the main material.*

| dPCflux | | Expert | | | | | SDM | | | | |
|---|---|---|---|---|---|---|---|---|---|---|---|
| | | Euclid | LCP length | LCP cost | Circuit | SMS | Euclid | LCP length | LCP cost | Circuit | SMS |
| Expert | Euclid | - | *p-value*<br>$(\rho)$ 0.00<br>$(r)$ 0.00<br>$(\tau)$ 0.00 | *p-value*<br>$(\rho)$ 0.00<br>$(r)$ 0.00<br>$(\tau)$ 0.00 | *p-value*<br>$(\rho)$ 0.00<br>$(r)$ 0.01<br>$(\tau)$ 0.01 | *p-value*<br>$(\rho)$ 0.00<br>$(r)$ 0.00<br>$(\tau)$ 0.00 | *p-value*<br>$(\rho)$ 0.00<br>$(r)$ 0.00<br>$(\tau)$ 0.00 | *p-value*<br>$(\rho)$ 0.00<br>$(r)$ 0.00<br>$(\tau)$ 0.00 | *p-value*<br>$(\rho)$ 0.00<br>$(r)$ 0.00<br>$(\tau)$ 0.00 | *p-value*<br>$(\rho)$ 0.00<br>$(r)$ 0.00<br>$(\tau)$ 0.00 | *p-value*<br>$(\rho)$ 0.10<br>$(r)$ 0.02<br>$(\tau)$ 0.02 |
| | LCP length | *coef*<br>$(\rho)$ 1.00<br>$(r)$ 1.00<br>$(\tau)$ 0.99 | - | *p-value*<br>$(\rho)$ 0.00<br>$(r)$ 0.00<br>$(\tau)$ 0.00 | *p-value*<br>$(\rho)$ 0.00<br>$(r)$ 0.01<br>$(\tau)$ 0.01 | *p-value*<br>$(\rho)$ 0.00<br>$(r)$ 0.00<br>$(\tau)$ 0.00 | *p-value*<br>$(\rho)$ 0.00<br>$(r)$ 0.00<br>$(\tau)$ 0.00 | *p-value*<br>$(\rho)$ 0.00<br>$(r)$ 0.00<br>$(\tau)$ 0.00 | *p-value*<br>$(\rho)$ 0.00<br>$(r)$ 0.00<br>$(\tau)$ 0.00 | *p-value*<br>$(\rho)$ 0.00<br>$(r)$ 0.00<br>$(\tau)$ 0.00 | *p-value*<br>$(\rho)$ 0.09<br>$(r)$ 0.02<br>$(\tau)$ 0.03 |
| | LCP cost | *coef*<br>$(\rho)$ 0.97<br>$(r)$ 0.97<br>$(\tau)$ 0.91 | *coef*<br>$(\rho)$ 0.97<br>$(r)$ 0.97<br>$(\tau)$ 0.89 | - | *p-value*<br>$(\rho)$ 0.00<br>$(r)$ 0.00<br>$(\tau)$ 0.00 | *p-value*<br>$(\rho)$ 0.00<br>$(r)$ 0.00<br>$(\tau)$ 0.00 | *p-value*<br>$(\rho)$ 0.00<br>$(r)$ 0.00<br>$(\tau)$ 0.00 | *p-value*<br>$(\rho)$ 0.00<br>$(r)$ 0.00<br>$(\tau)$ 0.00 | *p-value*<br>$(\rho)$ 0.00<br>$(r)$ 0.00<br>$(\tau)$ 0.00 | *p-value*<br>$(\rho)$ 0.00<br>$(r)$ 0.00<br>$(\tau)$ 0.01 | *p-value*<br>$(\rho)$ 0.15<br>$(r)$ 0.05<br>$(\tau)$ 0.06 |
| | Circuit | *coef*<br>$(\rho)$ 0.63<br>$(r)$ 0.53<br>$(\tau)$ 0.38 | *coef*<br>$(\rho)$ 0.63<br>$(r)$ 0.54<br>$(\tau)$ 0.39 | *coef*<br>$(\rho)$ 0.71<br>$(r)$ 0.58<br>$(\tau)$ 0.41 | - | *p-value*<br>$(\rho)$ 0.00<br>$(r)$ 0.00<br>$(\tau)$ 0.01 | *p-value*<br>$(\rho)$ 0.32<br>$(r)$ 0.13<br>$(\tau)$ 0.19 | *p-value*<br>$(\rho)$ 0.32<br>$(r)$ 0.12<br>$(\tau)$ 0.17 | *p-value*<br>$(\rho)$ 0.39<br>$(r)$ 0.18<br>$(\tau)$ 0.21 | *p-value*<br>$(\rho)$ 0.27<br>$(r)$ 0.20<br>$(\tau)$ 0.23 | *p-value*<br>$(\rho)$ 0.56<br>$(r)$ 0.22<br>$(\tau)$ 0.35 |
| | SMS | *coef*<br>$(\rho)$ 0.97<br>$(r)$ 0.95<br>$(\tau)$ 0.83 | *coef*<br>$(\rho)$ 0.97<br>$(r)$ 0.96<br>$(\tau)$ 0.84 | *coef*<br>$(\rho)$ 0.95<br>$(r)$ 0.94<br>$(\tau)$ 0.79 | *coef*<br>$(\rho)$ 0.67<br>$(r)$ 0.57<br>$(\tau)$ 0.41 | - | *p-value*<br>$(\rho)$ 0.01<br>$(r)$ 0.01<br>$(\tau)$ 0.01 | *p-value*<br>$(\rho)$ 0.01<br>$(r)$ 0.00<br>$(\tau)$ 0.01 | *p-value*<br>$(\rho)$ 0.01<br>$(r)$ 0.00<br>$(\tau)$ 0.01 | *p-value*<br>$(\rho)$ 0.00<br>$(r)$ 0.01<br>$(\tau)$ 0.01 | *p-value*<br>$(\rho)$ 0.19<br>$(r)$ 0.07<br>$(\tau)$ 0.08 |
| SDM | Euclid | *coef*<br>$(\rho)$ 0.62<br>$(r)$ 0.69<br>$(\tau)$ 0.52 | *coef*<br>$(\rho)$ 0.62<br>$(r)$ 0.67<br>$(\tau)$ 0.50 | *coef*<br>$(\rho)$ 0.57<br>$(r)$ 0.64<br>$(\tau)$ 0.45 | *coef*<br>$(\rho)$ 0.22<br>$(r)$ 0.33<br>$(\tau)$ 0.20 | *coef*<br>$(\rho)$ 0.53<br>$(r)$ 0.55<br>$(\tau)$ 0.40 | - | *p-value*<br>$(\rho)$ 0.00<br>$(r)$ 0.00<br>$(\tau)$ 0.00 | *p-value*<br>$(\rho)$ 0.00<br>$(r)$ 0.00<br>$(\tau)$ 0.00 | *p-value*<br>$(\rho)$ 0.00<br>$(r)$ 0.00<br>$(\tau)$ 0.00 | *p-value*<br>$(\rho)$ 0.00<br>$(r)$ 0.00<br>$(\tau)$ 0.00 |
| | LCP length | *coef*<br>$(\rho)$ 0.62<br>$(r)$ 0.72<br>$(\tau)$ 0.53 | *coef*<br>$(\rho)$ 0.62<br>$(r)$ 0.69<br>$(\tau)$ 0.51 | *coef*<br>$(\rho)$ 0.57<br>$(r)$ 0.66<br>$(\tau)$ 0.45 | *coef*<br>$(\rho)$ 0.22<br>$(r)$ 0.29<br>$(\tau)$ 0.21 | *coef*<br>$(\rho)$ 0.52<br>$(r)$ 0.58<br>$(\tau)$ 0.41 | *coef*<br>$(\rho)$ 1.00<br>$(r)$ 1.00<br>$(\tau)$ 0.98 | - | *p-value*<br>$(\rho)$ 0.00<br>$(r)$ 0.00<br>$(\tau)$ 0.00 | *p-value*<br>$(\rho)$ 0.00<br>$(r)$ 0.00<br>$(\tau)$ 0.00 | *p-value*<br>$(\rho)$ 0.00<br>$(r)$ 0.00<br>$(\tau)$ 0.00 |
| | LCP cost | *coef*<br>$(\rho)$ 0.63<br>$(r)$ 0.69<br>$(\tau)$ 0.51 | *coef*<br>$(\rho)$ 0.63<br>$(r)$ 0.67<br>$(\tau)$ 0.49 | *coef*<br>$(\rho)$ 0.57<br>$(r)$ 0.64<br>$(\tau)$ 0.44 | *coef*<br>$(\rho)$ 0.19<br>$(r)$ 0.27<br>$(\tau)$ 0.19 | *coef*<br>$(\rho)$ 0.54<br>$(r)$ 0.57<br>$(\tau)$ 0.41 | *coef*<br>$(\rho)$ 0.99<br>$(r)$ 0.98<br>$(\tau)$ 0.91 | *coef*<br>$(\rho)$ 0.98<br>$(r)$ 0.97<br>$(\tau)$ 0.91 | - | *p-value*<br>$(\rho)$ 0.00<br>$(r)$ 0.00<br>$(\tau)$ 0.00 | *p-value*<br>$(\rho)$ 0.00<br>$(r)$ 0.00<br>$(\tau)$ 0.00 |
| | Circuit | *coef*<br>$(\rho)$ 0.65<br>$(r)$ 0.66<br>$(\tau)$ 0.49 | *coef*<br>$(\rho)$ 0.65<br>$(r)$ 0.63<br>$(\tau)$ 0.47 | *coef*<br>$(\rho)$ 0.59<br>$(r)$ 0.61<br>$(\tau)$ 0.42 | *coef*<br>$(\rho)$ 0.24<br>$(r)$ 0.27<br>$(\tau)$ 0.19 | *coef*<br>$(\rho)$ 0.57<br>$(r)$ 0.52<br>$(\tau)$ 0.37 | *coef*<br>$(\rho)$ 0.98<br>$(r)$ 0.99<br>$(\tau)$ 0.94 | *coef*<br>$(\rho)$ 0.98<br>$(r)$ 0.98<br>$(\tau)$ 0.91 | *coef*<br>$(\rho)$ 0.98<br>$(r)$ 0.98<br>$(\tau)$ 0.88 | - | *p-value*<br>$(\rho)$ 0.00<br>$(r)$ 0.00<br>$(\tau)$ 0.00 |
| | SMS | *coef*<br>(ρ) 0.35<br>(*r*) 0.49<br>(τ) 0.35 | *coef*<br>(ρ) 0.36<br>(*r*) 0.47<br>(τ) 0.34 | *coef*<br>(ρ) 0.31<br>(*r*) 0.41<br>(τ) 0.28 | *coef*<br>(ρ) 0.13<br>(*r*) 0.27<br>(τ) 0.15 | *coef*<br>(ρ) 0.28<br>(*r*) 0.38<br>(τ) 0.26 | *coef*<br>(ρ) 0.87<br>(*r*) 0.89<br>(τ) 0.74 | *coef*<br>(ρ) 0.87<br>(*r*) 0.88<br>(τ) 0.73 | *coef*<br>(ρ) 0.82<br>(*r*) 0.83<br>(τ) 0.65 | *coef*<br>(ρ) 0.86<br>(*r*) 0.90<br>(τ) 0.72 | - |

# S7.d. Flux zonal average

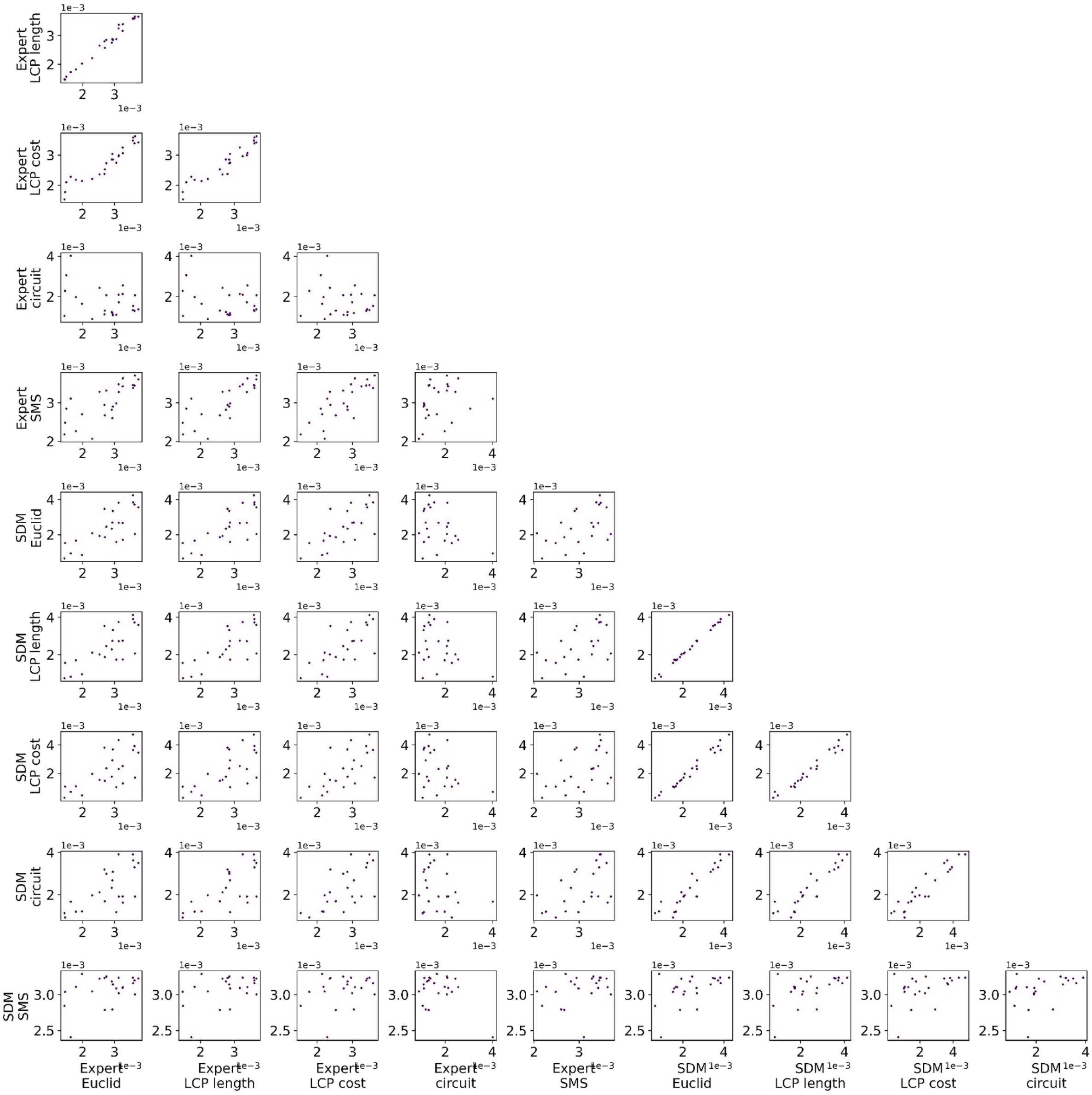

***Fig. S7d*** *Grid-cell average dPCflux correlations. Each point represents a grid cell.*

***Table S7d*** *Grid-average F correlation coefficients and p-values. ($\rho$) indicates Pearson correlation, ($r$) Spearman, and ($\tau$) Kendall. Highlighted values correspond to p-values above the statistical significance threshold of 0.01. See Supplementary Information Fig. S7d for corresponding scatterplots. An extract of the Spearman correlation coefficients of this table is presented in Table 4 of the main material.*

| F | | Expert | | | | | SDM | | | | |
|---|---|---|---|---|---|---|---|---|---|---|---|
| | | Euclid | LCP length | LCP cost | Circuit | SMS | Euclid | LCP length | LCP cost | Circuit | SMS |
| Expert | Euclid | - | *p-value*<br>($\rho$) 0.00<br>($r$) 0.00<br>($\tau$) 0.00 | *p-value*<br>($\rho$) 0.00<br>($r$) 0.00<br>($\tau$) 0.00 | *p-value*<br>($\rho$) 0.08<br>($r$) 0.74<br>($\tau$) 0.79 | *p-value*<br>($\rho$) 0.00<br>($r$) 0.00<br>($\tau$) 0.00 | *p-value*<br>($\rho$) 0.00<br>($r$) 0.00<br>($\tau$) 0.00 | *p-value*<br>($\rho$) 0.00<br>($r$) 0.00<br>($\tau$) 0.00 | *p-value*<br>($\rho$) 0.00<br>($r$) 0.00<br>($\tau$) 0.00 | *p-value*<br>($\rho$) 0.00<br>($r$) 0.00<br>($\tau$) 0.00 | *p-value*<br>($\rho$) 0.05<br>($r$) 0.30<br>($\tau$) 0.27 |
| | LCP length | *coef*<br>($\rho$) 0.99<br>($r$) 0.99<br>($\tau$) 0.92 | - | *p-value*<br>($\rho$) 0.00<br>($r$) 0.00<br>($\tau$) 0.00 | *p-value*<br>($\rho$) 0.17<br>($r$) 0.79<br>($\tau$) 0.83 | *p-value*<br>($\rho$) 0.00<br>($r$) 0.00<br>($\tau$) 0.00 | *p-value*<br>($\rho$) 0.00<br>($r$) 0.00<br>($\tau$) 0.00 | *p-value*<br>($\rho$) 0.00<br>($r$) 0.00<br>($\tau$) 0.00 | *p-value*<br>($\rho$) 0.00<br>($r$) 0.00<br>($\tau$) 0.00 | *p-value*<br>($\rho$) 0.00<br>($r$) 0.00<br>($\tau$) 0.00 | *p-value*<br>($\rho$) 0.04<br>($r$) 0.24<br>($\tau$) 0.25 |
| | LCP cost | *coef*<br>($\rho$) 0.95<br>($r$) 0.97<br>($\tau$) 0.88 | *coef*<br>($\rho$) 0.95<br>($r$) 0.96<br>($\tau$) 0.84 | - | *p-value*<br>($\rho$) 0.45<br>($r$) 0.87<br>($\tau$) 0.98 | *p-value*<br>($\rho$) 0.00<br>($r$) 0.00<br>($\tau$) 0.00 | *p-value*<br>($\rho$) 0.00<br>($r$) 0.00<br>($\tau$) 0.00 | *p-value*<br>($\rho$) 0.00<br>($r$) 0.00<br>($\tau$) 0.00 | *p-value*<br>($\rho$) 0.00<br>($r$) 0.00<br>($\tau$) 0.00 | *p-value*<br>($\rho$) 0.00<br>($r$) 0.00<br>($\tau$) 0.00 | *p-value*<br>($\rho$) 0.20<br>($r$) 0.53<br>($\tau$) 0.46 |
| | Circuit | *coef*<br>($\rho$) -0.36<br>($r$) -0.07<br>($\tau$) -0.04 | *coef*<br>($\rho$) -0.29<br>($r$) -0.06<br>($\tau$) -0.03 | *coef*<br>($\rho$) -0.16<br>($r$) -0.03<br>($\tau$) 0.01 | - | *p-value*<br>($\rho$) 0.20<br>($r$) 0.06<br>($\tau$) 0.07 | *p-value*<br>($\rho$) 0.11<br>($r$) 0.42<br>($\tau$) 0.43 | *p-value*<br>($\rho$) 0.08<br>($r$) 0.41<br>($\tau$) 0.43 | *p-value*<br>($\rho$) 0.13<br>($r$) 0.39<br>($\tau$) 0.37 | *p-value*<br>($\rho$) 0.28<br>($r$) 0.50<br>($\tau$) 0.68 | *p-value*<br>($\rho$) 0.06<br>($r$) 0.83<br>($\tau$) 0.88 |
| | SMS | *coef*<br>($\rho$) 0.75<br>($r$) 0.81<br>($\tau$) 0.62 | *coef*<br>($\rho$) 0.80<br>($r$) 0.83<br>($\tau$) 0.64 | *coef*<br>($\rho$) 0.79<br>($r$) 0.78<br>($\tau$) 0.57 | *coef*<br>($\rho$) 0.27<br>($r$) 0.39<br>($\tau$) 0.27 | - | *p-value*<br>($\rho$) 0.01<br>($r$) 0.02<br>($\tau$) 0.02 | *p-value*<br>($\rho$) 0.01<br>($r$) 0.01<br>($\tau$) 0.01 | *p-value*<br>($\rho$) 0.02<br>($r$) 0.03<br>($\tau$) 0.02 | *p-value*<br>($\rho$) 0.01<br>($r$) 0.02<br>($\tau$) 0.01 | *p-value*<br>($\rho$) 0.18<br>($r$) 0.10<br>($\tau$) 0.17 |
| SDM | Euclid | *coef*<br>($\rho$) 0.75<br>($r$) 0.69<br>($\tau$) 0.51 | *coef*<br>($\rho$) 0.75<br>($r$) 0.71<br>($\tau$) 0.53 | *coef*<br>($\rho$) 0.71<br>($r$) 0.72<br>($\tau$) 0.55 | *coef*<br>($\rho$) -0.34<br>($r$) -0.18<br>($\tau$) -0.12 | *coef*<br>($\rho$) 0.51<br>($r$) 0.50<br>($\tau$) 0.36 | - | *p-value*<br>($\rho$) 0.00<br>($r$) 0.00<br>($\tau$) 0.00 | *p-value*<br>($\rho$) 0.00<br>($r$) 0.00<br>($\tau$) 0.00 | *p-value*<br>($\rho$) 0.00<br>($r$) 0.00<br>($\tau$) 0.00 | *p-value*<br>($\rho$) 0.02<br>($r$) 0.03<br>($\tau$) 0.02 |
| | LCP length | *coef*<br>($\rho$) 0.76<br>($r$) 0.71<br>($\tau$) 0.54 | *coef*<br>($\rho$) 0.76<br>($r$) 0.72<br>($\tau$) 0.55 | *coef*<br>($\rho$) 0.71<br>($r$) 0.73<br>($\tau$) 0.57 | *coef*<br>($\rho$) -0.37<br>($r$) -0.18<br>($\tau$) -0.12 | *coef*<br>($\rho$) 0.51<br>($r$) 0.52<br>($\tau$) 0.38 | *coef*<br>($\rho$) 1.00<br>($r$) 1.00<br>($\tau$) 0.97 | - | *p-value*<br>($\rho$) 0.00<br>($r$) 0.00<br>($\tau$) 0.00 | *p-value*<br>($\rho$) 0.00<br>($r$) 0.00<br>($\tau$) 0.00 | *p-value*<br>($\rho$) 0.01<br>($r$) 0.02<br>($\tau$) 0.02 |

| | | | | | | | | | | |
|---|---|---|---|---|---|---|---|---|---|---|
| | LCP cost | *coef*<br>($\rho$) 0.69<br>($r$) 0.63<br>($\tau$) 0.46 | *coef*<br>($\rho$) 0.70<br>($r$) 0.63<br>($\tau$) 0.45 | *coef*<br>($\rho$) 0.65<br>($r$) 0.65<br>($\tau$) 0.49 | *coef*<br>($\rho$)-0.33<br>($r$)-0.19<br>($\tau$)-0.14 | *coef*<br>($\rho$) 0.47<br>($r$) 0.46<br>($\tau$) 0.34 | *coef*<br>($\rho$) 0.99<br>($r$) 0.98<br>($\tau$) 0.92 | *coef*<br>($\rho$) 0.98<br>($r$) 0.98<br>($\tau$) 0.91 | - | *p-value*<br>($\rho$) 0.00<br>($r$) 0.00<br>($\tau$) 0.00 | *p-value*<br>($\rho$) 0.03<br>($r$) 0.03<br>($\tau$) 0.02 |
| | Circuit | *coef*<br>($\rho$) 0.66<br>($r$) 0.58<br>($\tau$) 0.42 | *coef*<br>($\rho$) 0.68<br>($r$) 0.60<br>($\tau$) 0.45 | *coef*<br>($\rho$) 0.64<br>($r$) 0.61<br>($\tau$) 0.46 | *coef*<br>($\rho$)-0.24<br>($r$)-0.15<br>($\tau$)-0.07 | *coef*<br>($\rho$) 0.52<br>($r$) 0.49<br>($\tau$) 0.38 | *coef*<br>($\rho$) 0.93<br>($r$) 0.94<br>($\tau$) 0.82 | *coef*<br>($\rho$) 0.92<br>($r$) 0.94<br>($\tau$) 0.80 | *coef*<br>($\rho$) 0.94<br>($r$) 0.93<br>($\tau$) 0.79 | - | *p-value*<br>($\rho$) 0.05<br>($r$) 0.06<br>($\tau$) 0.01 |
| | SMS | *coef*<br>($\rho$) 0.41<br>($r$) 0.23<br>($\tau$) 0.17 | *coef*<br>($\rho$) 0.42<br>($r$) 0.25<br>($\tau$) 0.18 | *coef*<br>($\rho$) 0.28<br>($r$) 0.14<br>($\tau$) 0.11 | *coef*<br>($\rho$)-0.40<br>($r$) 0.05<br>($\tau$)-0.03 | *coef*<br>($\rho$) 0.29<br>($r$) 0.35<br>($\tau$) 0.21 | *coef*<br>($\rho$) 0.49<br>($r$) 0.44<br>($\tau$) 0.35 | *coef*<br>($\rho$) 0.51<br>($r$) 0.47<br>($\tau$) 0.36 | *coef*<br>($\rho$) 0.44<br>($r$) 0.46<br>($\tau$) 0.36 | *coef*<br>($\rho$) 0.42<br>($r$) 0.56<br>($\tau$) 0.42 | - |

# S8. SMS Methodological considerations

## S8.a. Maximal displacement capacity

To compute SMS maximal displacement capacity, we used the formula:

$$d_{sms.max} = \left( \frac{d_{euclid.max}}{pixelsize} \right)^2 \times R_{mean}$$

With $d_{euclid.max} = 5700$ the maximal Euclidean dispersal distance, $pixelsize = 10$ the spatial resolution, and $R_{mean}$ the landscape mean resistance: $R_{\exp.mean} = 6.91$, and $R_{sdm.mean} = 75.75$. The maximal displacement capacity obtained were: $d_{\exp.sms.max} = 2{,}245{,}059$ and $d_{sdm.sms.max} = 24{,}611{,}175$.

These maximal displacement capacities translate into different effective dispersal distances (Fig. S8.1a). The expert and SDM effective distances of 0.99 quantiles (q0.99) are 4,620.34 m and 2.087,99 m. These values are both below the reference Euclidean distance $d_{euclid.ref}$ = 5700 m corresponding to $p_{euclid.ref}$ = 0.01. However, as shown in Fig. S8.1b-c, the increase in SMS displacement capacity does not go along with an increase in q0.99 distances. This pattern could be explained by the spatial configuration of the matrix and suitable habitats.

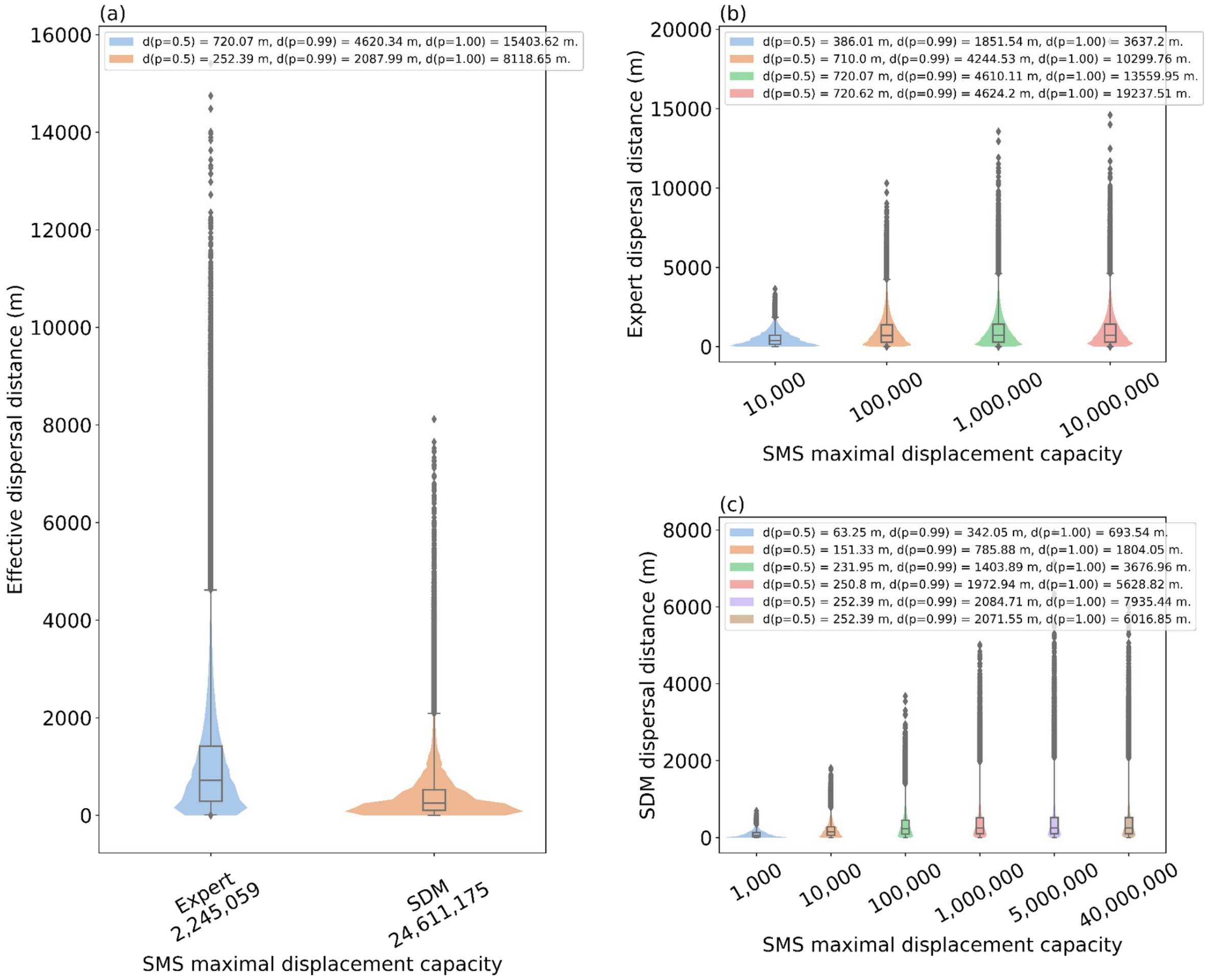

***Fig. S8a*** *Effective dispersal distance resulting from SMS successful dispersal events. Expert and SMS effective dispersal distances for maximal displacement capacity used in the article (a); Evolution of*

*expert effective dispersal distances according to the maximal displacement capacity (b); Evolution of SMS effective dispersal distance according to the maximal displacement capacity (c).*

## S8.b. Dispersal success rates

The conceptual differences between isolation-based models and stochastic SMS resulted in different formulas of connectivity probabilities (Eq. 1 & 4).

Isolation-based probabilities rely exclusively on the degree of isolation between the patches (Eq. 1). In contrast, SMS probabilities (Eq. 4) depend on many implicit factors, including the number of dispersal individuals, patch accessibility, and the maximum number of steps. The mean rate of successful dispersal events per patch is 42% in the expert approach and 49% in SDM (Fig S8.2a). As shown in Fig. S8.2b-c, the increase in maximal dispersal capacity does not go along with an increase in success rate. Unsuccessful dispersal events are not considered in the probabilities of connectivity.

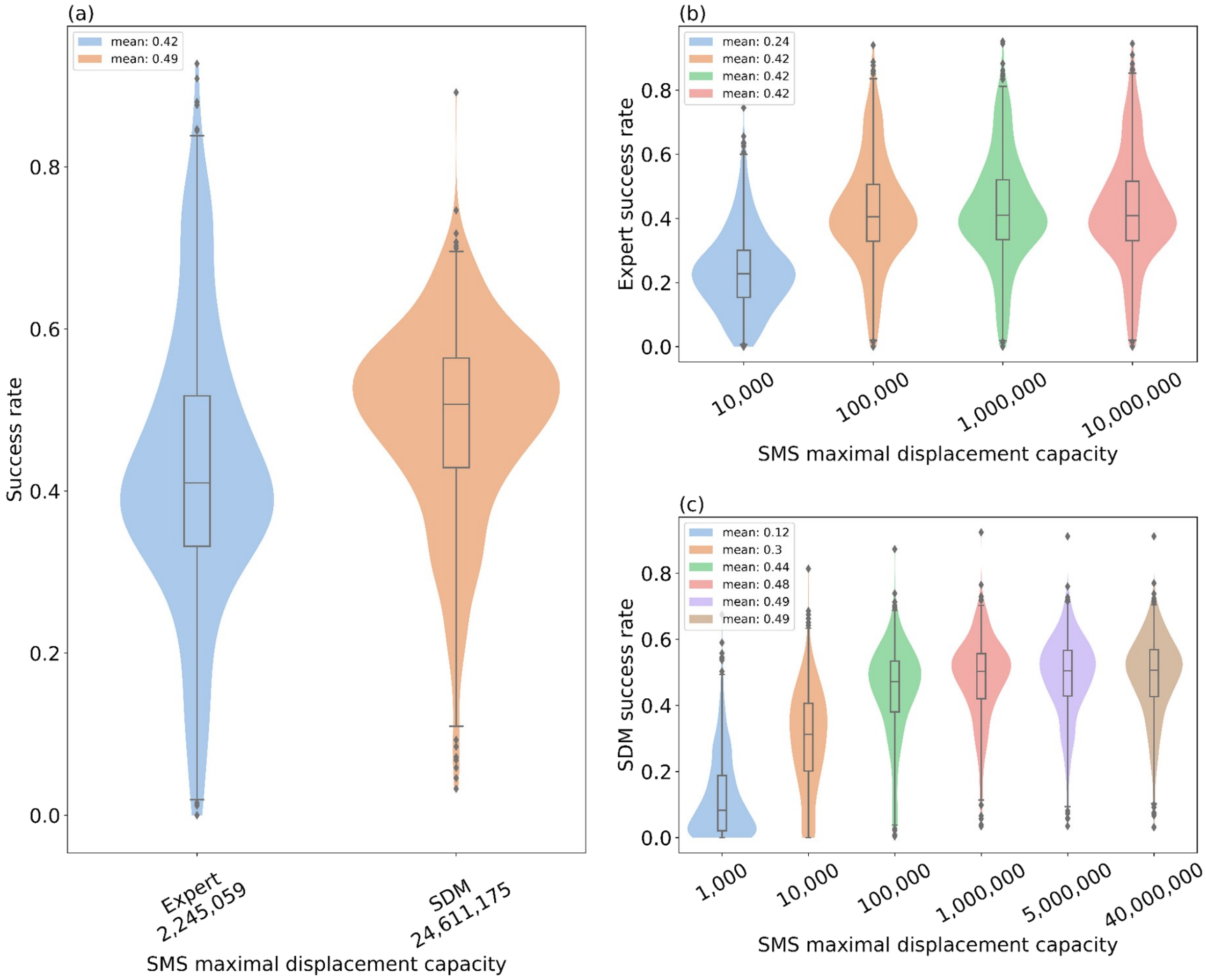

***Fig. S8b*** *Dispersal success rate distribution. Expert and SMS success rates for maximal displacement capacity used in the article (a); Evolution of expert success rate according to the maximal displacement capacity (b); Evolution of SMS success rate according to the maximal displacement capacity (c)*

## S8.c. *dPCflux* underestimation

Moreover, unlike isolation-based approaches, SMS does not produce a complete graph; this implies $p^*_{ij} \geq p_{ij}$, the probabilities of direct connectivity probably produced an estimation of dPClfux underestimated compared to the maximal product probability of all possible paths $dPCflux(p_{ij}^*) \geq dPCflux(p_{ij})$.

# S9. RCI

## S9.a. Examples

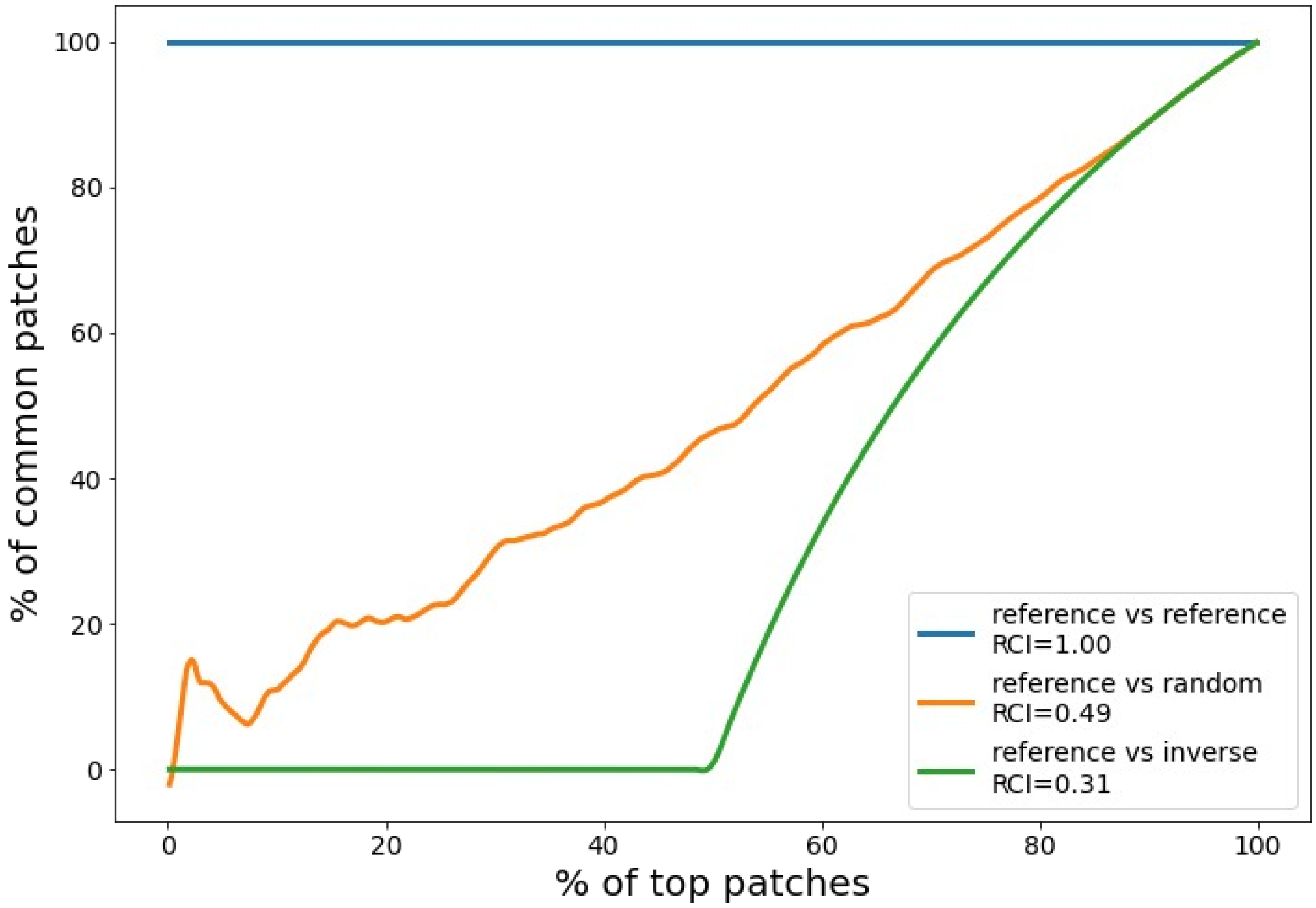

***Fig. S9a*** *Simplified RCI example. RCI = 1 corresponds to a comparison with the same ranking. RCI = 0.49 corresponds to a comparison with a random ranking. RCI = 0.31 corresponds to a comparison with an inverse ranking.*

The range of the RCI index is defined by theoretical extreme case scenarios.

In the **worst-case scenario**, the reference ranking is compared to its exact opposite. In this situation, the first common patches in the two rankings only appear once more than half of the top-ranked patches are considered.

In the **best-case scenario**, the reference ranking is compared to itself. Here, any percentage of top-ranked patches is identical between the two rankings, since they are the same.

## S9.b. RCI curves

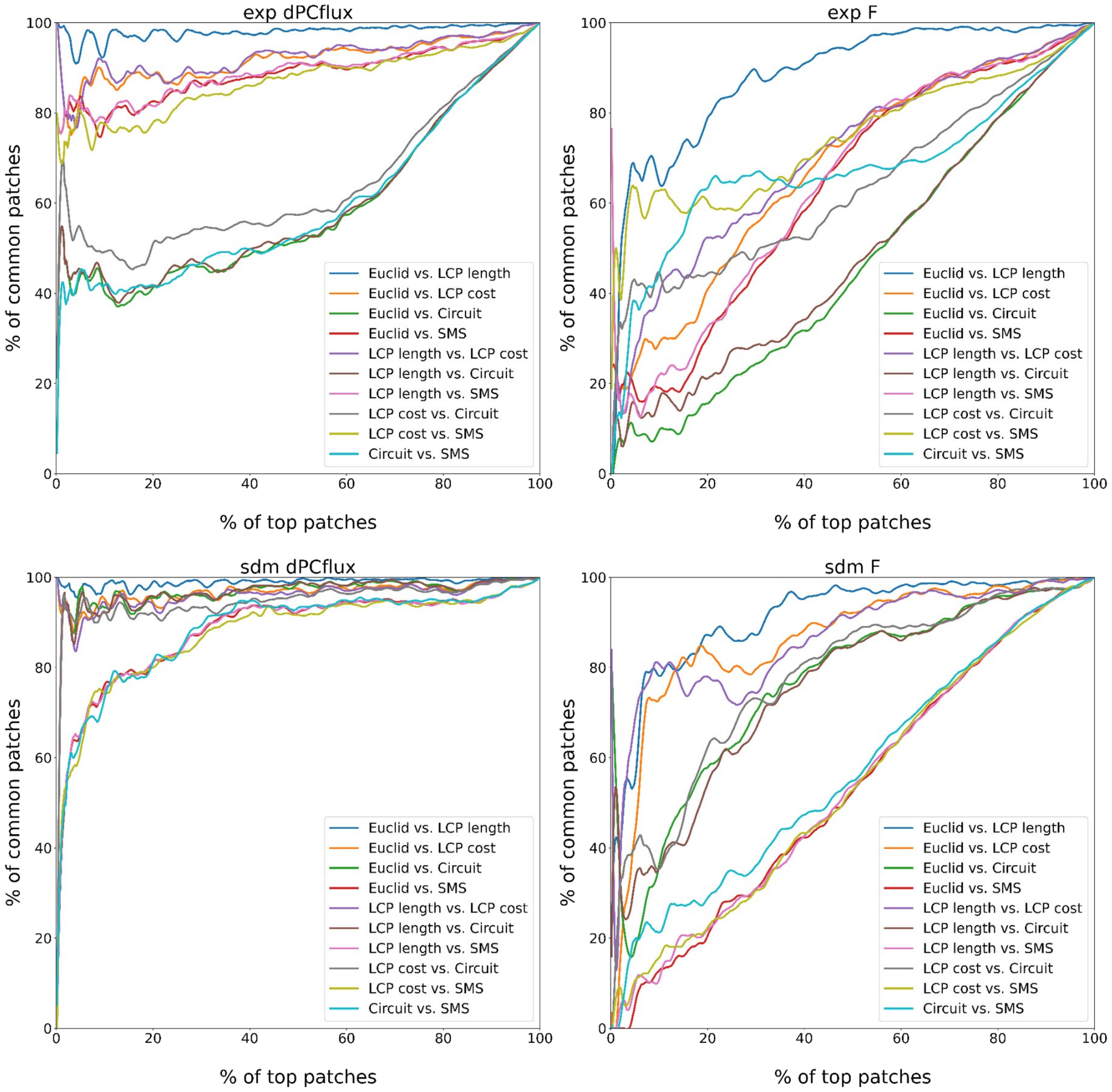

***Fig. S9b*** *Percentage of patches in common in the top x%. Each frame represents a ranking: The first row corresponds to the expert approach, the second to the SDM; the first column corresponds to dPCflux index, the second to F. Each curve represents a pairwise combination of connectivity modelling approaches.*

# S10. Regression computation

## S10.a. Regression formula

### Linear regression

*(Eq S10.1)* $y = ax + b$

We used logarithmic and exponential computation rules as well as variable change to reduce the formulas of power and logistic regression to linear problems.

### Power regression

*(Eq S10.2)* $y = \beta x^a \Leftrightarrow \log y = a \log x + \log \beta$

*(Eq S10.1)* $\Leftrightarrow Y = AX+B$

*With* $A = a$, $B = \log \beta$, $X = \log x$ *and* $Y = \log y$

### Logistic regression

*(Eq S10.3)* $y = K / (1 + e^{ax+b}) \Leftrightarrow y / K = 1 / (1 + e^{ax+b})$

$\Leftrightarrow K / y = 1 + e^{ax+b}$

$\Leftrightarrow (K/y) - 1 = e^{ax+b}$

$\Leftrightarrow \ln((K/y)-1) = ax+b$

*(Eq S10.1)* $\Leftrightarrow Y = ax + b$

*With* $Y = \ln((K/y)-1)$

## S10.b. Custom Regression class

We incorporated the previous variable transformations into a custom Python regression class, which was used to perform all regression analyses. The corresponding code is provided below.

```
import numpy as np
from sklearn.linear_model import LinearRegression

class Regression():
    def __init__(self, x, y, mode):
        self.mode = mode
        self.x = self.get_x(x)
        self.k = self.get_k(y)
        self.y = self.get_y(y)
        self.model = self.fit()
        self.r2 = self.score()
        self.a = self.get_a()
        self.b = self.get_b()

    def get_x(self, x):
        if self.mode == 'power':
            return np.log(x).reshape((-1,1))
        else:
            return x.reshape((-1,1))

    def get_k(self,y):
        if self.mode == 'logistic':
            return max(y) + 1
        else:
            return None

    def get_y(self, y):
        if self.mode == 'linear':
            return y
        elif self.mode == 'power':
            return np.log(y)
        elif self.mode == 'logistic':
            return np.log((self.k/y)-1)

    def fit(self):
        if self.mode == 'linear':
            return LinearRegression(fit_intercept=False).fit(self.x, self.y)
        elif self.mode == 'power' or self.mode == 'logistic':
            return LinearRegression(fit_intercept=True).fit(self.x, self.y)

    def score(self):
        return self.model.score(self.x, self.y)

    def get_a(self):
        return self.model.coef_[0]

    def get_b(self):
        if self.mode == 'linear':
            return self.model.intercept_
        elif self.mode == 'power':
            return np.exp(self.model.intercept_)
```

```
        elif self.mode == 'logistic':
            return self.model.predict([[0]])[0]

    def apply(self, x):
        if self.mode == 'linear':
            return self.a * x.reshape((-1,1))
        elif self.mode == 'power':
            return self.b * np.power(x.reshape((-1,1)), self.a)
        elif self.mode == 'logistic':
            return self.k / (1 + np.exp(self.a*x.reshape((-1,1)) + self.b))

    def reciproque(self, y):
        if self.mode == 'linear':
            return y / self.a
        elif self.mode == 'power':
            return np.exp((np.log(y)-np.log(self.b))/self.a)
        elif self.mode == 'logistic':
            return (np.log((self.k/y)-1)-self.b)/self.a
```

The complete file is publicly available on GitLab at:
https://oikolab.terroiko.fr:10001/publications/implication-of-modelling-choices-on-connectivity-estimation/-/blob/main/B_Connectivity_modelling/B123_EUC_LCP_CIR_probabilities_of_connectivity.ipynb

# S11. Maps of connectivity indices variations

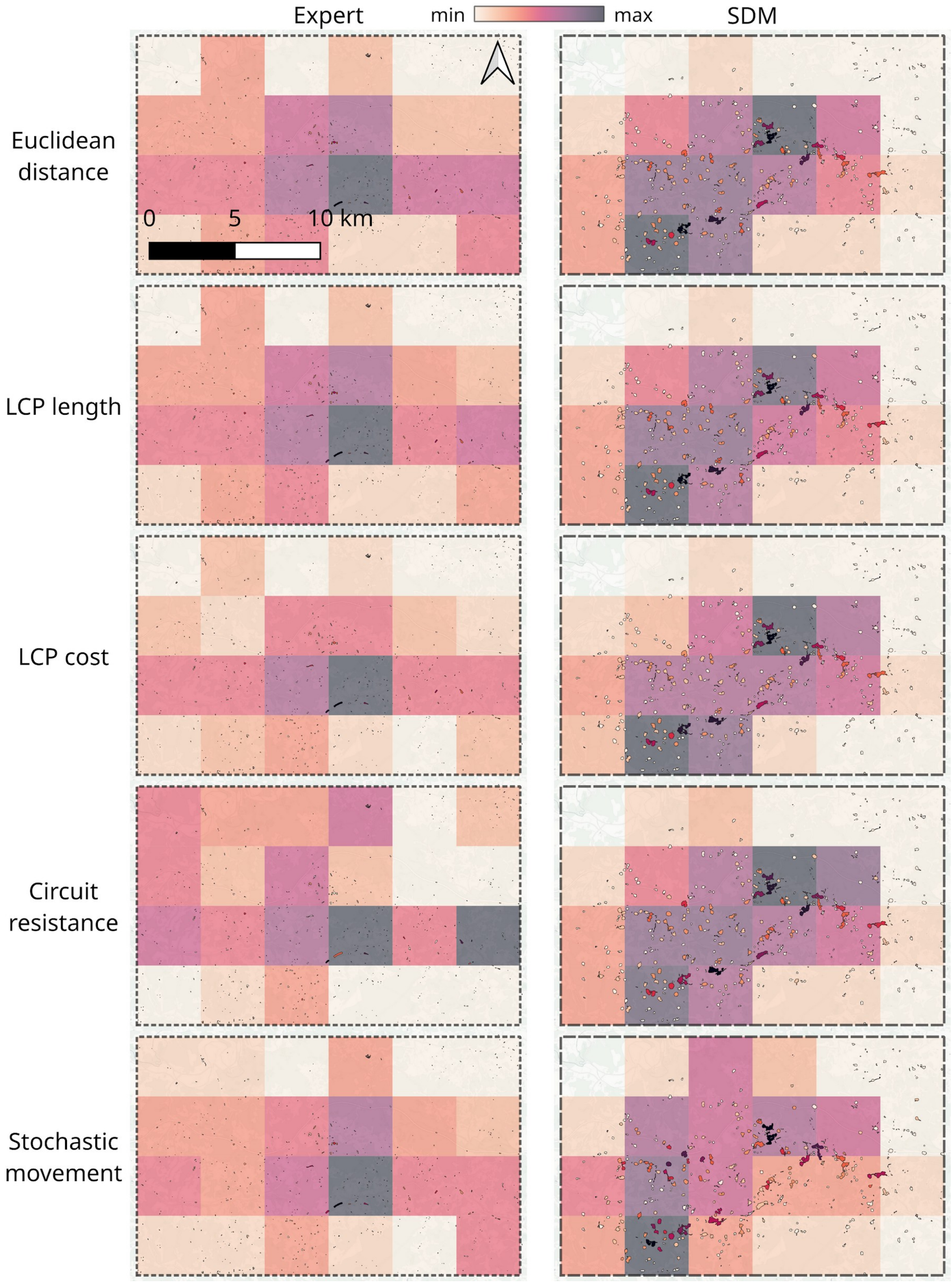

***Fig. S11a*** *Suitable habitats according to dPCflux. The background colour represents the dPCflux average per grid cell.*

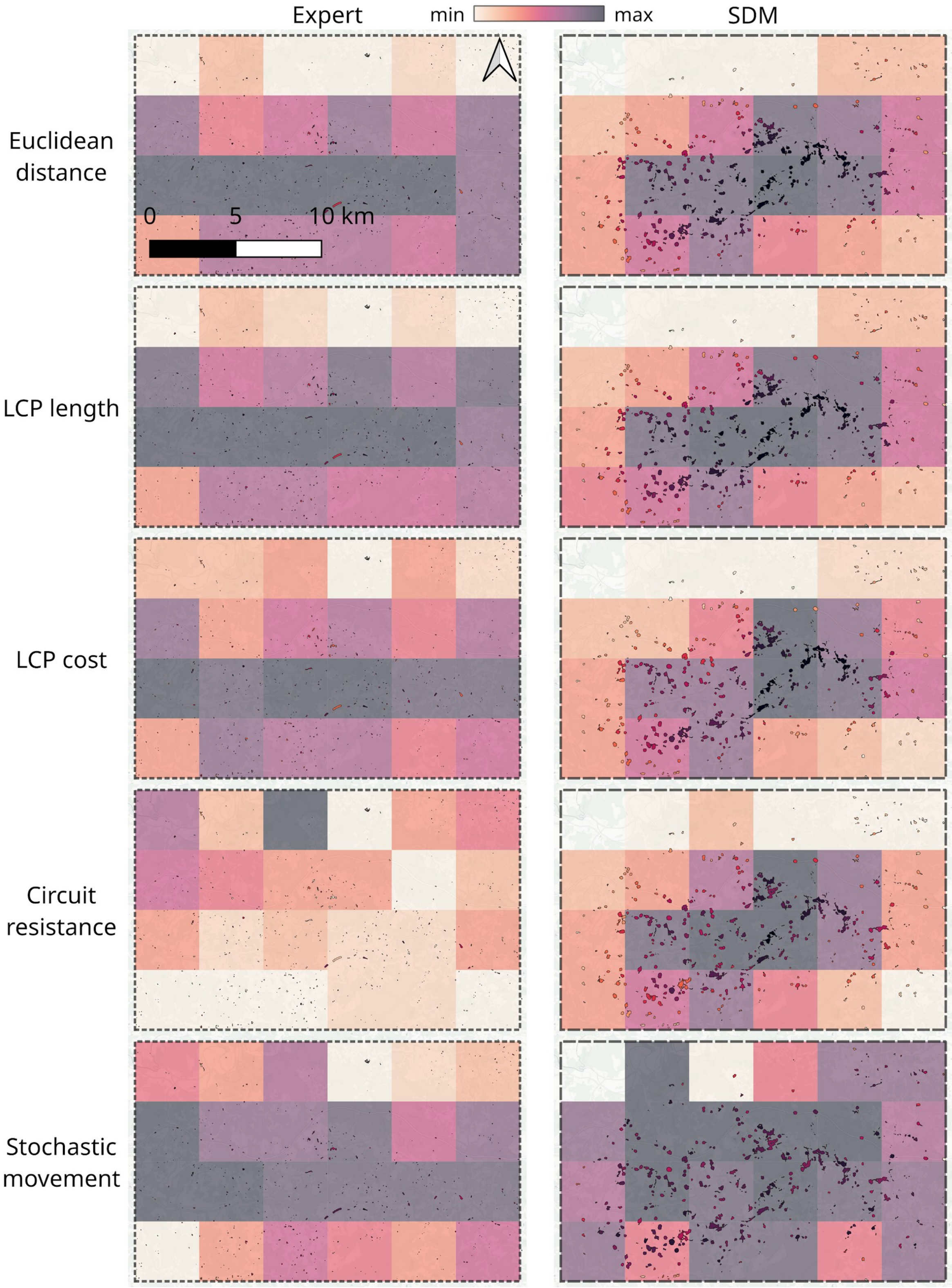

***Fig. S11b*** *Suitable habitats according to F. The background colour represents the F average per grid cell.*

# Supplementary References